\documentclass[fleqn,10pt]{wlscirep}
\usepackage[utf8]{inputenc}
\usepackage[T1]{fontenc}
\usepackage{bm}

\usepackage{booktabs}
\usepackage{adjustbox}
\usepackage{longtable}
\usepackage{caption}
\usepackage{mathtools}
\usepackage{amssymb}
\usepackage{amsmath}
\usepackage{etoolbox}
\usepackage{amsthm}
\usepackage{tocloft}
\usepackage{float}

\addtocontents{toc}{\protect\setcounter{tocdepth}{-1}}

\usepackage{titlesec}
\titleformat{\section}
  {\normalfont\Large\bfseries}
  {\thesection}
  {0em} % 这里设置编号后的间距
  {#1}
  
\allowdisplaybreaks

\title{Evolutionary dynamics of any multiplayer game on regular graphs}

\author[1,*]{Chaoqian Wang}
\author[2,3,4,5]{Matja\v z Perc}
\author[6]{Attila Szolnoki}
\affil[1]{Department of Computational and Data Sciences, George Mason University, Fairfax, VA 22030, USA}
\affil[2]{Faculty of Natural Sciences and Mathematics, University of Maribor, Koro{\v s}ka cesta 160, 2000 Maribor, Slovenia}
\affil[3]{Community Healthcare Center Dr. Adolf Drolc Maribor, Vo{\v s}njakova ulica 2, 2000 Maribor, Slovenia}
\affil[4]{Complexity Science Hub Vienna, Josefst{\"a}dterstra{\ss}e 39, 1080 Vienna, Austria}
\affil[5]{Department of Physics, Kyung Hee University, 26 Kyungheedae-ro, Dongdaemun-gu, Seoul, Republic of Korea}
\affil[6]{Institute of Technical Physics and Materials Science, Centre for Energy Research, P.O. Box 49, H-1525 Budapest, Hungary}
\affil[*]{e-mail: CqWang814921147@outlook.com~(C.~Wang)}

\begin{abstract}
Multiplayer games on graphs are at the heart of theoretical descriptions of key evolutionary processes that govern vital social and natural systems. However, a comprehensive theoretical framework for solving multiplayer games with an arbitrary number of strategies on graphs is still missing. Here, we solve this by drawing an analogy with the Balls-and-Boxes problem, based on which we show that the local configuration of multiplayer games on graphs is equivalent to distributing $k$ identical co-players among $n$ distinct strategies. We use this to derive the replicator equation for any $n$-strategy multiplayer game under weak selection, which can be solved in polynomial time. As an example, we revisit the second-order free-riding problem, where costly punishment cannot truly resolve social dilemmas in a well-mixed population. Yet, in structured populations, we derive an accurate threshold for the punishment strength, beyond which punishment can either lead to the extinction of defection or transform the system into a rock-paper-scissors-like cycle. The analytical solution also qualitatively agrees with the phase diagrams that were previously obtained for non-marginal selection strengths. Our framework thus allows an exploration of any multi-strategy multiplayer game on regular graphs.
\end{abstract}
\begin{document}

\flushbottom
\maketitle

\maketitle

\section*{}
Multi-strategy evolutionary dynamics in nature often lead to diverse and complex phenomena, such as cyclic dominance that is captured by the well-known rock-paper-scissors game~\cite{szolnoki2014cyclic}. Experimental evidence from diverse contexts, ranging from the three-morph mating system of the side-blotched lizard~\cite{sinervo1996rock} and {\it Escherichia coli} populations~\cite{kerr2002local}, to human economic behaviors~\cite{hauert2002volunteering}, demonstrates the occurrence of the rock-paper-scissors cycle in various real-world scenarios. Theoretical models of the rock-paper-scissors cycle have been explored in both two-player~\cite{hofbauer_98} and multiplayer game frameworks~\cite{semmann2003volunteering}, contributing to an understanding of its underlying properties --- as a consequence of strategy diversity, the intransitive interaction may emerge spontaneously. This phenomenon can be illustrated when we extend the basic two-strategy model of the evolution of cooperation by adding additional strategies that punish defectors~\cite{fehr2002altruistic,szolnoki2011phase} or reward cooperators~\cite{sigmund2001reward,wang2021tax}. The additional strategies are necessary when considering more realistic models, which underlines the importance of a multi-strategy approach.

Previous research in evolutionary dynamics primarily focused on two-strategy systems, where the unconditional cooperator and defector strategies represent the fundamental conflict of individual and collective interests~\cite{sigmund2010calculus}. 
While cooperation can maximize mutual benefits, defection, despite offering higher personal payoff, reduces overall benefits to others. Consequently, defection often appears as the dominant strategy. 
An escape route from this dilemma could be a spatially structured population~\cite{nowak1992evolutionary,nowak2006five}, where individuals interact with fixed neighbors but still adopt the strategies of those with higher payoffs. This setting allows cooperation to form clusters, utilizing the advantage of collective payoffs thus resisting the invasion of defection, a concept known as spatial reciprocity~\cite{nowak2010evolutionary}. It is recognized that no simple closed-form solution exists for general evolutionary dynamics in structured populations, unless by chance P=NP, {\it Polynomial time} equals to {\it Nondeterministic Polynomial time}~\cite{ibsen2015computational}. However, in the weak selection limit, where the influence of the game on strategy updates is marginal, analytical solutions have been obtained from infinite~\cite{lieberman2005evolutionary} to finite populations~\cite{taylor2007evolution}, and from regular~\cite{allen2014games,debarre2014social} to arbitrary graphs~\cite{allen2017evolutionary,mcavoy2020social,su2022evolution,su2023strategy}. This line of research has led to the development of evolutionary graph theory~\cite{nowak2010evolutionary}.

In evolutionary graph theory, a widely used mathematical technique is the pair approximation~\cite{gutowitz1987local,matsuda1987lattice,szabo1991correlations,matsuda1992statistical, szabo2007evolutionary}. This method applies to infinite populations on regular graphs and has revealed the well-known `$b/c>k$'
rule, which states that evolution favors cooperation when the benefit-to-cost ratio exceeds the number of neighbors~\cite{ohtsuki2006simple}. Pair approximation is also capable of analyzing more complex models, including unequal interaction and dispersal graphs~\cite{ohtsuki2007breaking}, asymmetric networks~\cite{su2022evolutionasym}, and stochastic games~\cite{su2019evolutionary}, predicting simulation outcomes with high accuracy. Notably, pair approximation has been applied to multi-strategy two-player games~\cite{ohtsuki2006replicator}, leading to the replicator equations for arbitrary $n$-strategy two-player games on a regular graph, as an important extension of the traditional replicator equations used in well-mixed populations~\cite{taylor1978evolutionary}.

Unlike two-player games, multiplayer games exhibit much greater complexity, primarily due to their potentially nonlinear payoff functions~\cite{perc_jrsi13}. In a structured population, multiplayer games require each individual to organize a game within their neighbors and themselves, which implies that individuals participate in games organized by both themselves and their neighbors, thereby interacting with second-order neighbors. Such interactions lead to higher-order interactions~\cite{alvarez2021evolutionary,battiston2021physics}, which cannot be simply reduced to a superposition of pairwise interactions. The complexity of multiplayer games can also be illustrated from the perspective of structure coefficients on graphs: a two-strategy two-player game needs only one structure coefficient~\cite{tarnita2009strategy}, a multi-strategy two-player game requires three~\cite{tarnita2011multiple,mcavoy2022evaluating}, but a two-strategy ($k+1$)-player game needs as many as $k$ structure coefficients~\cite{wu2013dynamic,mcavoy2016structure}. The number of potential equilibrium points in general multiplayer games also indicates their complexity~\cite{duong2016expected,duong2016analysis}. Even so, in the absence of triangle motifs, two-strategy multiplayer games can still be theoretically analyzed using pair approximation~\cite{li2014cooperation,li2016evolutionary}, whose results are consistent with predictions obtained by other more precise methods~\cite{su2019spatial,wang2023inertia}.

With two-strategy two-player~\cite{ohtsuki2006simple}, multi-strategy two-player~\cite{ohtsuki2006replicator}, and two-strategy multiplayer games~\cite{li2016evolutionary} all thoroughly studied, the analytical solution for multi-strategy multiplayer games on graphs remains unexplored. The range of potential models for multiplayer games with more than two strategies is vast, drawing from co-evolutionary strategies such as punishment~\cite{sigmund2007punish,helbing2010punish,szolnoki2011phase,szolnoki2011competition}, reward~\cite{sigmund2001reward,rand2009positive,hilbe2010incentives}, and the loner strategy~\cite{szabo2002phase,hauert2002volunteering}. Multi-strategy systems in multiplayer games have unique characteristics that multi-strategy two-player games do not capture: the payoff function can be nonlinear. For example, in pool punishment, the payoff structure depends solely on whether there is at least one punishing player among the $k+1$ players. This uniqueness reinforces the significance of studying multi-strategy multiplayer games.

\begin{figure}
	\centering
		\includegraphics[width=\textwidth]{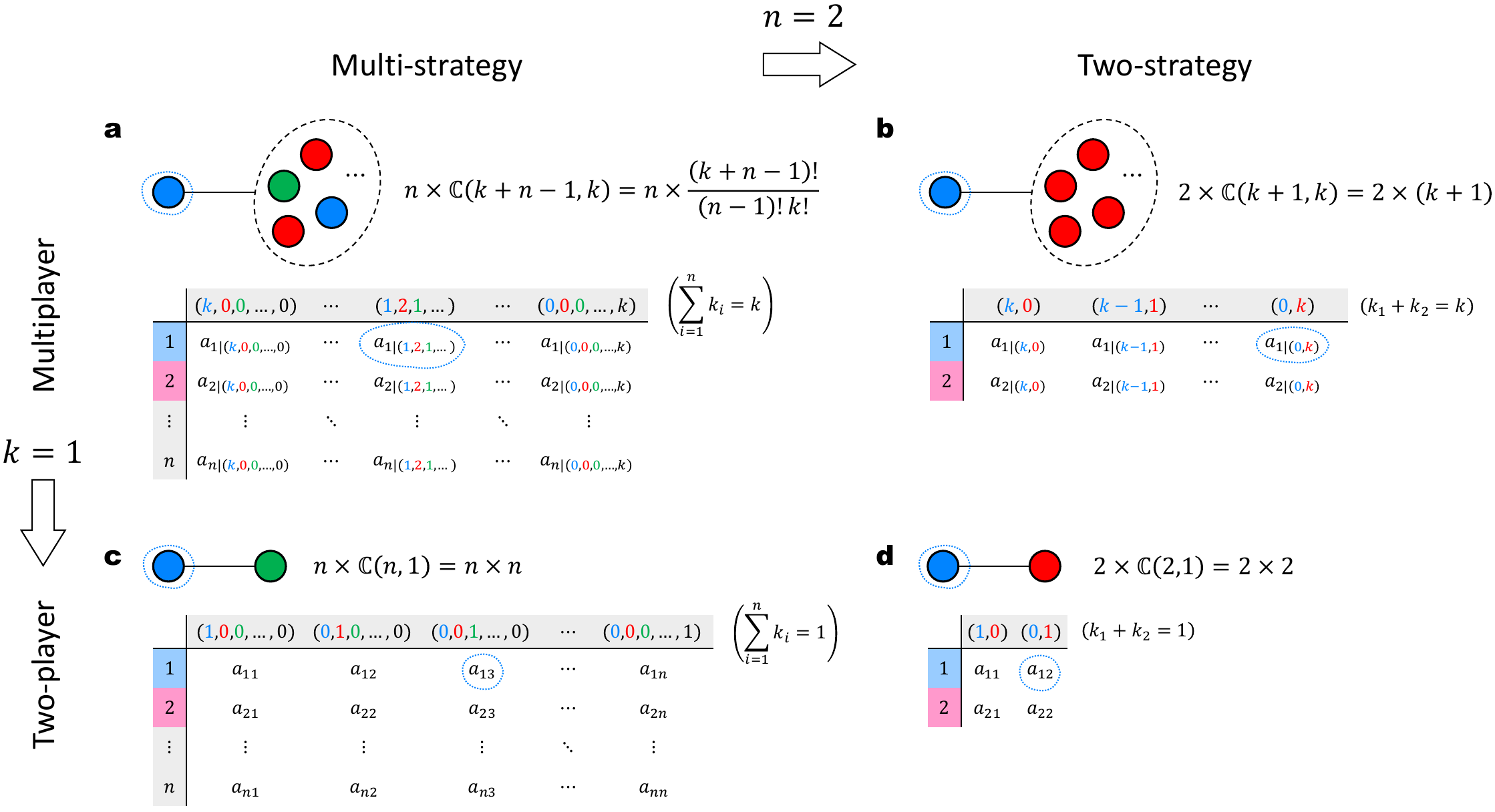}
	\caption{\textbf{Generalized payoff matrix for multi-strategy multiplayer games, reducible to two-strategy or two-player formats.}
 \textbf{a}, Payoff matrix for $n$-strategy ($k+1$)-player games: with $n$ strategies and $k$ co-players, the matrix size is $n\times \mathbb{C}(k+n-1,k)$.
 \textbf{b}, Payoff matrix for $2$-strategy ($k+1$)-player games~\cite{li2016evolutionary}: with $n=2$ strategies and $k$ co-players, the matrix size is $2\times \mathbb{C}(k+1,k)$.
 \textbf{c}, Payoff matrix for $n$-strategy $2$-player games~\cite{ohtsuki2006replicator}: with $n$ strategies and $k=1$ co-players, the matrix size is $n\times \mathbb{C}(n,1)$.
 \textbf{d}, Payoff matrix for $2$-strategy $2$-player games~\cite{ohtsuki2006simple}: with $n=2$ strategies and $k=1$ co-players, the matrix size is $2\times \mathbb{C}(2,1)$.} \label{fig_matrixdemo}
\end{figure}

However, previous research on these games on graphs has largely been limited to numerical simulations, which do not allow for the exploration of the complete parameter space. In the absence of mathematical tools for evolutionary graph theory in multi-strategy multiplayer games, recent studies have attempted to bypass this challenge by incorporating the third strategy within the existing two strategies. For instance, punishing or rewarding behaviors have been added to the existing cooperation strategy in the traditional two-strategy system~\cite{wang2022decentralized,sun2023state}. This approach allows for the examination of additional mechanisms like punishment and reward within the two-strategy system's framework. Yet, these alternative attempts still could not capture further rich dynamics, such as cyclic dominance, which is only possible in systems with at least three strategies.

\begin{figure}
	\centering
		\includegraphics[width=.9\textwidth]{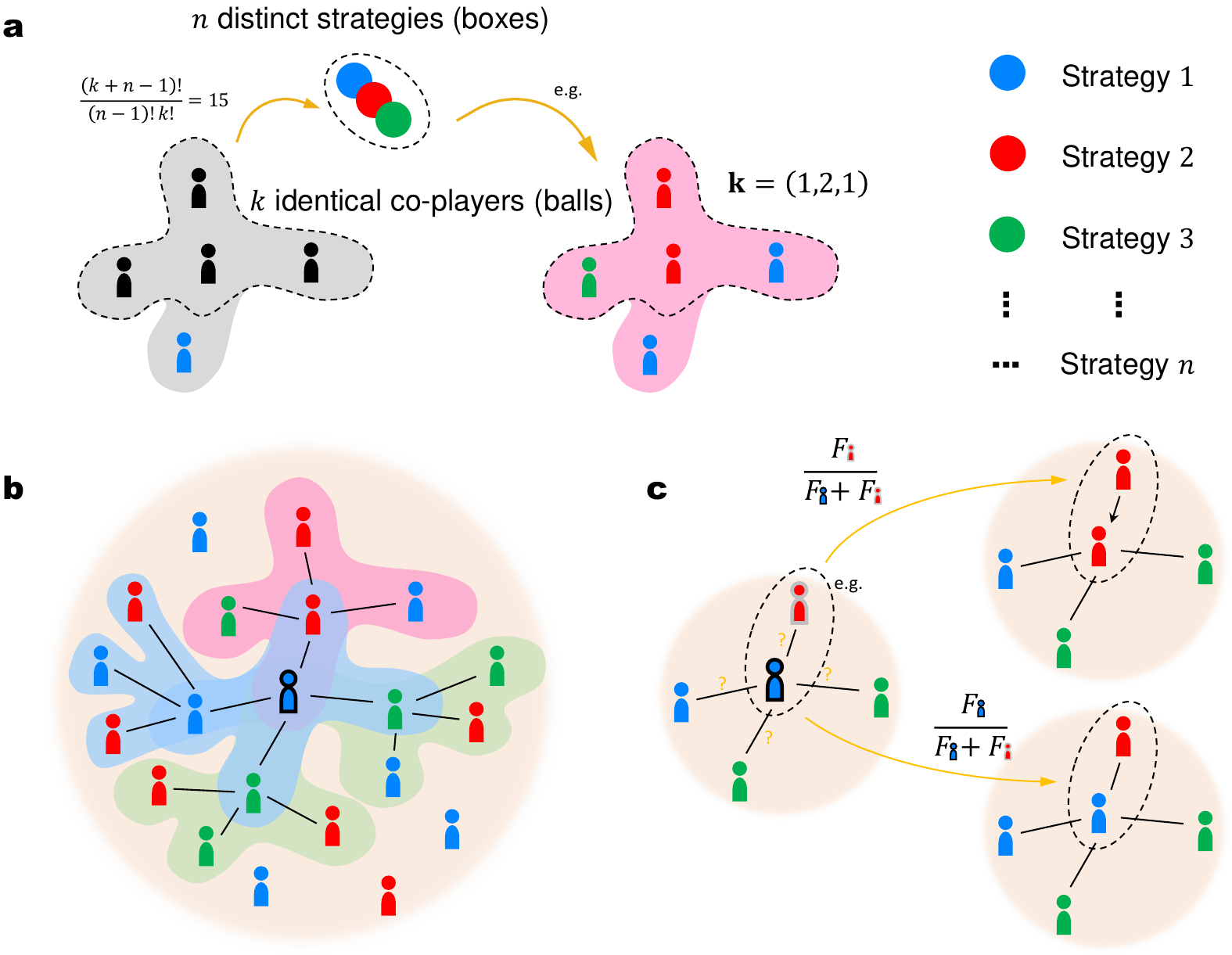}
	\caption{\textbf{The Balls-and-Boxes problem, payoff calculation, and strategy updates.}
 \textbf{a}, To determine a co-player configuration, we distribute $k$ identical co-players (balls) into $n$ distinct strategies (boxes), corresponding to the classic Balls-and-Boxes problem. For instance, with $k=4$ and $n=3$, there are $15$ possible co-player configurations.
 \textbf{b}, An individual accumulates payoffs from $1+k$ multiplayer games, organized by itself and its $k$ neighbors. Each multiplayer game involves $k+1$ players, including the organizer and its neighbors.
 \textbf{c}, Strategy updates in an individual are governed by the pairwise comparison rule. A random neighbor is selected, and the focal individual either adopts the neighbor's strategy or maintains its original strategy, based on a probability proportional to the fitness (transformed from payoffs) in the pair.} \label{fig_demo}
\end{figure}

In this work, we provide an analytical framework that addresses the gaps in multi-strategy multiplayer games in the realm of evolutionary graph theory. Inspired by the Balls-and-Boxes problem, we demonstrate that for a given multi-strategy multiplayer game, counting the co-player configurations of a focal individual is equivalent to distributing $k$ identical co-players into $n$ distinct strategies (Fig.~\ref{fig_matrixdemo}). On this basis, we develop a bottom-up approach for calculating the group-based payoff of individuals on regular graphs, deriving replicator equations on regular graphs in the weak selection limit and the absence of triangle motifs. Our results include two commonly used update rules, namely pairwise comparison (PC)~\cite{szabo1998evolutionary} and death-birth (DB)~\cite{ohtsuki2006simple}, applicable to arbitrary multiplayer games with any multi-strategy space in structured populations, where each individual has the same number of neighbors. Using the punishment mechanism~\cite{fehr2002altruistic,ohtsuki2009indirect,sigmund2010social} in the context of the tragedy of the commons~\cite{hardin1968tragedy} as an example, we explore the well-known second-order free-riding problem analytically, obtaining an accurate threshold of punishment strength necessary to resolve the social dilemma in structured populations. Additionally, our theoretical solutions can qualitatively reproduce the phase diagrams observed in previous numerical simulation studies under non-marginal selection strength.

\section*{Results}\label{sec_model}

\subsection*{Model overview}
We consider an infinite population on a regular graph, where each individual has $k$ neighbors. An individual can adopt one of $n$ strategies, labeled by the numbers $1, 2, \dots, n$. On a regular graph, the number of co-players in every multiplayer game is equivalent to the constant number $k$ of neighbors. For a given individual, suppose that there are $k_1$ co-players employing strategy $1$, $k_2$ co-players employing strategy $2$, and so on, up to $k_n$ co-players employing strategy $n$. In this context, the co-player configuration of an individual can be represented by $\mathbf{k}=(k_1,k_2,\dots,k_n)$, which satisfies the condition $\sum_{l=1}^n k_l=k$. As illustrated in Fig.~\ref{fig_demo}\textbf{a}, counting the number of possible configurations of $\mathbf{k}$ is analogous to the classic Balls-and-Boxes problem, distributing $k$ identical balls (i.e., co-players) into $n$ distinct boxes (i.e., strategies), allowing for the possibility of empty boxes (e.g., $k_1=0$). Hence, there are $\mathbb{C}(k+n-1,k)=(k+n-1)!/[(n-1)!k!]$ possible configurations of co-player strategy configurations $\mathbf{k}$.

% \begin{figure}
% 	\centering
% 		\includegraphics[width=.9\textwidth]{demo.pdf}
% 	\caption{\textbf{The Balls-and-Boxes problem, payoff calculation, and strategy updates.}
%  \textbf{a}, To determine a co-player configuration, we distribute $k$ identical co-players (balls) into $n$ distinct strategies (boxes), corresponding to the classic Balls-and-Boxes problem. For instance, with $k=4$ and $n=3$, there are $15$ possible co-player configurations.
%  \textbf{b}, An individual accumulates payoffs from $1+k$ multiplayer games, organized by itself and its $k$ neighbors. Each multiplayer game involves $k+1$ players, including the organizer and its neighbors.
%  \textbf{c}, Strategy updates in an individual are governed by the pairwise comparison rule. A random neighbor is selected, and the focal individual either adopts the neighbor's strategy or maintains its original strategy, based on a probability proportional to the fitness (transformed from payoffs) in the pair.} \label{fig_demo}
% \end{figure}

Interaction occurs between an individual and its $k$ co-players. In a multiplayer game involving the focal individual and $k$ co-players, the payoff is uniquely determined by the strategy of the focal individual and the strategy configuration of the co-players. For a focal individual employing strategy $i$ with the co-player configuration $\mathbf{k}$, its payoff is denoted by $a_{i|\mathbf{k}}$. It can be observed that the `generalized payoff matrix' comprises $n \times \mathbb{C}(k+n-1,k)$ elements represented by $a_{i|\mathbf{k}}$ through all possible focal strategies $i=1,2,\dots,n$ and co-player strategy configurations $\mathbf{k}$. For two-strategy two-player games ($n=2$, $k=1$), the number of elements in the payoff matrix reduces to $2\times \mathbb{C}(2,1)=4$; for multi-strategy two-player games ($k=1$), it reduces to $n \times \mathbb{C}(n,1)=n^2$; for two-strategy multiplayer games ($n=2$), it reduces to $2 \times \mathbb{C}(k+1,k)=2(k+1)$ (Fig.~\ref{fig_matrixdemo}).

The accumulated payoff of a focal individual is collected from the $1+k$ games organized by itself and its neighbors, as depicted in Fig.~\ref{fig_demo}\textbf{b}. Upon obtaining the accumulated payoffs $\pi$, we convert them into fitness, denoted as $F=\exp{(\delta\pi)}$~\cite{mcavoy2020social,wang2023inertia,wang2023conflict,wang2023greediness}. Strategies that yield higher fitness are more likely to reproduce. Here, $\delta\to 0^+$ represents a weak selection limit. The rationale behind weak selection is that, in reality, many factors other than the investigated game influence the probability of reproduction~\cite{ohtsuki2006simple}.

There are various commonly used strategy update rules. For simplicity, we focus on the pairwise comparison (PC) rule~\cite{szabo1998evolutionary} in the main text (another well-known rule, the death-birth, is discussed in Supplementary Information). During each elementary step, an individual $A$ and one of its neighbors $B$ are randomly selected from the population. Their payoffs are computed as $\pi_A$ and $\pi_B$ and then transformed into fitness values $F_A$ and $F_B$. Individual $A$ adopts the strategy of individual $B$ with a probability proportional to their fitness in the pair,
\begin{equation}\label{eq_Fermi}
    W=\frac{F_B}{F_A+F_B}=\frac{1}{1+\exp{[-\delta(\pi_B-\pi_A)]}}.
\end{equation}
Or, individual $A$ keeps its own strategy with the remaining probability $F_A/(F_A+F_B)$. Eq.~(\ref{eq_Fermi}) indicates that individual $A$ has a marginal tendency to either maintain its own strategy or adopt the one of individual $B$, depending on who has higher fitness. The evolution of strategies under the PC update process is illustrated in Fig.~\ref{fig_demo}\textbf{c}.

\subsection*{Group-based payoff with any number of strategies}
To formally analyze the evolutionary dynamics, we construct the system as described in Supplementary Note~\ref{sec_system}. According to pair approximation~\cite{ohtsuki2006simple}, there are two key concepts, the frequency of $i$-players (i.e., individuals employing strategy $i$), denoted as $x_i$, where $\sum_{i=1}^n x_i=1$, and the probability of an $i$-player being adjacent to a $j$-player, denoted by $q_{i|j}$, with $\sum_{j=1}^n q_{j|i}=1$. By separating different time scales, we find that $q_{i|j}=x_i(k-2)/(k-1) +\theta_{ij}/(k-1)$, where $\theta_{ij}=1$ if $i=j$ and $\theta_{ij}=0$ otherwise (Supplementary Note~\ref{sec_edgePC}). In other words, the value of $q_{i|j}$ can be determined by the value of $x_i$.

To express necessary computations, we introduce a variation of $\mathbf{k}$, denoted as $\mathbf{k}_{+l}=(k_1,k_2,\dots,k_l+1,\dots,k_n)$,
where $\sum_{l=1}^n k_l=k-1$. This represents a co-player configuration in which there is at least one $l$-player. Among the remaining $k-1$ co-players, the numbers of players adopting strategies $1,2,\dots,n$ are $k_1,k_2,\dots,k_n$, respectively. 

We label the payoff that an individual obtains in a multiplayer game as the single-game payoff. $\langle a_{X|\mathbf{k}}\rangle_Y$ is used to denote the expected single-game payoff for an $X$-player over the possible co-player configurations $\mathbf{k}$, where the $k$ members in $\mathbf{k}$ are neighbors of a $Y$-player, as defined by Eq.~(\ref{eq_<a>}) in the Methods. Similarly, the notation $\langle a_{X|\mathbf{k}_{+l}}\rangle_Y$ differs in that it is over $k-1$ unknown members in the possible co-player configurations $\mathbf{k}_{+l}$, with one known $l$-player.

Furthermore, we use the notation $\langle \pi_X^\mathbf{k} \rangle$ to represent the expected accumulated payoff of an $X$-player obtained in the $1+k$ games organized by the player and its neighbors, across all possible neighbor configurations $\mathbf{k}$ of the $X$-player, defined by Eq.~(\ref{eq_<pi_j>}) in the Methods. Similarly, $\langle \pi_X^{\mathbf{k}_{+i}} \rangle$ denotes the expected accumulated payoff over the configurations where the remaining $k-1$ neighbors are unknown besides a known $i$-player, as defined by Eq.~(\ref{eq_<pi_i|j>}) in the Methods.

Through bottom-up calculations from the microscopic level (Methods), we establish the following relationship between the expected accumulated and single-game payoffs. For $i$-players, the relation is given by
\begin{equation}\label{eq_<pi_i>_<a>}
    \langle \pi_i^\mathbf{k} \rangle
    =\langle a_{i|\mathbf{k}}\rangle_i+
    k\sum_{l=1}^n q_{l|i} \langle a_{i|\mathbf{k}'_{+l}}\rangle_l.
\end{equation}
Intuitively, the expected accumulated payoff of $i$-players, $\langle \pi_i^\mathbf{k} \rangle$, is composed by the expected single-game payoff from the game they organize, $\langle a_{i|\mathbf{k}}\rangle_i$, and the games organized by their $k$ neighbors, $k\sum_{l=1}^n q_{l|i} \langle a_{i|\mathbf{k}'_{+l}}\rangle_l$. Here, the different notation $\mathbf{k}'=(k'_1,k'_2,\dots,k'_n)$ from $\mathbf{k}$ is an independent configuration to clarify the priority in the summation.

A further concept is the expected accumulated payoff of a $j$-player who has at least one $i$-player as a neighbor, which is related to the expected single-game payoff as follows:
\begin{equation}\label{eq_<pi_j>+i_<a>}
    \langle \pi_j^{\mathbf{k}_{+i}} \rangle
    =\langle a_{j|\mathbf{k}_{+i}}\rangle_j+
    \langle a_{j|\mathbf{k}'_{+i}}\rangle_i+
    (k-1)\sum_{l=1}^n q_{l|j}\langle a_{j|\mathbf{k}'_{+l}}\rangle_l.
\end{equation}
Here, $\langle a_{j|\mathbf{k}_{+i}}\rangle_j$, $\langle a_{j|\mathbf{k}'_{+i}}\rangle_i$, and $(k-1)\sum_{l=1}^n q_{l|j}\langle a_{j|\mathbf{k}'_{+l}}\rangle_l$ are the expected single-game payoff from the game organized by the $j$-player itself, the game organized by the fixed $i$-player neighbor, and the games organized by the remaining $k-1$ neighbors of the $j$-player.

\subsection*{General replicator equations}
The evolution of frequencies $x_1,x_2,\dots,x_n$ can be deduced through the microscopic strategy update process. Specifically, in an infinite population, i.e., $N\to\infty$, a single unit of time comprises $N$ elementary steps, ensuring that each individual has an opportunity to update their strategy. During each elementary step, the frequency of $i$-players increases by $1/N$ when a focal $j$-player (where $j\neq i$) is chosen to update its strategy and is replaced by an $i$-player. Similarly, the frequency of $i$-players decreases by $1/N$ when a focal $i$-player is selected to update its strategy and the player who takes the position is not an $i$-player. Based on this perception, we derive a simple form of the replicator equations for $i=1,2,\dots,n$ in the weak selection limit (\ref{sec_PC}):
\begin{equation}\label{eq_replicatorPC}
    \dot{x}_i=\frac{\delta}{2}x_i \left(
    \langle \pi_i^\mathbf{k} \rangle-\sum_{j=1}^n q_{j|i}\langle \pi_j^{\mathbf{k}_{+i}} \rangle
    \right).
\end{equation}

We find that Eq.~(\ref{eq_replicatorPC}) offers an intuitive understanding, if we introduce the following two concepts: (1) $\pi_i^{(0)}=\langle \pi_i^\mathbf{k} \rangle$, the expected accumulated payoff of the $i$-player (zero steps away on the graph), and (2) $\pi_i^{(1)}=\sum_{j=1}^n q_{j|i}\langle \pi_j^{\mathbf{k}_{+i}} \rangle$, the expected accumulated payoff of the $i$-player's neighbors (one step away on the graph). These concepts suggest that $\dot{x}_i\propto x_i(\pi_i^{(0)}-\pi_i^{(1)})$. Under pairwise comparison, the reproduction rate of $i$-players is dependent on how their accumulated payoff exceeds that of their neighbors. In essence, the evolution of $x_i$ is the competition between an individual and its first-order neighbors, which aligns with the results obtained by a different theoretical framework in two-strategy systems~\cite{allen2014games,allen2017evolutionary,su2019evolutionary}. We further extend it to $n$-strategy systems in the framework of pair approximation. We also verify that the death-birth rule is essentially the competition between an individual and its second-order neighbors for $n$-strategy systems (Supplementary Information).

Applying Eqs.~(\ref{eq_<pi_i>_<a>}) and (\ref{eq_<pi_j>+i_<a>}) to Eq.~(\ref{eq_replicatorPC}), we can transform the expected accumulated payoff in the replicator equations into the expected single-game payoff, as shown in Eq.~(\ref{eq_alevel3_PC}) in the Methods, which keeps the simplest irreducible computational complexity given the payoff structure $a_{i|\mathbf{k}}$. In particular, we only need to calculate two types of quantities, $\langle a_{i|\mathbf{k}_{+j}}\rangle_i$ and $\langle a_{i|\mathbf{k}_{+j}}\rangle_j$ for $i,j=1,2,\dots,n$, based on the given payoff structure $a_{i|\mathbf{k}}$. The diagonal elements of these quantities coincide, as demonstrated in Eqs.~(\ref{eq_aiji_type}) and (\ref{eq_aijj_type}) in the Methods. Therefore, for any given payoff structure $a_{i|\mathbf{k}}$, there are at most $(2n-1)n$ distinct quantities to calculate manually when determining the replicator equations. The computational complexity is thus $\mathrm{O}(n^2)$, square of the number of strategies, which can be solved within polynomial time. We also find that the computational complexity under the death-birth rule is $\mathrm{O}(n^3)$, cubic of the number of strategies, which can also be solved within polynomial time (Supplementary Information).

For specific payoff structures, the computational complexity can be further reduced. A common example is linear systems. In such systems, the payoff function includes at most linear terms in $k_1,k_2,\dots,k_n$. This allows us to express the general payoff function as $a_{i|\mathbf{k}}=\sum_{j=1}^n b_{ij}k_j+c_i$, where $b_{ij}$ represents the coefficient of the linear term and $c_i$ is the constant term for $i,j=1,2,\dots,n$. Applying this special payoff structure to Eq.~(\ref{eq_alevel3_PC}) in the Methods, we can obtain a simplified form of the replicator equation for linear systems,
\begin{equation}\label{eq_linear_PC}
    \dot{x}_i=\frac{\delta(k-2)}{2(k-1)}x_i\left(
    (k+1)(\Bar{\pi}_i-\Bar{\pi})
    +3\sum_{j=1}^n x_j (b_{ii}-b_{ij}-b_{ji}-b_{jj})
    +6\sum_{j=1}^n \sum_{l=1}^n x_j x_l b_{jl}
    \right).
\end{equation}
Here, $\Bar{\pi}_i$ and $\Bar{\pi}$ denote the mean payoff of $i$-players and all players in a well-mixed population, which can be directly calculated using the traditional replicator dynamics approach (Methods). 

As a frequently studied example, the public goods game involves $n=2$ strategies within a linear payoff structure. Strategy 1, cooperation ($C$), pays a cost $c$ which is multiplied by a synergy factor $r$ and distributed among all $k+1$ players, while strategy 2, defection ($D$), pays nothing. The payoff structure can be expressed as $b_{11}=b_{21}=rc/(k+1)$, $b_{12}=b_{22}=0$, $c_1=rc/(k+1)-c$, $c_2=0$. Consequently, $\dot{x}_i\propto x_i (\Bar{\pi}_i-\Bar{\pi})$, indicating that evolution favors cooperation when $r>k+1$ (Supplementary Note~\ref{sec_PGG}). Coincidentally, the public goods game exhibits an equivalence between well-mixed and structured populations under pairwise comparison~\cite{zhang2023cooperation}, a phenomenon not necessarily observed under other update rules~\cite{li2014cooperation}. This equivalence provides a unique opportunity: when introducing additional strategies into the public goods game, the distinct effects of these new strategies in structured populations can be isolated without interference from the existing two strategies. For a general condition when pairwise comparison equates well-mixed and structured populations, we refer to the Supplementary Note~\ref{sec_PCequateWMandST}. 

We apply the multi-strategy multiplayer framework to various additional mechanisms in public goods games, including punishment~\cite{szolnoki2011competition,szolnoki2011phase} ($n=3$), reward~\cite{szolnoki2010reward} ($n=3$), and multi-stage investment~\cite{szolnoki2022tactical} ($n=4$) (\ref{sec_appli}). Here, we present the applications to two punishment types, peer and pool punishment, by which we revisit the well-known second-order free-rider problem in structured populations.

\subsection*{Peer punishment in public goods games}

\begin{figure}
	\centering
		\includegraphics[width=\textwidth]{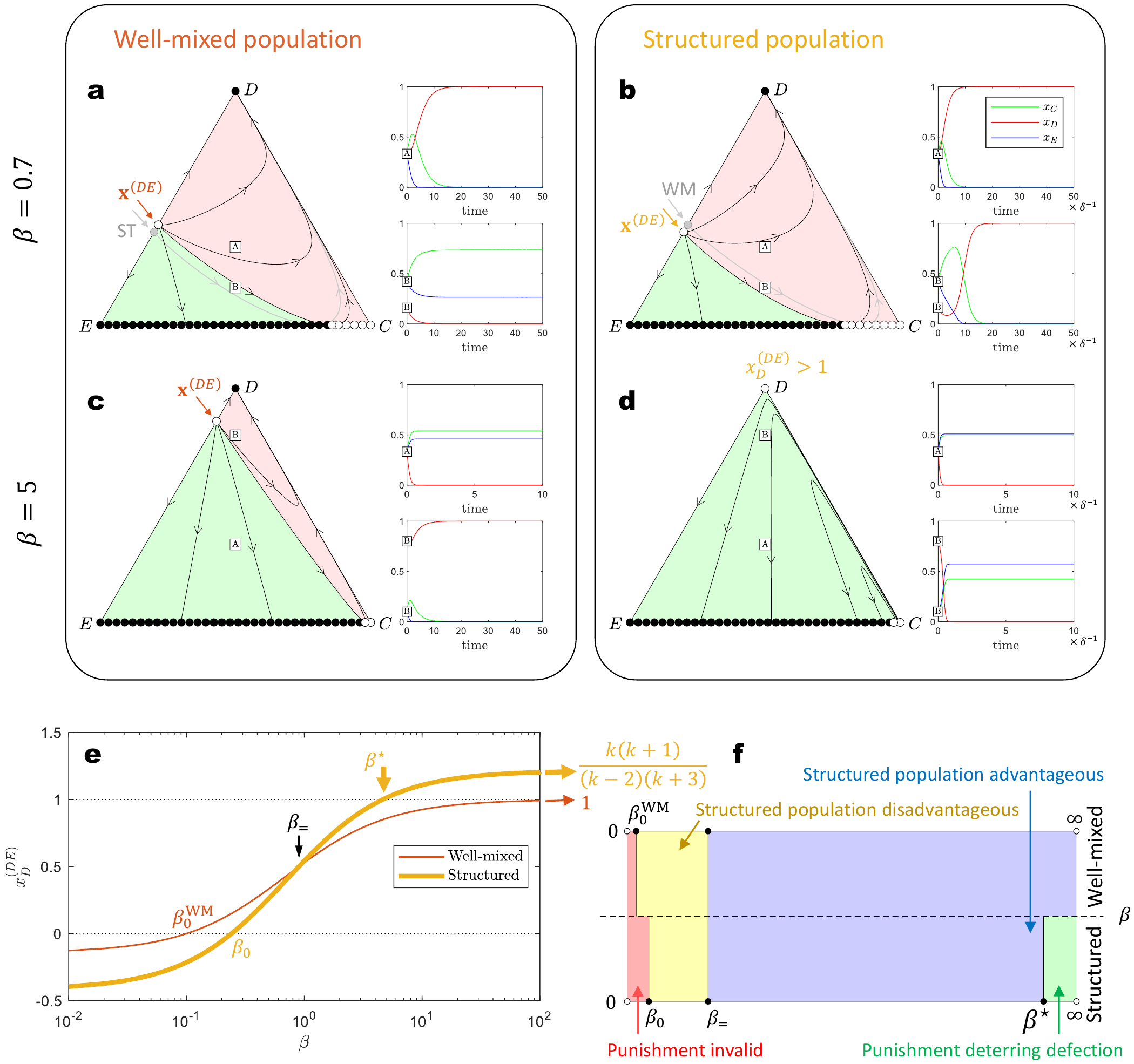}
	\caption{\textbf{Peer punishment can resolve the social dilemma of public goods game in structured populations.} The traditional replicator dynamics produce results for well-mixed populations, and our framework allows for exploring the dynamics in structured populations. \textbf{a} and \textbf{b}, The state space is bifurcated by $\mathbf{x}^{(DE)}$ and $\mathbf{x}_\star^{(CE)}$. The final state, either $D$ or $(C+E)_\text{V}$, is determined by the initial conditions. Under mild punishment ($\beta=0.7$), a structured population hinders cooperation by reducing the state space leading to the $(C+E)_\text{V}$ outcome. \textbf{c} and \textbf{d}, Conversely, with strong punishment ($\beta=5$), structured populations consistently result in the extinction of defection, thereby resolving the social dilemma in public goods games. In contrast, the state space in well-mixed populations remains divided into two distinct basins. \textbf{e}, As the punishment strength $\beta$ increases, $x_D^{(DE)}$ increases, expanding the initial space leading to the $(C+E)_\text{V}$ outcome. In well-mixed populations, $x_D^{(DE)}\to 1$ as $\beta\to \infty$, and the basin leading to defection cannot be completely eliminated. However, in structured populations, $x_D^{(DE)}\to k(k+1)/[(k-2)(k+3)]>1$ when $\beta>\beta^\star$, invariably resulting in the extinction of defection. \textbf{f}, The diagram of the different effects of punishment in well-mixed versus structured populations. Structured populations are advantageous in promoting cooperation under strong punishment but are less effective when the punishment is mild. \textbf{Input parameters}: $r=3$, $c=1$, $\alpha=0.7$, $k=4$.} 
	\label{fig_peer}
\end{figure}

\begin{figure}[t]
	\centering
		\includegraphics[width=\textwidth]{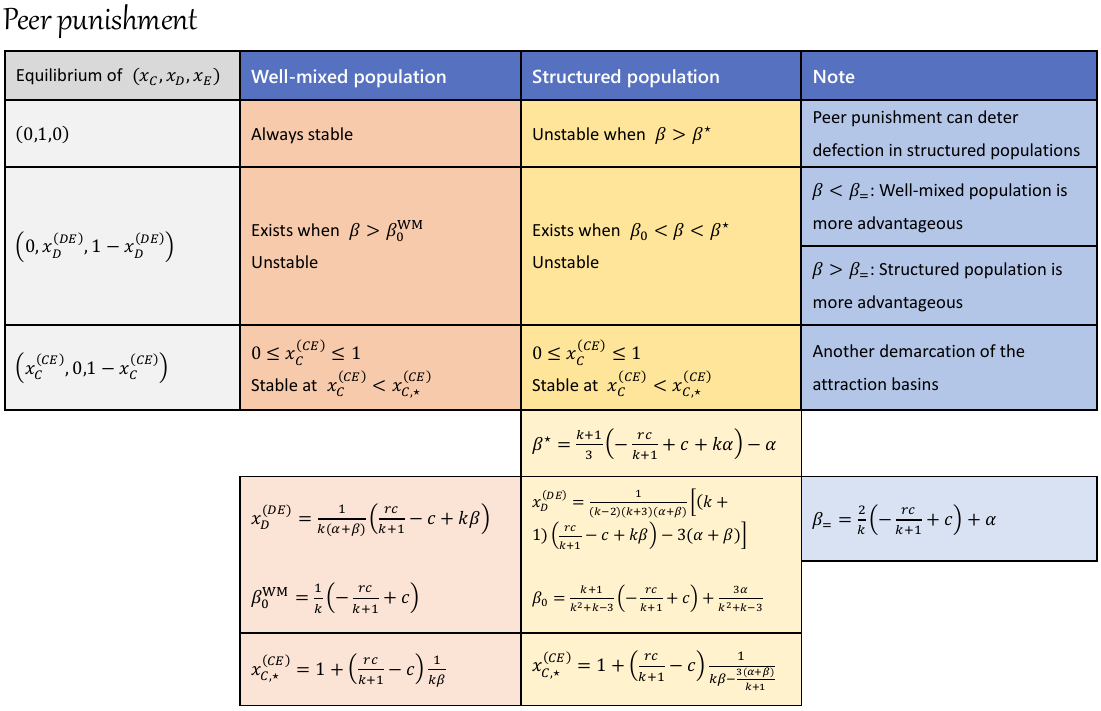}
	\caption{\textbf{Analytical results of public goods game with peer punishment in both well-mixed and structured populations.}
    This table summarizes the equilibrium points and their stability by analyzing the peer punishment system. The analytical forms of all key concepts are presented. See Supplementary Note~\ref{sec_peer} for details of the analysis.} \label{fig_peer_table}
\end{figure}

\begin{figure}
	\centering
		\includegraphics[width=.95\textwidth]{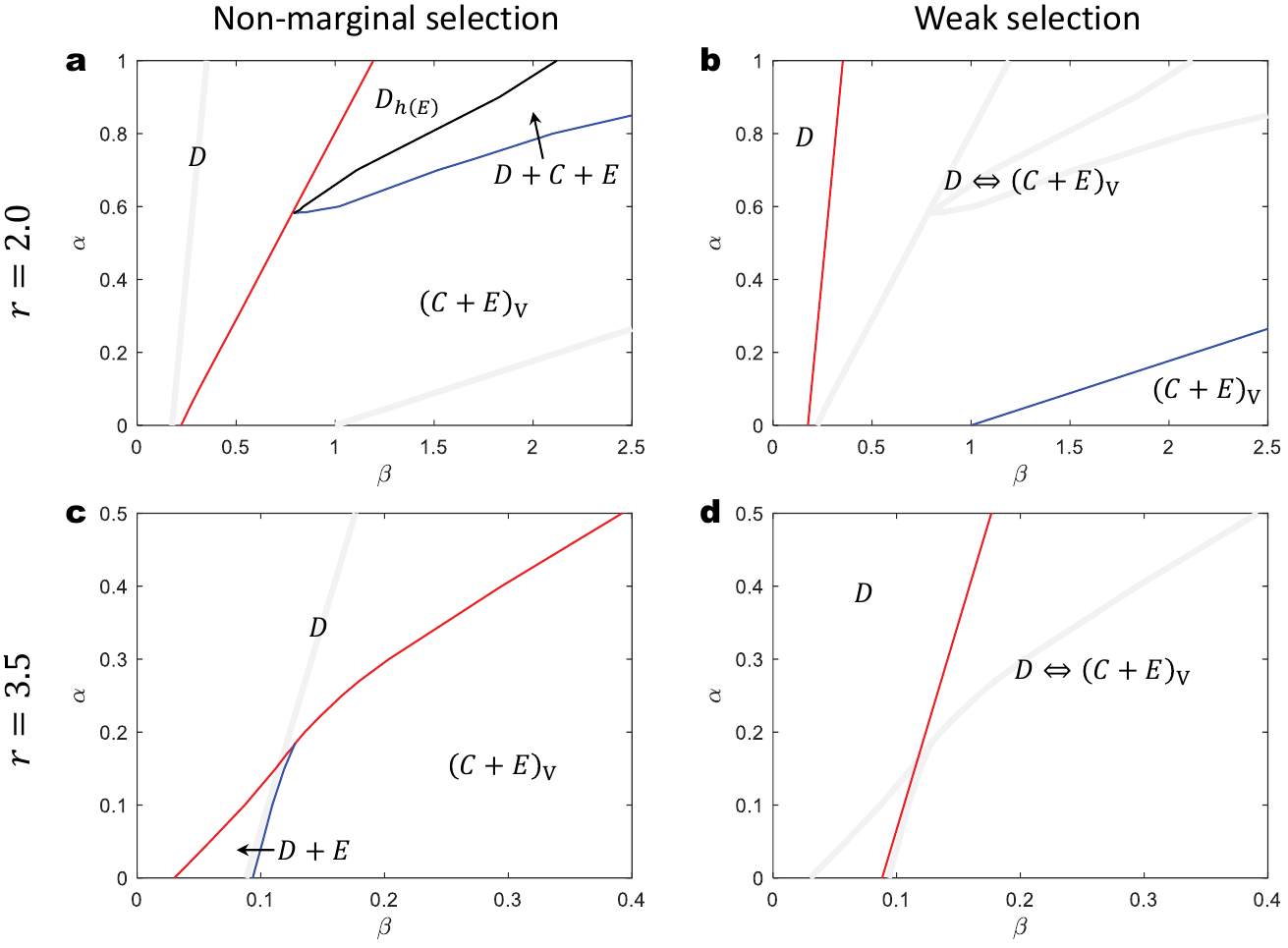}
	\caption{\textbf{Phase diagrams of the system behavior with respect to punishing cost $\alpha$ and fine $\beta$ are qualitatively similar under non-marginal and weak selection strength.} 
    \textbf{a} and \textbf{c} (the data are in agreement with those published in Figs.~3 and 4 from ref.~\cite{szolnoki2011competition}), Numerical simulations under non-marginal selection ($\delta=2$). Here, the different phases are defined as follows: $D$---only $D$ exists; $D_{h(E)}$---only $D$ or $E$ exists based on certain probabilities; $D+C+E$---$D$, $C$, and $E$ coexist; $(C+E)_\text{V}$---$C$ and $E$ coexist like in the Voter model. 
    \textbf{b} and \textbf{d}, The phase diagram created by analytical $\beta_0$ and $\beta^\star$ under weak selection ($\delta\to 0^+$), where $\beta_0$ divides the $D$ and $D\Leftrightarrow (C+E)_\text{V}$ phases, $\beta^\star$ separates the $D\Leftrightarrow (C+E)_\text{V}$ and $(C+E)_\text{V}$ phases. Specifically, in \textbf{b}, $\beta_0=3/17+(3/17)\alpha$ (red), $\beta^\star=1+(17/3)\alpha$ (blue); in \textbf{d}, $\beta_0=3/34+(3/17)\alpha$. The definition of the $D\Leftrightarrow (C+E)_\text{V}$ phase---the system finally evolves to either the $D$ phase or the $(C+E)_\text{V}$ phase, depending on initial conditions. \textbf{Other parameters:} $c=1$, $k=4$.} 
	\label{fig_phase_peer}
\end{figure}

In public goods games with peer punishment~\cite{brandt_prsb03,helbing2010evolutionary}, the payoff structure is linear (Supplementary Note~\ref{sec_peer}), which allows us to utilize Eq.~(\ref{eq_linear_PC}) directly.

There are $n=3$ strategies: $1=\text{Cooperation ($C$)}$, $2=\text{Defection ($D$)}$, and $3=\text{Peer punishment ($E$)}$. Besides the two strategies in the public goods game, the third strategy, peer punishment, pays a cost $\alpha$ for punishing a co-player who defects. A defector, when punished, incurs a fine $\beta$. Thus, given $k_2$ defective co-players, a punishing player has $\alpha k_2$ paid, and given $k_3$ punishment co-players, a defector has $\beta k_3$ charged. Furthermore, it is assumed that punishing players also perform the cooperative behavior, investing $c$ to the common pool. This makes the strategy $C$ the second-order free-rider who exploits the effort in punishment of strategy $E$.

The first question is how the behaviors of peer punishment in structured populations, obtained by our framework (Supplementary Note~\ref{sec_peer}), differ from the ones in a well-mixed population. We find that peer punishment introduces a bi-stable space of the system state, as seen in Fig.~\ref{fig_peer}\textbf{a}, \textbf{b}. Even when $r<k+1$, the system can either evolve to a final state where strategies $E$ and $C$ coexist, or to a state dominated by strategy $D$, depending on the initial conditions. As the punishing fine $\beta$ increases, the basin of attraction for strategy $D$ diminishes. In a well-mixed population, strategy $D$ maintains a basin of attraction regardless of the punishment strength (Fig.~\ref{fig_peer}\textbf{c}). This aligns with previous findings that peer punishment does not truly resolve social dilemmas in well-mixed populations~\cite{sigmund2010social}. However, in structured populations, we observe that the basin of attraction for strategy $D$ can be entirely eliminated if the punishing fine $\beta$ exceeds a critical threshold, $\beta>\beta^\star$, where
\begin{equation}
    \beta^\star
    =\frac{k+1}{3}\left(-\frac{rc}{k+1}+c+k\alpha\right)-\alpha.
\end{equation}
Consequently, in such scenarios, the system consistently converges to a coexistence of strategies $E$ and $C$ (Fig.~\ref{fig_peer}\textbf{d}). The numerical observation from previous research suggests that peer punishment can effectively resolve social dilemmas in structured populations. Our analysis adds an analytical perspective to this conclusion.

The distinct roles of peer punishment in well-mixed and structured populations can be attributed to the fraction of defectors, $x_D^{(DE)}$, in an unstable edge equilibrium, $\mathbf{x}^{(DE)}=(0,x_D^{(DE)},1-x_D^{(DE)})$, as presented in Fig.~\ref{fig_peer}\textbf{e}. When $x_D^{(DE)}>1$, this unstable equilibrium disappears, rendering the $D$-vertex equilibrium unstable. In a well-mixed population, $x_D^{(DE)}<1$ and $x_D^{(DE)}\to 1$ as $\beta\to \infty$, indicating that the described scenario is unattainable. However, in structured populations, $x_D^{(DE)}>1$ becomes feasible once $\beta>\beta^\star$. Additionally, peer punishment acts as a double-edged sword. When $x_D^{(DE)}<0$, the system invariably converges to the full defection state, signifying ineffective punishment. As the punishing fine $\beta$ increases, peer punishment first becomes effective in well-mixed populations when $\beta>\beta_0^{\text{WM}}$. Structured populations, in contrast, require a higher $\beta_0$ value for punishment to be effective. In particular, peer punishment is less advantageous in structured populations than in well-mixed populations when $\beta<\beta_=$ (Fig.~\ref{fig_peer}\textbf{a}, \textbf{b}). However, structured populations gain an advantage when $\beta>\beta_=$, and can eventually lead to the extinction of defection at sufficient high $\beta>\beta^\star$ values. The comparison between well-mixed and structured populations in relation to the punishing fine is illustrated in Fig.~\ref{fig_peer}\textbf{f}, and the expressions for key $\beta$ values are listed in Fig.~\ref{fig_peer_table}.

We also compare the analytical predictions by our framework to the results from previous work, which was only at a numerical level. As shown in Fig.~\ref{fig_phase_peer}, we find our analytical results align qualitatively with the $\alpha$-$\beta$ phase diagrams presented in previous research~\cite{szolnoki2011competition}. Although there are differences in detail between the results obtained from non-marginal selection through numerical simulations (Fig.~\ref{fig_phase_peer}\textbf{a}, \textbf{c}) and those derived under weak selection via analytical solutions (Fig.~\ref{fig_phase_peer}\textbf{b}, \textbf{d}), both approaches consistently predict unique behaviors in structured populations that are absent in well-mixed populations. For instance, both the non-marginal and weak selection strengths indicate the existence of a $(C+E)_\text{V}$ phase at low $\alpha$ and high $\beta$, where strategy $D$ becomes extinct and strategies $C$ and $E$ coexist, equivalent to the Voter model~\cite{clifford1973model,liggett_85}. Moreover, at moderate levels of $\alpha$ and $\beta$, we anticipate a $D\Leftrightarrow (C+E)_\text{V}$ phase under weak selection. In this phase, the system evolves towards either $D$ or $(C+E)_\text{V}$ based on the initial state, although strategy $C$ may eventually become extinct due to the continuous introduction of a small number of defectors~\cite{helbing_pre10c}. A similar phase, named $D_{h(E)}$, is detected under non-marginal selection. The term `$h$' denotes `homoclinic instability', implying that strategy $E$ can overcome $D$ through a nucleation mechanism, particularly if a small colony of $E$ players survives after the extinction of cooperators. This likelihood increases with larger populations.

\subsection*{Pool punishment in public goods games}

\begin{figure}
	\centering
		\includegraphics[width=\textwidth]{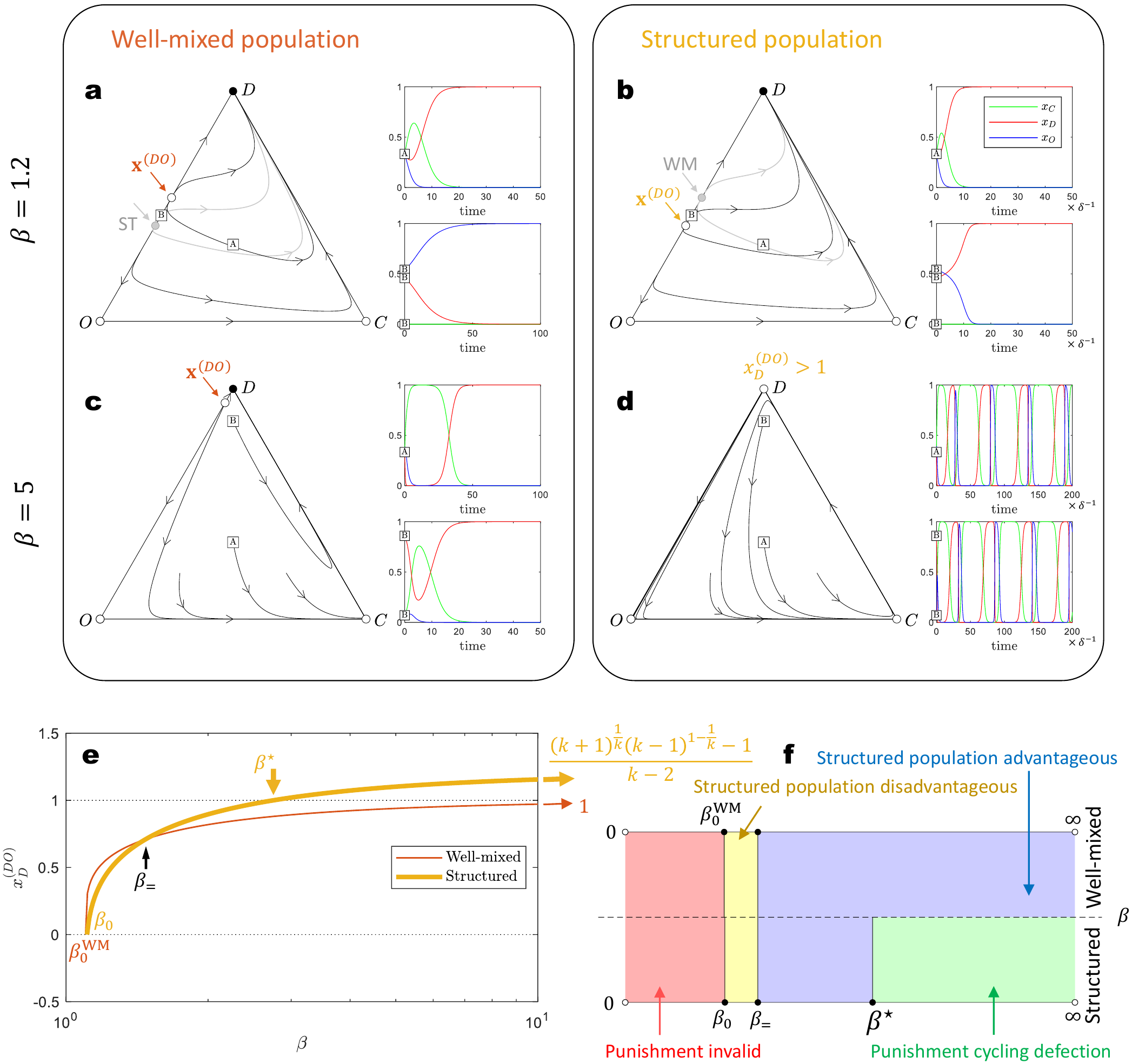}
	\caption{\textbf{Pool punishment can resolve the social dilemma of public goods game in structured populations.} 
    \textbf{a} and \textbf{b}, In the three-strategy system space, the state consistently converges to full $D$. However, along the $DO$ edge, an unstable equilibrium point, $\mathbf{x}^{(DO)}$, creates a bi-stable space. In the $D$ versus $O$ dynamics, the final state, either $D$ or $O$, is determined by the initial conditions. Under mild punishment ($\beta=1.2$), a structured population tends to favor defection, reducing the basin leading to the $O$ outcome. 
    \textbf{c} and \textbf{d}, Conversely, with strong punishment ($\beta=5$), structured populations result in the cyclic dominance of the three strategies, thereby preventing the full $D$ state in public goods games. In contrast, the state space in well-mixed populations remains two distinct basins on the $DO$ edge, preventing cyclic dominance. 
    \textbf{e}, As the punishment strength $\beta$ increases, $x_D^{(DO)}$ increases, expanding the initial space leading to the $O$ outcome. In well-mixed populations, $x_D^{(DO)}\to 1$ as $\beta\to \infty$, and the basin leading to defection cannot be completely eliminated. However, in structured populations, $x_D^{(DO)}\to [(k+1)^{1/k}(k-1)^{1-1/k}-1]/(k-2)>1$ when $\beta>\beta^\star$, invariably resulting in the cyclic dominance of the three strategies. 
    \textbf{f}, The diagram of the different effects of punishment in well-mixed versus structured populations. Structured populations are advantageous in promoting cooperation under strong punishment but are a bit less effective when the punishment is mild. 
    \textbf{Input parameters}: $r=3$, $c=1$, $\alpha=0.7$, $k=4$.} 
	\label{fig_pool}
\end{figure}

\begin{figure}[t]
	\centering
		\includegraphics[width=\textwidth]{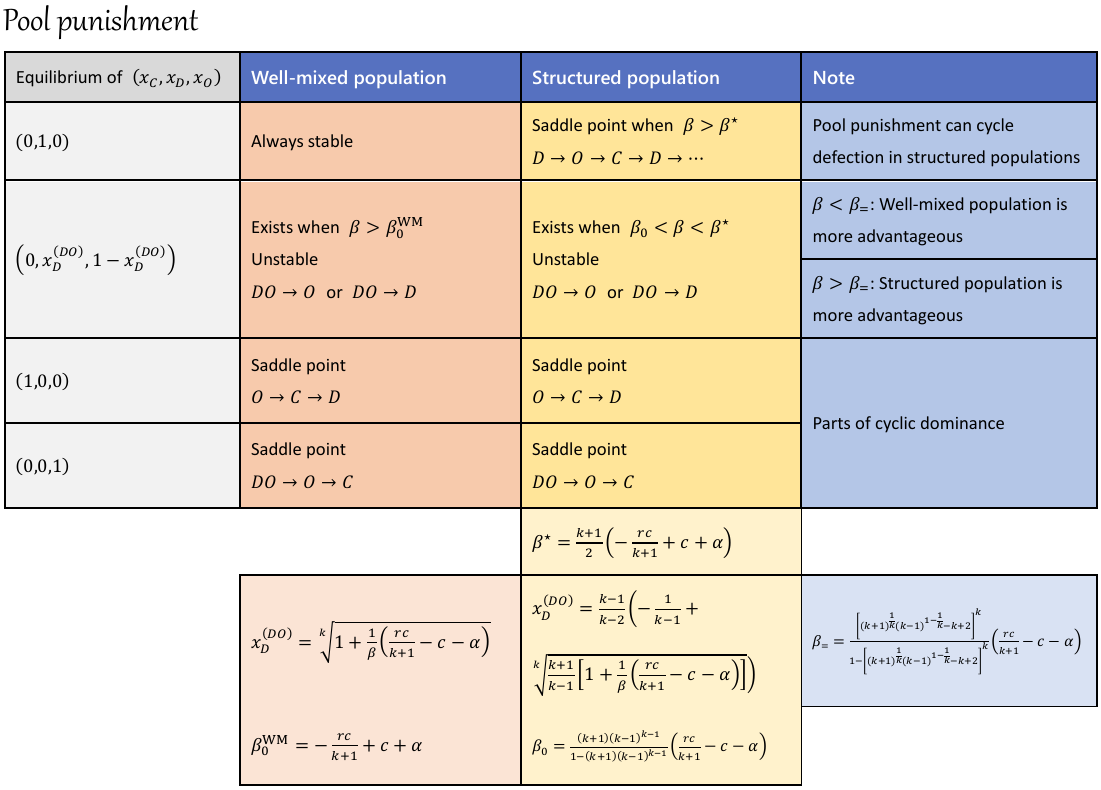}
	\caption{\textbf{Analytical results of public goods game with pool punishment in both well-mixed and structured populations.}
    This table summarizes the equilibrium points and their stability by analyzing the pool punishment system. The analytical forms of all key concepts are presented. See Supplementary Note~\ref{sec_pool} for details of the analysis.} \label{fig_pool_table}
\end{figure}

\begin{figure}
	\centering
		\includegraphics[width=.95\textwidth]{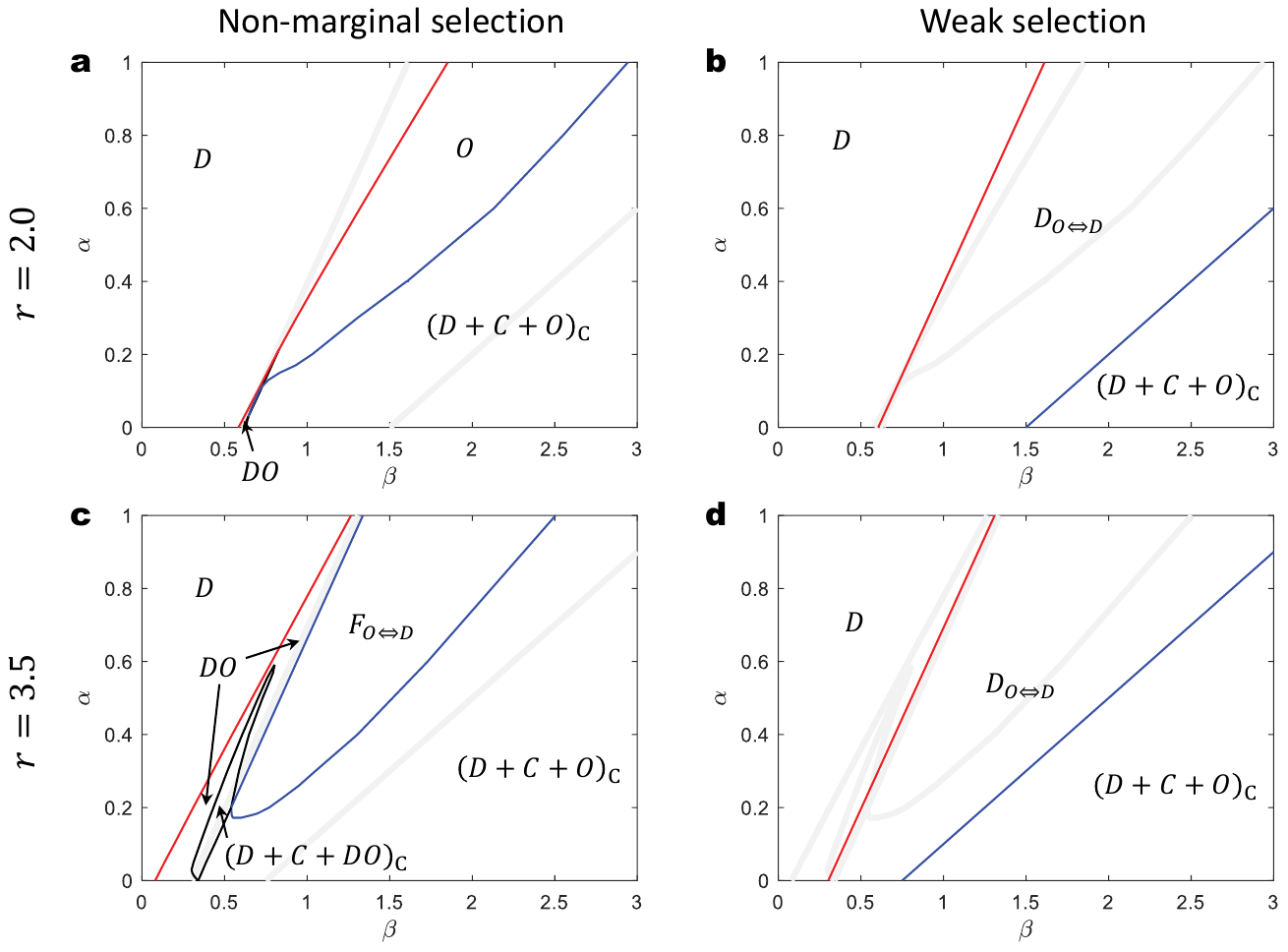}
	\caption{\textbf{Phase diagrams of the system behavior with pool punishment are qualitatively similar under non-marginal and weak selection strength.} 
    \textbf{a} and \textbf{c} (the data are in agreement with those published in Figs.~5 and 10 from ref.~\cite{szolnoki2011phase}), Numerical simulations under non-marginal selection ($\delta=2$). The phases are defined as follows: $D$---only $D$ exists; $O$---only $O$ exists; $(D+C+O)_\text{C}$---cyclic dominance among $D$, $C$, and $O$; $F_{O\Leftrightarrow D}$---fixation of either $O$ or $D$; $DO$---$D$ and $O$ coexist; $(D+C+DO)_\text{C}$---cyclic dominance among $D$, $C$, and $DO$. 
    \textbf{b} and \textbf{d}, The phase diagram is divided by analytical $\beta_0$ and $\beta^\star$ under weak selection ($\delta\to 0^+$). Here, $\beta_0$ divides the $D$ and $D_{O\Leftrightarrow D}$ phases, while $\beta^\star$ separates the $D_{O\Leftrightarrow D}$ and $(D+C+O)_\text{C}$ phases. Specifically, in \textbf{b}, $\beta_0=81/134+(135/134)\alpha$ (red), $\beta^\star=3/2+(5/2)\alpha$ (blue); in \textbf{d}, $\beta_0=81/268+(135/134)\alpha$, $\beta^\star=3/4+(5/2)\alpha$. The definition of the $D_{O\Leftrightarrow D}$ phase---the system finally evolves to full $D$ if cooperation is initially present, or to the fixation of either $O$ or $D$ in the absence of initial cooperators. \textbf{Other parameters:} $c=1$, $k=4$.} 
	\label{fig_phase_pool}
\end{figure}

Another example is pool punishment in public goods games~\cite{szolnoki2011phase,sasaki_srep15}. From the perspective of computational complexity, pool punishment differs from peer punishment in its nonlinear payoff structure, which requires utilizing Eq.~(\ref{eq_alevel3_PC}).

Similarly, there are $n=3$ strategies: $1=\text{Cooperation ($C$)}$, $2=\text{Defection ($D$)}$, and $3=\text{Pool punishment ($O$)}$. Again, based on the 2-strategy public goods game, the third strategy, pool punishment, contributes a cost $\alpha$ to the public pool for punishment. A defector is punished with a fine $\beta$ if the public pool for punishment has funds (i.e., there is at least one punisher among the co-players); if no punishers are present, the defector incurs no charge. Irrespective of the number of defecting co-players $0\leq k_2\leq k$, a punishing player pays $\alpha$. It is also assumed that those employing pool punishment engage in cooperative behavior, investing $c$ to the common pool, making the strategy $C$ a second-order free-rider.

Our analysis in structured populations (Supplementary Note~\ref{sec_pool}) and the traditional analysis for well-mixed populations reveal that pool punishment does not change the fact that the system cannot converge to a defection-free state when $r<k+1$, as demonstrated in Fig.~\ref{fig_pool}\textbf{a}, \textbf{b}. However, along the $DO$ edge (i.e., without the presence of strategy $C$), the system can evolve to a final state of either full $D$ or full $O$, depending on the initial conditions. As the punishing fine $\beta$ increases, the attraction basin for strategy $D$ shrinks. In well-mixed populations, strategy $D$ retains an attraction basin regardless of the punishment strength (Fig.~\ref{fig_pool}\textbf{c}). This is consistent with previous findings that pool punishment does not effectively resolve social dilemmas in well-mixed populations~\cite{sigmund2010social}. However, in structured populations, the attraction basin for strategy $D$ can be completely eliminated if the punishing fine $\beta$ exceeds a critical threshold, $\beta>\beta^\star$, where
\begin{equation}
    \beta^\star=\frac{k+1}{2}\left(-\frac{rc}{k+1}+c+\alpha\right).
\end{equation}
Given that $O$ and $C$ are still unstable, the system consequently enters a cyclic dominance pattern among $D$, $O$, and $C$ in such scenarios (Fig.~\ref{fig_pool}\textbf{d}). The cyclic dominance follows the sequence $D\to O\to C\to D\to \cdots$. Numerical observations from previous studies suggest that pool punishment can resolve social dilemmas in structured populations by inducing a cycle of defection~\cite{szolnoki2011phase}. Our theoretical approach provides accurate insight into this phenomenon.

Similarly, the distinct impacts of pool punishment in well-mixed and structured populations can be identified by the fraction of defectors, $x_D^{(DO)}$, in an unstable edge equilibrium, $\mathbf{x}^{(DO)}=(0,x_D^{(DO)},1-x_D^{(DO)})$, as shown in Fig.~\ref{fig_pool}\textbf{e}. When $x_D^{(DO)}>1$, this unstable equilibrium vanishes, leading to instability of the $D$-vertex equilibrium. In well-mixed populations, $x_D^{(DO)}<1$ and $x_D^{(DO)}\to 1$ as $\beta\to \infty$, suggesting that the described scenario is unfeasible. Conversely, in structured populations, $x_D^{(DO)}>1$ becomes true once $\beta>\beta^\star$. Pool punishment also presents a paradoxical effect: when $x_D^{(DO)}<0$, the system consistently converges to the full defection state, even along the $DO$ edge, indicating ineffective punishment. As the punishing fine $\beta$ increases, pool punishment first becomes effective in well-mixed populations at $\beta>\beta_0^{\text{WM}}$. Structured populations, in contrast, require a bit higher $\beta_0$ threshold for effective punishment. Pool punishment is less advantageous in structured populations than in well-mixed populations when $\beta<\beta_=$ (Fig.~\ref{fig_pool}\textbf{a}, \textbf{b}). Nevertheless, structured populations gain an advantage when $\beta>\beta_=$, and can eventually prevent the fixation of defection by inducing cyclic dominance among the three strategies at sufficient high $\beta>\beta^\star$. The comparison between well-mixed and structured populations in relation to the punishing fine is shown in Fig.~\ref{fig_pool}\textbf{f}, with the expressions for key $\beta$ values listed in Fig.~\ref{fig_pool_table}.

Again, we compare our analytical predictions to the results from previous numerical work. The analytical results are in qualitative agreement with the $\alpha$-$\beta$ phase diagrams from previous research~\cite{szolnoki2011phase}, as shown in Fig.~\ref{fig_phase_pool}. Again, while there are detailed differences between outcomes derived from non-marginal selection through numerical simulation (Fig.~\ref{fig_phase_pool}\textbf{a}, \textbf{c}) and those obtained under weak selection with analytical methods (Fig.~\ref{fig_phase_pool}\textbf{b}, \textbf{d}), both approaches indicate distinct behavioral patterns in structured populations that are not observed in well-mixed populations. For example, both non-marginal and weak selection indicate the existence of a cyclic dominance phase, $(D+C+O)_\text{C}$, at low $\alpha$ and high $\beta$, where strategy $D$ invades $C$, strategy $C$ invades $O$, and strategy $O$ invades $D$. Moreover, at moderate levels of $\alpha$ and $\beta$, we predict a $D_{O\Leftrightarrow D}$ phase under weak selection. In this phase, the system consistently evolves towards full $D$ in the three-strategy space; however, in the absence of strategy $C$, the system instead evolves towards either full $O$ or full $D$ based on the initial state. A comparable phase, named $F_{O\Leftrightarrow D}$, is detected under non-marginal selection. The term `$F$' denotes `fixation', which means that system evolves towards either full $O$ or full $D$.

\section*{Discussion}
Spatial evolutionary dynamics under weak selection can be considered as the incorporation of a marginal game effect ($\delta\to 0^+$) on the Voter model~\cite{clifford1973model,liggett_85} ($\delta=0$). In structured populations, identical strategies naturally become adjacent to each other, forming clusters through neutral drift, a process independent of the game, as described by the first-order Taylor expansion in edge dynamics. This inherent tendency for the same strategies to cluster together leads to what is known as spatial reciprocity, a phenomenon captured by the second-order Taylor expansion. Simply put, under weak selection, clusters of the same strategy, caused by spatial structures, unilaterally affect the emergence of cooperation. Conversely, the evolution of cooperation does not influence the spatial pattern of these clusters. This character under weak selection reduces computational complexity, making the closed solution for various evolutionary dynamics such as multi-strategy systems on graphs possible.

In the family of evolutionary graph theory with weak selection and pair approximation, which covers two-strategy two-player, multi-strategy two-player, and two-strategy multiplayer games, we fill in the last piece of the puzzle: the multi-strategy multiplayer games. For a focal individual, we illustrate every possible configuration in which $k$ identical co-players are distributed among $n$ distinct strategies. On this basis, we calculate the group-based payoff for any number of strategies via a bottom-up approach. While we identify each co-player by pair approximation, the ($k+1$)-player game is treated as a whole and the smallest indivisible unit in our statistical analysis. In this way, the payoff computation for the focal individual is not merely a sum of pairwise interactions, but rather an $n$-element function of the configuration~$\mathbf{k}=(k_1,k_2,\dots,k_n)$, determined by all co-players simultaneously. The nonlinearity of payoff functions cannot be derived from the superposition pairwise interactions, which reflects the higher-order properties of multi-strategy multiplayer games that are different from multi-strategy two-player games~\cite{perc_jrsi13}.

Building on the group-based payoff calculation, we develop strategy update dynamics on regular graphs using the standard pair approximation method~\cite{ohtsuki2006simple,ohtsuki2006replicator} under two common update rules: pairwise comparison and death-birth. Interestingly, our general findings are in line with those previously obtained through a different theoretical approach for two-strategy systems~\cite{allen2014games}. In particular, our $n$-strategy replicator equations imply that pairwise comparison equates to competition among all $n$ strategies between first-order neighbors, while death-birth is equivalent to competition among second-order neighbors. While this is consistent with the previous conclusions for two-strategy systems~\cite{allen2014games}, our results further extend them to the generalized $n$-strategy space.

It is worth mentioning that by contrasting a profile of our results with the other approach in two-strategy public goods games~\cite{su2019spatial,wang2023inertia}, we can see the limitations of pair approximation: unlike their approach, which can account for triangle motifs, our pair approximation cannot. According to previous works on pair approximation~\cite{li2014cooperation,li2016evolutionary}, we see that under the death-birth rule (i.e., second-order neighbor competition), pair approximation results align with the other approach only in the absence of triangle motifs. Under the pairwise comparison rule (i.e., first-order neighbor competition), however, triangle motifs appear to have no effect on multiplayer games~\cite{wang2023inertia}, where the results of pair approximation always match those of the other approach, which considers triangle motifs. This is one reason why pairwise comparison is the primary focus of this paper. For rigor, applying our framework to a specific network structure is best followed by our basic assumption: the absence of triangle motifs. We look forward to a new theory in the future that will cancel this assumption.

For any multi-strategy multiplayer game in our framework, we need only input the payoff function $a_{i|\mathbf{k}}$ for each strategy $i$ across all $(k+n-1)!/[(n-1)!k!]$ co-player strategy configurations $\mathbf{k}$. Then, we can apply the general formula provided in this work to obtain the replicator equations on a regular graph. For general payoff functions, we have decomposed the general replicator equation into sums of expected single-game payoffs, as shown in Eq.~(\ref{eq_alevel3_PC}) (for PC updates) and Supplementary Eq.~(\ref{sieq_alevel2_DB}) (for DB updates). From there, it simplifies the problem to calculating the single games under different strategy configurations and then summing them up. The computation is feasible in polynomial time, which is related to the number of strategies $n$. We find the computational complexity is $\mathrm{O}(n^2)$ for pairwise comparison and $\mathrm{O}(n^3)$ for death-birth. For certain specific payoff functions, the general formula may be further simplified, depending on whether the expected payoff across different strategy configurations has a simple primitive functional form. As an example, we provide a simple general formula for linear payoff functions in both pairwise comparison and death-birth updates, as shown in Eq.~(\ref{eq_linear_PC}) and Supplementary Eq.~(\ref{sieq_linear_DB}).

As an application of our theoretical framework, we revisit the second-order free-riding problem. In a simple three-strategy system of cooperation, defection, and cooperative punishment, the defection strategy is a free-rider from cooperation, while the original cooperation is also a free-rider from cooperative punishment. Prior research has shown that costly punishment in well-mixed populations cannot truly resolve social dilemmas~\cite{sigmund2010social}, although in structured populations it can~\cite{szolnoki2011competition,szolnoki2011phase}. We further interpret the conclusion within our analytical framework, revealing an accurate threshold for punishment strength $\beta^\star$ in both linear peer punishment and nonlinear pool punishment systems. When the punishment strength $\beta>\beta^\star$, costly punishment can resolve the social dilemma in structured populations. In peer punishment, a sufficiently strong punishment eliminates the attraction basin of full defection in the bi-stable state space. In pool punishment, a strong enough punishment leads the system to a rock-paper-scissors-like cyclic dominance. The results obtained under weak selection also qualitatively reproduce the phase diagrams found in earlier numerical studies under non-marginal selection~\cite{szolnoki2011competition,szolnoki2011phase}, identifying unique phases observable only in structured populations.

In addition, our general $n$-strategy dynamics framework can reduce to classic two-strategy multiplayer game dynamics at $n=2$. First, although some prior work has explored specific models under pairwise comparison~\cite{wang2022decentralized,luo2021evolutionary}, to our knowledge, no work has provided a general replicator equation and discussion for two-strategy multiplayer games under pairwise comparison. As a complement to this, we discuss the general replicator equation when $n=2$ for two-strategy multiplayer games under pairwise comparison in Supplementary Note~\ref{sec_PCn=2}. Second, the general replicator equations for two-strategy multiplayer games under death-birth have been discussed by Li~{\it et~al.}~\cite{li2016evolutionary}. We show that our results obtained under death-birth are identical to theirs at $n=2$ (Supplementary Note~\ref{sec_DBn=2}).

Our theoretical framework is widely applicable, yielding analytical solutions for numerous multi-strategy multiplayer game models previously proposed. Besides the two punishment mechanisms investigated in the main text, we also explore the reward mechanism~\cite{sigmund2001reward,rand2009positive,hilbe2010incentives,szolnoki2010reward} (a mirror mechanism to punishment) and multi-stage public goods game~\cite{szolnoki2022tactical} (a four-strategy system) in Supplementary Information. Classic three-strategy games remaining unexplored include tax-based reward and punishment systems~\cite{wang2021tax}, the loner strategy~\cite{szabo2002phase,hauert2002volunteering}, and so on. Moreover, a pair of additional strategies can be introduced together to create four-strategy systems, such as the competition between peer and pool punishment~\cite{szolnoki2011competition}. In fact, provided coevolutionary factors expressed as payoff functions of co-player configurations, any multi-strategy multiplayer game system can be analyzed within our framework.

\section*{Methods}
\subsection*{Bottom-up statistical quantities}

\begin{figure}
	\centering
		\includegraphics[width=\textwidth]{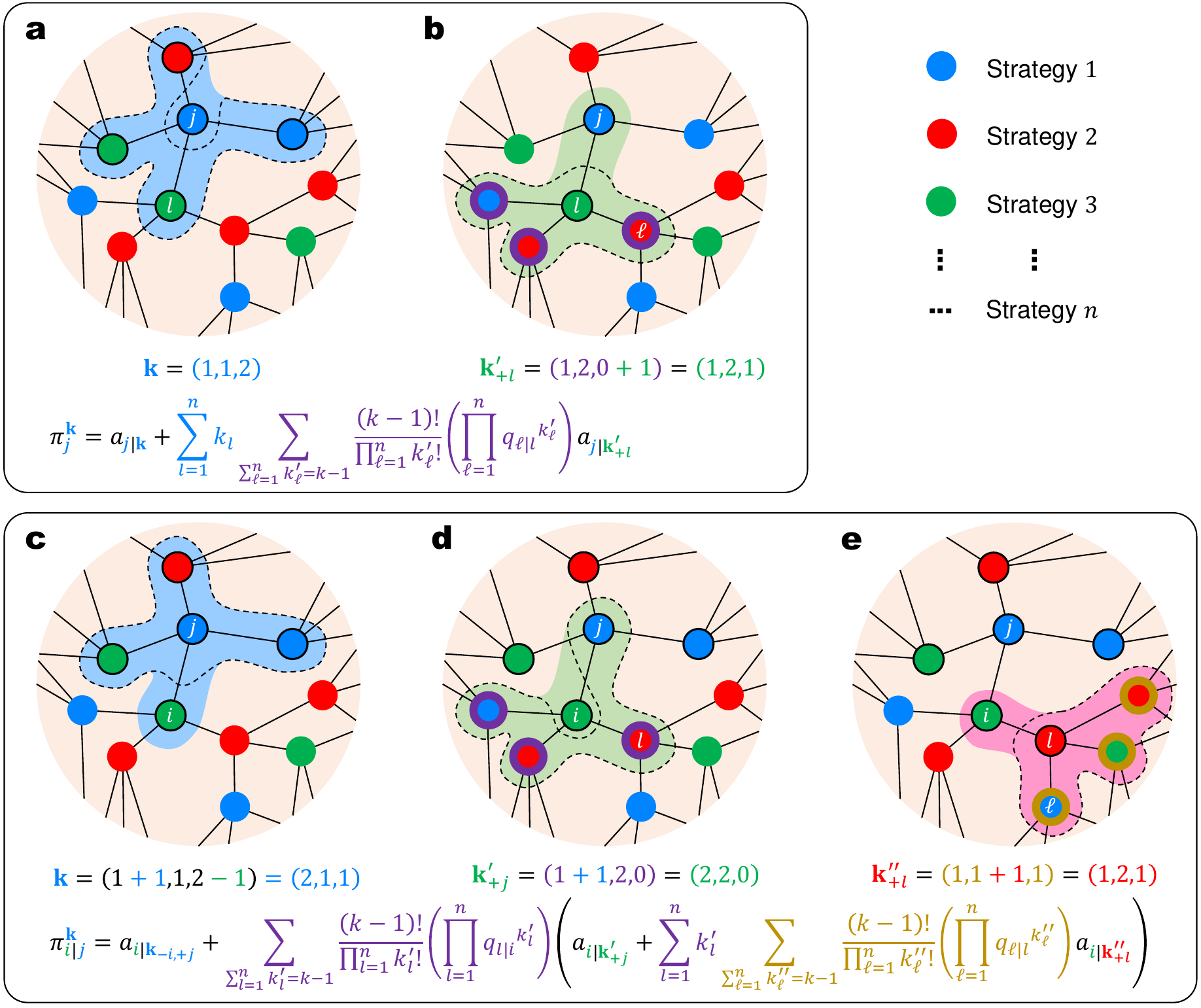}
	\caption{\textbf{Visualization of the bottom-up statistical payoff calculation when every individual has $k=4$ neighbors.}
 \textbf{a}, To calculate the group-based payoff of a focal $j$-player with neighbor configuration $\mathbf{k}$, we first have the game organized by the $j$-player itself, with $\mathbf{k}$ as the co-player configuration. 
 \textbf{b}, Then, we have $\sum_{l=1}^n k_l=k$ games organized by the $j$-player's neighbors who adopt strategy $l=1,2,\dots,n$. For the $j$-player, the co-player configuration contains $k-1$ undetermined co-players neighboring the $l$-player, plus the $l$-player.
 \textbf{c}, To calculate the group-based payoff of an $i$-player neighboring a focal $j$-player, with $j$-player's neighbor configuration $\mathbf{k}$, we first have the game organized by the $j$-player. For the $i$-player, the co-player configuration is $\mathbf{k}_{-i,+j}$.
 \textbf{d}, Then, we have the game organized by the $i$-player itself, with $k-1$ undetermined co-players neighboring the $i$-player plus the $j$-player.
 \textbf{e}, Finally, we have the games organized by the $i$-player's remaining $\sum_{l=1}^n k'_l=k-1$ neighbors. For the $i$-player, the co-player configuration contains $k-1$ undetermined co-players neighboring the $l$-player, plus the $l$-player.} \label{fig_payoff}
\end{figure}

Here, we provide the microscopic details behind the expected payoffs. There are several necessary variations of $\mathbf{k}$ for expressing the details. One is $\mathbf{k}_{+l}$, which still contains $k$ co-players but describe a configuration with at least one $l$-player. The variables satisfy $\sum_{\ell=1}^n k_\ell=k-1$, with the number of $l$-players (when $\ell=l$) written as $k_l+1$. Similarly, $\mathbf{k}_{-i,+j}$ can describe a configuration where the variables satisfy $\sum_{\ell=1}^n k_\ell=k$, with the number of $i$-players as $k_i-1$ and $j$-players as $k_j+1$. Also, note that $\mathbf{k}'$ and $\mathbf{k}''$ are different variables and have no relation with $\mathbf{k}$. The primes are only to distinguish the sequence of summations: we do the computation on $\mathbf{k}''$, $\mathbf{k}'$, and finally $\mathbf{k}$.

We start from the level where $\mathbf{k}$ is given. Given neighbor configuration $\mathbf{k}$ for a focal $j$-player, its accumulated payoff can be expressed as
\begin{equation}\label{eq_pi_j}
    \pi_j^\mathbf{k}=a_{j|\mathbf{k}}+
    \sum_{l=1}^n k_l \sum_{\sum_{\ell=1}^n k'_\ell=k-1}\frac{(k-1)!}{\prod_{\ell=1}^n k'_\ell!} \left(\prod_{\ell=1}^n {q_{\ell|l}}^{k'_\ell}\right) a_{j|\mathbf{k}'_{+l}}.
\end{equation}
The $j$-player accumulates payoff from the games organized by itself and its $\sum_{l=1}^n k_l=k$ neighbors. The visualization of Eq.~(\ref{eq_pi_j}) is shown in Fig.~\ref{fig_payoff}\textbf{a} and \textbf{b}. Similarly, the accumulated payoff of an $i$-player neighboring a $j$-player, given the $j$-player's neighbor configuration $\mathbf{k}$, can be expressed and calculated by
\begin{equation}\label{eq_pi_i|j}
    \pi_{i|j}^\mathbf{k}=a_{i|\mathbf{k}_{-i,+j}}+
    \sum_{\sum_{l=1}^n k'_l=k-1}\frac{(k-1)!}{\prod_{l=1}^n k'_l!} \left(\prod_{l=1}^n {q_{l|i}}^{k'_l}\right) \left(
    a_{i|\mathbf{k}'_{+j}}+\sum_{l=1}^n k'_l \sum_{\sum_{\ell=1}^n k''_\ell=k-1}\frac{(k-1)!}{\prod_{\ell=1}^n k''_\ell!} \left(\prod_{\ell=1}^n {q_{\ell|l}}^{k''_\ell}\right) a_{i|\mathbf{k}''_{+l}}
     \right).
\end{equation}
The $i$-player accumulates payoff from the games organized by the $j$-player (Fig.~\ref{fig_payoff}\textbf{c}), itself (Fig.~\ref{fig_payoff}\textbf{d}), and its remaining $\sum_{l=1}^n k'_l=k-1$ neighbors (Fig.~\ref{fig_payoff}\textbf{e}). Further explanations of Eqs.~(\ref{eq_pi_j}) and (\ref{eq_pi_i|j}) can be found in Supplementary Information.

Based on the microscopic quantities given specific $\mathbf{k}$, we can further express expected values over all possible $\mathbf{k}$. The expected payoff of a focal $X$-player over all possible $\mathbf{k}$ can be statistically computed as
\begin{equation}\label{eq_<pi_j>}
    \langle \pi_X^{\mathbf{k}}\rangle
    =\sum_{\sum_{i'=1}^n k_{i'}=k}\frac{k!}{\prod_{i'=1}^n k_{i'}!} \left(\prod_{i'=1}^n {q_{{i'}|X}}^{k_{i'}}\right)
    \pi_X^{\mathbf{k}}.
\end{equation}
The possible neighbor configurations satisfying $\sum_{i'=1}^n k_{i'}=k$ are found around the $X$-player as identified by $q_{i'|X}$. Here, $i'$ is an independent count, which has no relation with $i$. 

Similarly, the expected payoff of an $i$-player neighboring an $X$-player over all possible neighbor configuration $\mathbf{k}_{+i}$ of the $X$-player can be expressed as 
\begin{equation}\label{eq_<pi_i|j>}
    \langle \pi_{i|X}^{\mathbf{k}_{+i}} \rangle
    =\sum_{\sum_{i'=1}^n k_{i'}=k-1}\frac{(k-1)!}{\prod_{i'=1}^n k_{i'}!} \left(\prod_{i'=1}^n {q_{{i'}|X}}^{k_{i'}}\right)
    \pi_{i|X}^{\mathbf{k}_{+i}}.
\end{equation}
Here, $\mathbf{k}_{+i}$ is because we have a specific $i$-player in the neighbor configuration of the $X$-player. The remaining $k-1$ neighbors $\sum_{i'=1}^n k_{i'}=k-1$ of the $X$-player are found around the $X$-player, as identified by $q_{i'|X}$.

Eqs.~(\ref{eq_pi_i|j}) and (\ref{eq_<pi_i|j>}) may seem redundant because they do not appear directly in the final results. However, they are crucial in the process of deductions for both pairwise comparison and death-birth rules.

We also have the similar notation for expected single-game payoff. To specify, the expected payoff of an $i$-player in a single game over all co-player configuration $\mathbf{k}$, where $\mathbf{k}$ is found neighboring an $X$-player, is expressed as
\begin{equation}\label{eq_<a>}
    \langle a_{i|\mathbf{k}}\rangle_X=\sum_{\sum_{i'=1}^n k_{i'}=k}\frac{k!}{\prod_{i'=1}^n k_{i'}!} \left(\prod_{i'=1}^n {q_{{i'}|X}}^{k_{i'}}\right)
    a_{i|\mathbf{k}}.
\end{equation}

The concepts in Eqs.~(\ref{eq_pi_j})--(\ref{eq_<a>}) are sufficient to identify the difference between this work and the previous literature~\cite{ohtsuki2006simple,ohtsuki2006replicator,li2016evolutionary}. In particular, they emphasize that the minimal unit to refer is the co-player configuration $\mathbf{k}$, based on which the payoff of the multiplayer game is computed. We do not try to decompose the multi-body interaction identified by $\mathbf{k}$ into multiple pairwise interactions. 

Combined with the pair approximation method~\cite{ohtsuki2006replicator} and detailed calculations in the strategy evolution dynamics, we can then obtain the master equation (\ref{eq_replicatorPC}) in the main text (see Supplementary Note~\ref{sec_linear} for the approach of pair approximation deduction).

\subsection*{The decomposition to single games}
The following form of the master equation is important, which holds for any $n$-strategy multiplayer game and allows us to obtain the replicator dynamics by summing the expected payoff calculations in a series of single games:
\begin{equation}\label{eq_alevel3_PC}
    \dot{x}_i=\frac{\delta(k-2)}{2(k-1)}x_i \sum_{j=1}^n x_j
    \Bigg(
    \langle a_{i|\mathbf{k}_{+j}}\rangle_i+
    (k-1) \langle a_{i|\mathbf{k}_{+j}}\rangle_j
    +\langle a_{i|\mathbf{k}_{+i}}\rangle_i
    -\langle a_{j|\mathbf{k}_{+i}}\rangle_j
    -\langle a_{j|\mathbf{k}_{+i}}\rangle_i
    -(k-2)\sum_{l=1}^n x_l 
    \langle a_{j|\mathbf{k}_{+l}}\rangle_l
    -\langle a_{j|\mathbf{k}_{+j}}\rangle_j
    \Bigg).
\end{equation}
In application, given the payoff functions $a_{i|\mathbf{k}}$ where $i=1,2,\dots, n$, we can compute all $\langle \cdot\rangle$ terms and then ensemble them to obtain the replicator equations. The result of each $\langle \cdot\rangle$ should be a function of $x_1, x_2, \dots, x_n$ (transformed from $q_{j|i}$ manually), degree $k$, and game parameters.

The advantage of Eq.~(\ref{eq_alevel3_PC}) is that we have attributed everything about $\langle \cdot\rangle$ into two types, the `$\langle a_{i|\mathbf{k}_{+j}}\rangle_i$ type' and the `$\langle a_{i|\mathbf{k}_{+j}}\rangle_j$ type'. They can be expressed by matrices through $i$ and $j$:

\begin{itemize}
    \item The $\langle a_{i|\mathbf{k}_{+j}}\rangle_i$ type: 
    \begin{equation}\label{eq_aiji_type}
        \left[\langle a_{i|\mathbf{k}_{+j}}\rangle_i\right]_{ij}=
        \displaystyle{\begin{pmatrix}
        \langle a_{1|\mathbf{k}_{+1}}\rangle_1 & \langle a_{1|\mathbf{k}_{+2}}\rangle_1 & \cdots & \langle a_{1|\mathbf{k}_{+n}}\rangle_1 \\
        \langle a_{2|\mathbf{k}_{+1}}\rangle_2 & \langle a_{2|\mathbf{k}_{+2}}\rangle_2 & \cdots & \langle a_{2|\mathbf{k}_{+n}}\rangle_2 \\
        \vdots & \vdots & \ddots & \vdots \\
        \langle a_{n|\mathbf{k}_{+1}}\rangle_n & \langle a_{n|\mathbf{k}_{+2}}\rangle_n & \cdots & \langle a_{n|\mathbf{k}_{+n}}\rangle_n
    \end{pmatrix}}.
    \end{equation}

    \item The $\langle a_{i|\mathbf{k}_{+j}}\rangle_j$ type: 
    \begin{equation}\label{eq_aijj_type}
        \left[\langle a_{i|\mathbf{k}_{+j}}\rangle_j\right]_{ij}=
        \displaystyle{\begin{pmatrix}
        \langle a_{1|\mathbf{k}_{+1}}\rangle_1 & \langle a_{1|\mathbf{k}_{+2}}\rangle_2 & \cdots & \langle a_{1|\mathbf{k}_{+n}}\rangle_n \\
        \langle a_{2|\mathbf{k}_{+1}}\rangle_1 & \langle a_{2|\mathbf{k}_{+2}}\rangle_2 & \cdots & \langle a_{2|\mathbf{k}_{+n}}\rangle_n \\
        \vdots & \vdots & \ddots & \vdots \\
        \langle a_{n|\mathbf{k}_{+1}}\rangle_1 & \langle a_{n|\mathbf{k}_{+2}}\rangle_2 & \cdots & \langle a_{n|\mathbf{k}_{+n}}\rangle_n
    \end{pmatrix}}.
    \end{equation}
\end{itemize}
There are $n^2$ elements in each matrix. Their diagonals are equal, meaning we can compute $n$ fewer elements. Therefore, the total amount of computation is $n^2+n^2-n=(2n-1)n$ elements. The computational complexity is $\mathrm{O}(n^2)$, which can be accomplished in polynomial time (we also see that the death-birth rule's computational complexity is $\mathrm{O}(n^3)$, as specified in Supplementary Information).

\subsection*{Special linear system}
Although Eq.~(\ref{eq_alevel3_PC}) allows general calculations of any multiplayer game, we do not have to employ it directly every time. For some special payoff structures, we can deduce simplified general forms in advance. Here, we present the general results of a commonly studied subclass, the linear multiplayer games. 

The linear multiplayer games in this work are defined as those whose payoff structure can be expressed as linear functions of co-player configuration $\mathbf{k}=(k_1,k_2,\dots,k_n)$. That is, $a_{i|\mathbf{k}}=\sum_{j=1}^n b_{ij}k_j+c_i$, where 
\begin{equation}\label{eq_b_c}
    \mathbf{b}=\begin{pmatrix}
        b_{11} & b_{12} & \cdots & b_{1n} \\
        b_{21} & b_{22} & \cdots & b_{2n} \\
        \vdots & \vdots & \ddots & \vdots \\
        b_{n1} & b_{n2} & \cdots & b_{nn}
    \end{pmatrix},~
    \mathbf{c}=\begin{pmatrix}
        c_1 \\
        c_2 \\
        \vdots \\
        c_n
    \end{pmatrix}.
\end{equation}
For linear multiplayer games, the payoff structure is completely determined by the matrix $\mathbf{b}$ and $\mathbf{c}$.

Once we apply $a_{i|\mathbf{k}}=\sum_{j=1}^n b_{ij}k_j+c_i$ to compute the `$\langle a_{i|\mathbf{k}_{+j}}\rangle_i$ type' and the `$\langle a_{i|\mathbf{k}_{+j}}\rangle_j$ type' as shown in Eqs.~(\ref{eq_aiji_type}) and (\ref{eq_aijj_type}) and transform all $q_{j|i}$ quantities to $x_j$ quantities, we can obtain $\langle a_{i|\mathbf{k}_{+j}}\rangle_i=(k-2)\sum_{l=1}^n b_{il}x_l+b_{ii}+b_{ij}+c_i$, $\langle a_{i|\mathbf{k}_{+j}}\rangle_j=(k-2)\sum_{l=1}^n b_{il}x_l+2b_{ij}+c_i$. Then, substituting the results into Eq.~(\ref{eq_alevel3_PC}) leads to Eq.~(\ref{eq_linear_PC}) in the main text.

As we mentioned in the main text, in Eq.~(\ref{eq_linear_PC}), $\Bar{\pi}_i$ and $\Bar{\pi}$ are mean payoffs of $i$-players and all players in a well-mixed population. They can be calculated by the traditional replicator dynamics, but to specify, using the matrix $\mathbf{b}$ and $\mathbf{c}$, they can also be written as $\Bar{\pi}_i=k\sum_{l=1}^n x_l b_{il}+c_i$, $\Bar{\pi}=\sum_{i=1}^n x_i\Bar{\pi}_i=k\sum_{i=1}^n \sum_{l=1}^n x_i x_l b_{il}+\sum_{i=1}^n x_i c_i$. Logically, this is how we replace the corresponding terms in Eq.~(\ref{eq_linear_PC}) by $\Bar{\pi}_i$ and $\Bar{\pi}$.

\section*{Data availability}
All data generated or analysed during this study are included within the paper and its supplementary information files.

\section*{Acknowledgements}
M.P. was supported by the Slovenian Research and Innovation Agency (Javna agencija za znanstvenoraziskovalno in inovacijsko dejavnost Republike Slovenije) (Grant Nos. P1-0403 and N1-0232). A.S. was supported by the National Research, Development and Innovation Office (NKFIH) under Grant No. K142948.

\section*{Author contributions}
C.W. conceived and designed the research with contributions from M.P. and A.S.; C.W. performed the calculations; C.W. and A.S. analyzed the results; C.W., M.P., and A.S. wrote the paper and approved the submission.

\section*{Competing interests}
The authors declare no competing interests.

\newpage
\begin{center}
  \Huge Supplementary Information for
\end{center}

\noindent
\textbf{\Huge Evolutionary dynamics of any multiplayer game on regular graphs}
\vspace{2em}

\noindent
{\Large Chaoqian Wang, Matja\v z Perc, \& Attila Szolnoki}
\vspace{1em} % 添加额外的垂直空间

\noindent
e-mail: CqWang814921147@outlook.com (C. Wang)

% 改变目录的行为，从这里开始将内容添加到目录
\addtocontents{toc}{\protect\setcounter{tocdepth}{3}}
% 插入目录
\tableofcontents

\renewcommand{\theequation}{S\arabic{equation}}
\setcounter{equation}{0}

\let\oldsection\thesection
\renewcommand{\thesection}{Supplementary Note~\arabic{section}}
\renewcommand{\thesubsection}{\oldsection.\arabic{subsection}}

\renewcommand{\thefigure}{S\arabic{figure}}
\setcounter{figure}{0}

\newpage

\section{: System construction and payoff calculation}\label{sec_theory}
We employ the pair approximation method~\cite{ohtsuki2006simple,ohtsuki2006replicator,li2016evolutionary,gutowitz1987local,szabo1991correlations} to deduce the system dynamics. First, we express the system and explain the payoff calculation in detail.

\subsection{Expressing the system}\label{sec_system}
On the regular graph of degree $k$, the number of nodes is $N$ and the number of edges is $kN/2$, where the population size $N\to \infty$. In this infinite population, the proportion of individuals who choose strategy $i$ is denoted by $x_i$. The probability of finding a $j$-player neighboring an $i$ player is denoted by $q_{j|i}$. The proportion of indirected $ij$-edges connecting a pair of strategy $i$ and $j$ is denoted by $p_{ij}$. 

To sum up, there are $n+n^2+n^2$ variables to describe the system: $n$ for $x_i$, $n^2$ for $q_{j|i}$, and $n^2$ for $p_{ij}$. They yield the following constraints.
\begin{align}\label{sieq_constraint}
    \sum_{i=1}^n x_i &=1, \nonumber\\
    \sum_{j=1}^n q_{j|i} &=1,\quad \text{for $i=1,2,\dots,n$}, \nonumber\\
    p_{ij} &=x_i q_{j|i},\quad \text{for $i=1,2,\dots,n$, $j=1,2,\dots,n$}, \nonumber\\
    p_{ij} &=p_{ji},\quad \text{for $i=1,2,\dots,n$, $j
    =1,2,\dots,n$, $i\neq j$}.
\end{align}
According to the third line in Eq.~(\ref{sieq_constraint}), the system can be described by $x_i$ and $q_{j|i}$, eliminating the need to use $p_{ij}$.

\subsection{Payoff calculation of a $j$-player}\label{sec_payoffj}
As mentioned in the main text, to express a $j$-player's accumulated payoff, received from the multiplayer games organized by itself and neighbors, we need to introduce a variant of neighbor configuration $\mathbf{k}$ as follows.
\begin{equation}
    \mathbf{k}_{+j}=(k_1,k_2,\dots,k_j+1,\dots,k_n),
\end{equation}
where $\sum_{l=1}^n k_l=k-1$. This notation describes a configuration which has two components: (i) one individual chooses strategy $j$; (ii) among $k-1$ remaining individuals, $k_1,k_2,\dots,k_j,\dots,k_n$ individuals choose strategy $1,2,\dots,j,\dots,n$. Here, $j$ could be an integer satisfying $1\leq j\leq n$.

Consider a focal $j$-player with neighbor configuration $\mathbf{k}$. Then, the accumulated payoff of this $j$-player, denoted by $\pi_j^\mathbf{k}$, can be expressed as
\begin{equation}\label{sieq_pi_j}
    \pi_j^\mathbf{k}=a_{j|\mathbf{k}}+
    \sum_{l=1}^n k_l \sum_{\sum_{\ell=1}^n k'_\ell=k-1}\frac{(k-1)!}{\prod_{\ell=1}^n k'_\ell!} \left(\prod_{\ell=1}^n {q_{\ell|l}}^{k'_\ell}\right) a_{j|\mathbf{k}'_{+l}}.
\end{equation}
It is straightforward that $a_{j|\mathbf{k}}$ is the payoff received from the game organized by the $j$-player itself. Furthermore, in the $j$-player's neighbor configuration $\mathbf{k}$, there are $k_l$ individuals with strategy $l$. 

In the game organized by an $l$-player, there must be a $j$-player, the focal player. We denote the neighbor configuration of the $l$-player as $\mathbf{k}'_{+j}$, where the upper right prime ($'$) is only to distinguish from $\mathbf{k}$. Here, $\mathbf{k}'_{+j}$ contains the focal $j$-player and the remaining $k-1$ uncertain $\ell$-players, determined by going through $\sum_{\ell=1}^n k'_\ell=k-1$. Given that the $l$-player's co-player configuration in its own group is $\mathbf{k}'_{+j}$, the focal $j$-player's co-player configuration in the same group is then expressed as $\mathbf{k}'_{+l}$: removing itself and adding the $l$-player. As a result, $\mathbf{k}'_{+l}$ contains the $l$-player and the remaining $k-1$ uncertain $\ell$-players. Therefore, the $j$-player receives payoff $a_{j|\mathbf{k}'_{+l}}$ from the games organized by each $l$-player neighbor by determining the remaining $k-1$ uncertain $\ell$-co-players in $\mathbf{k}'_{+l}$.

\subsection{Payoff calculation of an $i$-player neighboring a $j$-player}
Now we consider a more complex case, the accumulated payoff of an $i$-player neighboring a focal $j$-player. To express the accumulated payoff of the $i$-player, we need to further introduce a variant of $\mathbf{k}$ as follows.
\begin{equation}
    \mathbf{k}_{-i,+j}=(k_1,k_2,\dots,k_i-1,\dots,k_j+1,\dots,k_n),
\end{equation}
where $\sum_{l=1}^n k_l=k$. This notation describes a variant to the configuration $\mathbf{k}$, with one less individual choosing strategy $i$, and one more individual choosing strategy $j$. Here, $i$ and $j$ could be integers, satisfying $1\leq i\leq n$ and $1\leq j\leq n$. Moreover, the order of $i$ and $j$ ($i<j$ or $i>j$) does not matter in $\mathbf{k}_{-i,+j}$.

Again, we consider a focal $j$-player with neighbor configuration $\mathbf{k}$. The payoff of an $i$-player neighboring this focal $j$-player can be expressed as
\begin{equation}\label{sieq_pi_i|j}
    \pi_{i|j}^\mathbf{k}=a_{i|\mathbf{k}_{-i,+j}}+
    \sum_{\sum_{l=1}^n k'_l=k-1}\frac{(k-1)!}{\prod_{l=1}^n k'_l!} \left(\prod_{l=1}^n {q_{l|i}}^{k'_l}\right) \left(
    a_{i|\mathbf{k}'_{+j}}+\sum_{l=1}^n k'_l \sum_{\sum_{\ell=1}^n k''_\ell=k-1}\frac{(k-1)!}{\prod_{\ell=1}^n k''_\ell!} \left(\prod_{\ell=1}^n {q_{\ell|l}}^{k''_\ell}\right) a_{i|\mathbf{k}''_{+l}}
     \right).
\end{equation}
In Eq.~(\ref{sieq_pi_i|j}), $a_{i|\mathbf{k}_{-i,+j}}$ is the payoff received from the game organized by the focal $j$-player. Note that $\mathbf{k}$ describes the neighbor configuration of the $j$-player. Therefore, $i$-player joining the game organized by the focal $j$-player, the remaining participants' configuration for this $i$-player becomes $\mathbf{k}_{-i,+j}$.

Let $\mathbf{k}'_{+j}$ denote the $i$-player's neighbor configuration. There is one neighbor employing strategy $j$, the focal $j$-player. The remaining $k-1$ neighbors are undetermined. Therefore, we go through $\sum_{l=1}^n k'_l=k-1$ for all possibilities of $\mathbf{k}'_{+j}$. In the game organized by the $i$-player itself, the payoff can be expressed as $a_{i|\mathbf{k}'_{+j}}$.

Among the $i$-player's remaining $k-1$ neighbors, the number of $l$-players is $k_l'$. We denote an $l$-player's neighbor configuration by $\mathbf{k}_{+i}''$: the $i$-player's strategy $i$ is determined, and the remaining $k-1$ players are undetermined. We go through $\sum_{\ell=1}^n k_{\ell}''=k-1$ for all possibilities. While the configuration of other players around the $l$-player is $\mathbf{k}_{+i}''$, the configuration around the $i$-player should be rewritten by $\mathbf{k}''_{+l}$: removing the $i$-player and adding the $l$-player. Therefore, in the game organized by an $l$-player neighboring the $i$-player, the payoff of the $i$-player is $a_{i|\mathbf{k}_{+l}''}$.

Knowing the payoff calculation, we can study the evolutionary dynamics. Here, we investigate two update rules for the evolution of strategies, pairwise comparison and death-birth, presented in \ref{sec_PC} and \ref{sec_DB}, respectively.

\section{: Pairwise comparison}\label{sec_PC}
In a unit time, a random focal individual is selected to update 
the strategy by comparing the fitness with a random neighbor. Suppose that the focal individual $A$ compares its fitness with the neighbor $B$. The individual $A$ adopts the strategy of $B$ with the following probability:
\begin{equation}
    W=\frac{F_B}{F_A+F_B}.
\end{equation}
Otherwise, individual $A$ keeps its own strategy with the remaining probability $F_A/(F_A+F_B)$. Here, $F_A$ and $F_B$ are the fitness of individuals $A$ and $B$. As mentioned in the main text, the transformation from payoff to fitness is $F=\exp{(\delta\pi)}$~\cite{mcavoy2020social,wang2023inertia}. In this way, the adopting probability $W$ has the following well-known form~\cite{szabo1998evolutionary}:
\begin{equation}\label{sieq_Fermi}
    W=\frac{1}{1+\exp{[-\delta(\pi_B-\pi_A)]}},
\end{equation}
where $\pi_A$ and $\pi_B$ are the payoff of individuals $A$ and $B$, and $\delta\to 0^+$ is the weak selection limit. According to Eq.~(\ref{sieq_Fermi}), if individual $B$ has a higher payoff, then its strategy has a slightly higher probability to be adopted by individual $A$.

Below, we analyze the strategy evolution under pairwise comparison. 

\subsection{The increase of $i$-players}\label{sec_PC_increasei}
The increase of $i$-players happens when a focal $j$-player ($j\neq i$) is selected to update its strategy and an $i$-player takes the position. Given the focal $j$-player's neighbor configuration $\mathbf{k}$, the probability that an $i$-player takes the $j$-player's position is
\begin{equation}\label{sieq_jgetPC}
    \mathcal{P}(j\gets i)=\frac{k_i}{k}
    \frac{F_{i|j}^\mathbf{k}}{F_j^\mathbf{k}+F_{i|j}^\mathbf{k}}
    =\frac{k_i}{2k}+\frac{k_i}{4k}
    (\pi_{i|j}^\mathbf{k}-\pi_j^\mathbf{k}) \delta +\mathcal{O}(\delta^2).
\end{equation}
The randomly selected neighbor from $\mathbf{k}$ is an $i$-player with probability $k_i/k$. Then, the focal $j$-player adopts the $i$-player's strategy with the probability of pairwise comparison. Taylor expansion at $\delta\to 0^+$ has been performed in Eq.~(\ref{sieq_jgetPC}).

Then, we apply Eq.~(\ref{sieq_jgetPC}) to all possibilities for $j\neq i$ and the neighbor configuration $\mathbf{k}$, obtaining the probability that the number of $i$-players increases by 1 (i.e., the frequency of $i$-players increases by $1/N$) during a unit time step,
\begin{align}\label{sieq_i+_PC}
    \mathcal{P}\left(\Delta x_i=\frac{1}{N}\right)
    &=\sum_{j=1,j\neq i}^n x_j
    \sum_{\sum_{i'=1}^n k_{i'}=k}\frac{k!}{\prod_{i'=1}^n k_{i'}!} \left(\prod_{i'=1}^n {q_{i'|j}}^{k_{i'}}\right)
    \mathcal{P}(j\gets i) \nonumber\\
    &=\frac{1}{2}\sum_{j=1,j\neq i}^n x_j q_{i|j}
    +\frac{1}{4}\sum_{j=1,j\neq i}^n x_j 
    \sum_{\sum_{i'=1}^n k_{i'}=k}\frac{k!}{\prod_{i'=1}^n k_{i'}!} \left(\prod_{i'=1}^n {q_{i'|j}}^{k_{i'}}\right)
    \frac{k_i}{k}\left(
    \pi_{i|j}^\mathbf{k}-\pi_j^\mathbf{k}
    \right) \delta +\mathcal{O}(\delta^2),
\end{align}
where the notation $i'$ is independent, only to distinguish from $i$.

\subsection{The decrease of $i$-players}\label{sec_PC_decreasei}
The decrease of $i$-players happens when a focal $i$-player is selected to update its strategy and the player who takes the position is not an $i$-player. Given the focal $i$-player's neighbor configuration $\mathbf{k}$, the probability that the player who takes the position is not an $i$-player is
\begin{equation}\label{sieq_igetPC}
    \sum_{j=1,j\neq i}^n \mathcal{P}(i\gets j)=\sum_{j=1,j\neq i}^n \frac{k_j}{k}\frac{F_{j|i}^\mathbf{k}}{F_i^\mathbf{k}+F_{j|i}^\mathbf{k}}
    =\frac{k-k_i}{2k}+\sum_{j=1,j\neq i}^n \frac{k_j}{4k}
    (\pi_{j|i}^\mathbf{k}-\pi_i^\mathbf{k}) \delta +\mathcal{O}(\delta^2).
\end{equation}
Applying it to all possibilities for the neighbor configuration $\mathbf{k}$ after selecting a focal $i$-player with probability $x_i$, we obtain the probability that the number of $i$-players decreases by 1 during a unit time step,
\begin{align}\label{sieq_i-_PC}
    \mathcal{P}\left(\Delta x_i=-\frac{1}{N}\right)
    &=x_i
    \sum_{\sum_{i'=1}^n k_{i'}=k}\frac{k!}{\prod_{i'=1}^n k_{i'}!} \left(\prod_{i'=1}^n {q_{i'|i}}^{k_{i'}}\right)
    \sum_{j=1,j\neq i}^n \mathcal{P}(i\gets j) \nonumber\\
    &=\frac{x_i (1-q_{i|i})}{2}
    +\frac{1}{4}x_i 
    \sum_{\sum_{i'=1}^n k_{i'}=k}\frac{k!}{\prod_{i'=1}^n k_{i'}!} \left(\prod_{i'=1}^n {q_{i'|i}}^{k_{i'}}\right)
    \sum_{j=1,j\neq i}^n
    \frac{k_j}{k}\left(
    \pi_{j|i}^\mathbf{k}-\pi_i^\mathbf{k}
    \right) \delta +\mathcal{O}(\delta^2).
\end{align}

\subsection{The replicator equation}\label{sec_repliactor_PC}
The instant change in the frequency $x_i$ of $i$-players consists of the increase and decrease of $i$-players. Applying Eqs.~(\ref{sieq_i+_PC}) and (\ref{sieq_i-_PC}), and considering that a full Monte Carlo step contains $N$ elementary steps, we have
\begin{align}\label{sieq_replicator_PC}
    \dot{x}_i
    =&~N\times \left\{\frac{1}{N}\mathcal{P}\left(\Delta x_i=\frac{1}{N}\right)
    +\left(-\frac{1}{N}\right)\mathcal{P}\left(\Delta x_i=-\frac{1}{N}\right)\right\} \nonumber\\
    =&~\frac{1}{4} \sum_{j=1,j\neq i}^n x_j \sum_{\sum_{i'=1}^n k_{i'}=k}\frac{k!}{\prod_{i'=1}^n k_{i'}!} \left(\prod_{i'=1}^n {q_{{i'}|j}}^{k_{i'}}\right) \frac{k_i}{k}\left(
    \pi_{i|j}^\mathbf{k}-\pi_j^\mathbf{k}
    \right)\delta \nonumber\\
    &-\frac{1}{4}x_i 
    \sum_{\sum_{i'=1}^n k_{i'}=k}\frac{k!}{\prod_{i'=1}^n k_{i'}!} \left(\prod_{i'=1}^n {q_{i'|i}}^{k_{i'}}\right)
    \sum_{j=1,j\neq i}^n
    \frac{k_j}{k}\left(
    \pi_{j|i}^\mathbf{k}-\pi_i^\mathbf{k}
    \right) \delta +\mathcal{O}(\delta^2).
\end{align}
In Eq.~(\ref{sieq_replicator_PC}), the $\delta^0$ term has been eliminated: applying Eq.~(\ref{sieq_constraint}), we know $x_j q_{i|j}=p_{ij}=x_i q_{j|i}$ and $\sum_{j=1}^n q_{j|i}=1$, such that the $\delta^0$ term in Eq.~(\ref{sieq_i+_PC}) can be expressed as $\frac{1}{2}\sum_{j=1,j\neq i}^n x_j q_{i|j}=\frac{1}{2}x_i\sum_{j=1,j\neq i}^n q_{j|i}=\frac{x_i (1-q_{i|i})}{2}$, which is equal to the one in Eq.~(\ref{sieq_i+_PC}).

The $\delta^0$ term being eliminated, the instant change in $x_i$ happens on the order of $\delta^1$ (which is the reason we must perform the Taylor expansion to $\delta^1$ in Supplementary Notes~\ref{sec_PC_increasei} and \ref{sec_PC_decreasei}). Meanwhile, the instant change in $q_{i|j}$ for $i,j=1,2,\dots,n$ happens on the order of $\delta^0$ since the $\delta^0$ term is non-zero (see Supplementary Note~\ref{sec_edgePC}). That is, the change in $q_{i|j}$ is much faster than $x_i$, so that $x_i$ changes on the basis of $q_{i|j}$ achieving equilibrium. According to Supplementary Note~\ref{sec_edgePC}, we have the following solution when $q_{i|j}$ achieves stability.
\begin{equation}\label{sieq_qij_PC}
    q_{i|j}=\begin{cases}
    \displaystyle{\frac{k-2}{k-1}x_i}, & j\neq i,\\
    \\
    \displaystyle{\frac{k-2}{k-1}x_i+\frac{1}{k-1}}, & j=i.
    \end{cases}
\end{equation}

The primitive replicator equation of $x_i$ is Eq.~(\ref{sieq_replicator_PC}),
where $\mathcal{O}(\delta^2)=0$, $\pi_j^\mathbf{k}$ and $\pi_i^\mathbf{k}$ is given by Eq.~(\ref{sieq_pi_j}), $\pi_{i|j}^\mathbf{k}$ and $\pi_{j|i}^\mathbf{k}$ is given by Eq.~(\ref{sieq_pi_i|j}), and $q_{{i'}|j}$ is given by Eq.~(\ref{sieq_qij_PC}). The degrees of freedom of the replicator dynamics system are $n-1$, represented by independent variables $x_i$ ($i=1,2,\dots,n$, cancel one of them by $\sum_i^n x_i=1$).

\subsection{Simplification and discussion}
The replicator equation given by Eq.~(\ref{sieq_replicator_PC}) can be further simplified and discussed. To do this, we need to introduce a useful equation, as given by Theorem~\ref{1} in \ref{sec_theorem}. For frequency quantities $0\leq z_j\leq 1$ and an arbitrary function $g(\mathbf{k})$ of vector $\mathbf{k}=(k_1,k_2,\dots,k_n)$, we have:
\begin{equation}\label{sieq_simplify}
    \sum_{\sum_{j=1}^n k_j=k}
    \frac{k!}{\prod_{j=1}^n k_j!} \left(\prod_{j=1}^n {z_j}^{k_j}\right)
    k_i g(\mathbf{k})
    =
    kz_i
    \sum_{\sum_{j=1}^n k_j=k-1}
    \frac{(k-1)!}{\prod_{j=1}^n k_j!} \left(\prod_{j=1}^n {z_j}^{k_j}\right)
    g(\mathbf{k}_{+i}),
\end{equation}
which simplifies the expression by canceling the quantity $k_i$ in the summation at the cost of changing $f(\mathbf{k})$ to $f(\mathbf{k}_{+i})$.

\subsubsection{Decomposition to accumulated payoff}
Using Eq.~(\ref{sieq_simplify}), we can simplify the replicator equation~(\ref{sieq_replicator_PC}) to expected accumulated payoffs. That is, we can obtain an equation to avoid specific calculations of $k_i$ in front of $\pi_{i|j}^\mathbf{k}$ and $\pi_j^\mathbf{k}$. The simplified replicator equation is
\begin{align}\label{sieq_pilevel1_PC}
    \dot{x}_i
    =&~\frac{\delta}{4}\sum_{j=1,j\neq i}^n x_j q_{i|j}\sum_{\sum_{i'=1}^n k_{i'}=k-1}\frac{(k-1)!}{\prod_{i'=1}^n k_{i'}!} \left(\prod_{i'=1}^n {q_{{i'}|j}}^{k_{i'}}\right) \left(
    \pi_{i|j}^{\mathbf{k}_{+i}}-\pi_j^{\mathbf{k}_{+i}}
    \right) \nonumber\\
    &~-\frac{\delta}{4}x_i 
    \sum_{\sum_{i'=1}^n k_{i'}=k-1}\frac{(k-1)!}{\prod_{i'=1}^n k_{i'}!} \left(\prod_{i'=1}^n {q_{i'|i}}^{k_{i'}}\right)
    \sum_{j=1,j\neq i}^n
    q_{j|i}\left(
    \pi_{j|i}^{\mathbf{k}_{+j}}-\pi_i^{\mathbf{k}_{+j}}
    \right)  \nonumber\\
    =&~\frac{\delta}{4}x_i\sum_{j=1,j\neq i}^n q_{j|i}\sum_{\sum_{i'=1}^n k_{i'}=k-1}\frac{(k-1)!}{\prod_{i'=1}^n k_{i'}!} \left(\prod_{i'=1}^n {q_{{i'}|j}}^{k_{i'}}\right) \left(
    \pi_{i|j}^{\mathbf{k}_{+i}}-\pi_j^{\mathbf{k}_{+i}}
    \right) \nonumber\\
    &~-\frac{\delta}{4}x_i 
    \sum_{j=1,j\neq i}^n q_{j|i}
    \sum_{\sum_{i'=1}^n k_{i'}=k-1}\frac{(k-1)!}{\prod_{i'=1}^n k_{i'}!} \left(\prod_{i'=1}^n {q_{i'|i}}^{k_{i'}}\right)
    \left(
    \pi_{j|i}^{\mathbf{k}_{+j}}-\pi_i^{\mathbf{k}_{+j}}
    \right).
\end{align}
In this way, we only need to consider the calculation of $\pi_{i|j}^{\mathbf{k}_{+i}}$, $\pi_j^{\mathbf{k}_{+i}}$, $\pi_{j|i}^{\mathbf{k}_{+j}}$, $\pi_i^{\mathbf{k}_{+j}}$ summations.

As mentioned in the main text, let us denote that
\begin{equation}\label{sieq_<pi_i|j>}
    \langle \pi_{i|X}^{\mathbf{k}_{+i}} \rangle
    =\sum_{\sum_{i'=1}^n k_{i'}=k-1}\frac{(k-1)!}{\prod_{i'=1}^n k_{i'}!} \left(\prod_{i'=1}^n {q_{{i'}|X}}^{k_{i'}}\right)
    \pi_{i|X}^{\mathbf{k}_{+i}}.
\end{equation}
The intuition of $\langle \pi_{i|X}^{\mathbf{k}_{+i}} \rangle$ is the expected accumulated payoff of an $i$-player neighboring an $X$-player. In this case, there must be an $i$-player in the $X$-player's neighbor configuration, which is the intuition of `$_{+i}$' in $\mathbf{k}_{+i}$.

We also denote that
\begin{equation}\label{sieq_<pi_j>}
    \langle \pi_X^{\mathbf{k}_{+i}} \rangle
    =\sum_{\sum_{i'=1}^n k_{i'}=k-1}\frac{(k-1)!}{\prod_{i'=1}^n k_{i'}!} \left(\prod_{i'=1}^n {q_{{i'}|X}}^{k_{i'}}\right)
    \pi_X^{\mathbf{k}_{+i}},
\end{equation}
which intuitively means the expected accumulated payoff of an $X$-player whose neighbor configuration contains at least one $i$-player. 

Using the notations of Eqs.~(\ref{sieq_<pi_i|j>}) and (\ref{sieq_<pi_j>}), the replicator equation of Eq.~(\ref{sieq_pilevel1_PC}) can be written as 
\begin{equation}\label{sieq_pilevel1.5_PC}
    \dot{x}_i=\frac{\delta}{4}x_i\sum_{j=1}^n q_{j|i} \left[
    \left(
    \langle \pi_{i|j}^{\mathbf{k}_{+i}} \rangle-\langle \pi_j^{\mathbf{k}_{+i}} \rangle
    \right)
    -
    \left(
    \langle \pi_{j|i}^{\mathbf{k}_{+j}} \rangle-\langle \pi_i^{\mathbf{k}_{+j}} \rangle
    \right)\right].
\end{equation}

According to Theorem~\ref{2} in \ref{sec_theorem}, we have $\langle \pi_{i|j}^{\mathbf{k}_{+i}} \rangle=\langle \pi_i^{\mathbf{k}_{+j}} \rangle$, which can be equivalently written as $\langle \pi_{j|i}^{\mathbf{k}_{+j}} \rangle=\langle \pi_j^{\mathbf{k}_{+i}} \rangle$. Therefore, we have $\langle \pi_{j|i}^{\mathbf{k}_{+j}} \rangle-\langle \pi_i^{\mathbf{k}_{+j}} \rangle=\langle \pi_j^{\mathbf{k}_{+i}} \rangle-\langle \pi_{i|j}^{\mathbf{k}_{+i}} \rangle$, and Eq.~(\ref{sieq_pilevel1.5_PC}) can be simplified as
\begin{equation}\label{sieq_pilevel2_PC}
    \dot{x}_i=\frac{\delta}{2}x_i\sum_{j=1}^n q_{j|i} \left(
    \langle \pi_{i|j}^{\mathbf{k}_{+i}} \rangle-\langle \pi_j^{\mathbf{k}_{+i}} \rangle
    \right).
\end{equation}

According to Theorem~\ref{theorem_sumpiij_pii} in \ref{sec_theorem}, we have $\sum_{j=1}^n q_{j|i}\langle \pi_{i|j}^{\mathbf{k}_{+i}} \rangle=\langle \pi_i^{\mathbf{k}} \rangle$. In this way, we can further write Eq.~(\ref{sieq_pilevel2_PC}) as 
\begin{equation}\label{sieq_pilevel3_PC}
    \dot{x}_i=\frac{\delta}{2}x_i \left(
    \langle \pi_i^\mathbf{k} \rangle-\sum_{j=1}^n q_{j|i}\langle \pi_j^{\mathbf{k}_{+i}} \rangle
    \right),
\end{equation}
which bears an intuitive understanding if we introduce the following concepts:
\begin{itemize}
    \item $\pi_i^{(0)}=\langle \pi_i^\mathbf{k} \rangle$, the expected accumulated payoff of the $i$-player itself (zero-step away on the graph).
    \item $\pi_i^{(1)}=\sum_{j=1}^n q_{j|i}\langle \pi_j^{\mathbf{k}_{+i}} \rangle$, the expected accumulated payoff of the $i$-player's first-order neighbors (one-step away on the graph).
\end{itemize}
Using these concepts, we know that $\dot{x}_i\propto x_i(\pi_i^{(0)}-\pi_i^{(1)})$ as mentioned in the main text. Under pairwise comparison, the reproduction rate of $i$-players depends on how much their expected accumulated payoff higher than neighbors, and the essence of replicator dynamics $\dot{x}_i$ is the competition between oneself and its first-order neighbors. This is consistent with the result obtained by the identity-by-descent idea~\cite{allen2014games,allen2017evolutionary}, but we further generalize it to $n$-strategy systems in replicator dynamics.

To sum up, this section simplifies and divides the replicator equation into expected accumulated payoffs, focusing on elucidating the intuitive insights of the equation.

\subsubsection{Decomposition to single-game payoff}\label{sec_PCsingle}
Next, we further divide the replicator equation of Eq.~(\ref{sieq_pilevel3_PC}) into expected single-game payoffs, stressing the convenience for actual calculation.

For convenience, we introduce a notation to the payoff in a single game as also mentioned in the main text,
\begin{equation}\label{sieq_<a>}
    \langle a_{i|\mathbf{k}}\rangle_j=\sum_{\sum_{i'=1}^n k_{i'}=k}\frac{k!}{\prod_{i'=1}^n k_{i'}!} \left(\prod_{i'=1}^n {q_{{i'}|j}}^{k_{i'}}\right)
    a_{i|\mathbf{k}},
\end{equation}
which means the expected payoff of an $i$-player in a single game with co-player configuration $\mathbf{k}$ found near a $j$-player. Please note that $\langle a_{i|\mathbf{k}}\rangle_j$ is only for introducing the concept and does not have actual physical sense, because an $i$-player cannot have all of its $k$ co-players found near a $j$-player on a graph if $j\neq i$. Only expressions such as $\langle a_{i|\mathbf{k}}\rangle_i$, $\langle a_{i|\mathbf{k}_{+j}}\rangle_i$, and $\langle a_{i|\mathbf{k}_{+j}}\rangle_j$ are meaningful, which will be applied.

Applying the notation of Eq.~(\ref{sieq_<a>}) to $\langle \pi_j^{\mathbf{k}_{+i}} \rangle$ in Eq.~(\ref{sieq_<pi_j>}) and also consider Eq.~(\ref{sieq_pi_j}), we have
\begin{align}\label{sieq_<pi_j>_<a>}
    \langle \pi_j^{\mathbf{k}_{+i}} \rangle
    =&~\sum_{\sum_{i'=1}^n k_{i'}=k-1}\frac{(k-1)!}{\prod_{i'=1}^n k_{i'}!} \left(\prod_{i'=1}^n {q_{{i'}|j}}^{k_{i'}}\right)
    \Bigg[a_{j|\mathbf{k}_{+i}}+
    \sum_{\sum_{\ell=1}^n k'_\ell=k-1}\frac{(k-1)!}{\prod_{\ell=1}^n k'_\ell!} \left(\prod_{\ell=1}^n {q_{\ell|i}}^{k'_\ell}\right) a_{j|\mathbf{k}'_{+i}}
    \nonumber\\
    &~+\sum_{l=1}^n k_l \sum_{\sum_{\ell=1}^n k'_\ell=k-1}\frac{(k-1)!}{\prod_{\ell=1}^n k'_\ell!} \left(\prod_{\ell=1}^n {q_{\ell|l}}^{k'_\ell}\right) a_{j|\mathbf{k}'_{+l}}
    \Bigg] \nonumber\\
    =&~\sum_{\sum_{i'=1}^n k_{i'}=k-1}\frac{(k-1)!}{\prod_{i'=1}^n k_{i'}!} \left(\prod_{i'=1}^n {q_{{i'}|j}}^{k_{i'}}\right) a_{j|\mathbf{k}_{+i}}
    +\sum_{\sum_{\ell=1}^n k'_\ell=k-1}\frac{(k-1)!}{\prod_{\ell=1}^n k'_\ell!} \left(\prod_{\ell=1}^n {q_{\ell|i}}^{k'_\ell}\right) a_{j|\mathbf{k}'_{+i}} \nonumber\\
    &~+(k-1)\sum_{l=1}^n q_{l|j} \sum_{\sum_{\ell=1}^n k'_\ell=k-1}\frac{(k-1)!}{\prod_{\ell=1}^n k'_\ell!} \left(\prod_{\ell=1}^n {q_{\ell|l}}^{k'_\ell}\right) a_{j|\mathbf{k}'_{+l}}
    \nonumber\\
    =&~\langle a_{j|\mathbf{k}_{+i}}\rangle_j+
    \langle a_{j|\mathbf{k}_{+i}}\rangle_i+
    (k-1)\sum_{l=1}^n q_{l|j}\langle a_{j|\mathbf{k}_{+l}}\rangle_l.
\end{align}
The last step writes $\mathbf{k}'$ as $\mathbf{k}$ because the configurations are calculated within each $\langle \cdot\rangle$ separately. Intuitively, the expected accumulated payoff of an $i$-player's neighboring $j$-player $\langle \pi_j^{\mathbf{k}_{+i}} \rangle$ consists of the following components: $\langle a_{j|\mathbf{k}_{+i}}\rangle_j$ is from the game organized by the $j$-player, $\langle a_{j|\mathbf{k}_{+i}}\rangle_i$ is from the game organized by the $i$-player, and $\langle a_{j|\mathbf{k}_{+l}}\rangle_l$ is from the game organized by the remaining $k-1$ possible $l$-players neighboring the $j$-player.

Similarly, we have
\begin{align}\label{sieq_<pi_i>_<a>}
    \langle \pi_i^\mathbf{k} \rangle
    =&~\sum_{\sum_{i'=1}^n k_{i'}=k}\frac{k!}{\prod_{i'=1}^n k_{i'}!} \left(\prod_{i'=1}^n {q_{{i'}|i}}^{k_{i'}}\right)
    \pi_i^{\mathbf{k}}
    \nonumber\\
    =&~\sum_{\sum_{i'=1}^n k_{i'}=k}\frac{k!}{\prod_{i'=1}^n k_{i'}!} \left(\prod_{i'=1}^n {q_{{i'}|i}}^{k_{i'}}\right)
    \left[
    a_{i|\mathbf{k}}+
    \sum_{l=1}^n k_l \sum_{\sum_{\ell=1}^n k'_\ell=k-1}\frac{(k-1)!}{\prod_{\ell=1}^n k'_\ell!} \left(\prod_{\ell=1}^n {q_{\ell|l}}^{k'_\ell}\right) a_{i|\mathbf{k}'_{+l}}
    \right]
    \nonumber\\
    =&~\sum_{\sum_{i'=1}^n k_{i'}=k}\frac{k!}{\prod_{i'=1}^n k_{i'}!} \left(\prod_{i'=1}^n {q_{{i'}|i}}^{k_{i'}}\right)a_{i|\mathbf{k}}
    +k\sum_{l=1}^n q_{l|i}\sum_{\sum_{\ell=1}^n k'_\ell=k-1}\frac{(k-1)!}{\prod_{\ell=1}^n k'_\ell!} \left(\prod_{\ell=1}^n {q_{\ell|l}}^{k'_\ell}\right) a_{i|\mathbf{k}'_{+l}}
    \nonumber\\
    =&~\langle a_{i|\mathbf{k}}\rangle_i+
    k\sum_{l=1}^n q_{l|i} \langle a_{i|\mathbf{k}'_{+l}}\rangle_l
    \nonumber\\
    =&~\langle a_{i|\mathbf{k}}\rangle_i+
    k\sum_{j=1}^n q_{j|i} \langle a_{i|\mathbf{k}_{+j}}\rangle_j.
\end{align}
The expected accumulated payoff of an $i$-player $\langle \pi_i^\mathbf{k} \rangle$ consists of the following components: $\langle a_{i|\mathbf{k}}\rangle_i$ is from the game organized by the $i$-player itself, and $\langle a_{i|\mathbf{k}_{+j}}\rangle_j$ is from the game organized by a neighboring $j$-player. 

Using Eqs.~(\ref{sieq_<pi_j>_<a>}) and (\ref{sieq_<pi_i>_<a>}), we can simplify the replicator equation of Eq.~(\ref{sieq_pilevel3_PC}) into expected single-game payoffs:
\begin{equation}\label{sieq_alevel1_PC}
    \dot{x}_i=\frac{\delta}{2}x_i \left[\left(
    \langle a_{i|\mathbf{k}}\rangle_i+
    k\sum_{j=1}^n q_{j|i} \langle a_{i|\mathbf{k}_{+j}}\rangle_j
    \right)
    -\sum_{j=1}^n q_{j|i}\left(\langle a_{j|\mathbf{k}_{+i}}\rangle_j+
    \langle a_{j|\mathbf{k}_{+i}}\rangle_i+
    (k-1)\sum_{l=1}^n q_{l|j}\langle a_{j|\mathbf{k}_{+l}}\rangle_l
    \right)\right].
\end{equation}

To decrease the number of elements, we use the relation $\langle a_{i|\mathbf{k}}\rangle_i=\sum_{j=1}^{n} q_{j|i} \langle a_{i|\mathbf{k}_{+j}}\rangle_i$ according to Theorem~\ref{theorem_aiki_qjiaiji} in \ref{sec_theorem}. In this way, Eq.~(\ref{sieq_alevel1_PC}) can be written as
\begin{equation}\label{sieq_alevel2_PC}
    \dot{x}_i=\frac{\delta}{2}x_i \sum_{j=1}^n q_{j|i}
    \left(
    \langle a_{i|\mathbf{k}_{+j}}\rangle_i+
    k \langle a_{i|\mathbf{k}_{+j}}\rangle_j
    -\langle a_{j|\mathbf{k}_{+i}}\rangle_j
    -\langle a_{j|\mathbf{k}_{+i}}\rangle_i
    -(k-1)\sum_{l=1}^n q_{l|j} 
    \langle a_{j|\mathbf{k}_{+l}}\rangle_l
    \right).
\end{equation}
Using the relation between $x_j$ and $q_{j|i}$ according to Eq.~(\ref{sieq_qij_PC}), we can write Eq.~(\ref{sieq_alevel2_PC}) again as
\begin{equation}\label{sieq_alevel3_PC}
    \dot{x}_i=\frac{\delta(k-2)}{2(k-1)}x_i \sum_{j=1}^n x_j
    \Bigg(
    \langle a_{i|\mathbf{k}_{+j}}\rangle_i+
    (k-1) \langle a_{i|\mathbf{k}_{+j}}\rangle_j
    +\langle a_{i|\mathbf{k}_{+i}}\rangle_i
    -\langle a_{j|\mathbf{k}_{+i}}\rangle_j
    -\langle a_{j|\mathbf{k}_{+i}}\rangle_i
    -(k-2)\sum_{l=1}^n x_l 
    \langle a_{j|\mathbf{k}_{+l}}\rangle_l
    -\langle a_{j|\mathbf{k}_{+j}}\rangle_j
    \Bigg).
\end{equation}

For specific applications, we can employ Eq.~(\ref{sieq_alevel2_PC}) or Eq.~(\ref{sieq_alevel3_PC}) according to the actual needs. On the one hand, adopting Eq.~(\ref{sieq_alevel2_PC}) cannot avoid subsequently transforming $q_{j|i}$ to $x_j$ manually. We always need to transform $q_{j|i}$ to $x_j$ to obtain the final replicator equation, but at which stage we compute it depends on the actual type of complexity. Adopting Eq.~(\ref{sieq_alevel3_PC}) may be a quick solution at the cost of substituting additional $\langle a_{i|\mathbf{k}_{+i}}\rangle_i$ and $\langle a_{j|\mathbf{k}_{+j}}\rangle_j$ manually. However, we should note that the calculation inside each $\langle \cdot\rangle$ contains $q_{j|i}$ and we still need to transform them manually.

As mentioned in the main text, the advantage of Eqs.~(\ref{sieq_alevel2_PC}) and (\ref{sieq_alevel3_PC}) is that we have attributed everything about $\langle \cdot\rangle$ into two types, the `$\langle a_{i|\mathbf{k}_{+j}}\rangle_i$ type' and the `$\langle a_{i|\mathbf{k}_{+j}}\rangle_j$ type'. They can be expressed by matrices by going through $i$ and $j$:
\begin{itemize}
    \item The $\langle a_{i|\mathbf{k}_{+j}}\rangle_i$ type: 
    \begin{equation}\label{sieq_aiji_type}
        \left[\langle a_{i|\mathbf{k}_{+j}}\rangle_i\right]_{ij}=
        \displaystyle{\begin{pmatrix}
        \langle a_{1|\mathbf{k}_{+1}}\rangle_1 & \langle a_{1|\mathbf{k}_{+2}}\rangle_1 & \cdots & \langle a_{1|\mathbf{k}_{+n}}\rangle_1 \\
        \langle a_{2|\mathbf{k}_{+1}}\rangle_2 & \langle a_{2|\mathbf{k}_{+2}}\rangle_2 & \cdots & \langle a_{2|\mathbf{k}_{+n}}\rangle_2 \\
        \vdots & \vdots & \ddots & \vdots \\
        \langle a_{n|\mathbf{k}_{+1}}\rangle_n & \langle a_{n|\mathbf{k}_{+2}}\rangle_n & \cdots & \langle a_{n|\mathbf{k}_{+n}}\rangle_n
    \end{pmatrix}}.
    \end{equation}

    \item The $\langle a_{i|\mathbf{k}_{+j}}\rangle_j$ type: 
    \begin{equation}\label{sieq_aijj_type}
        \left[\langle a_{i|\mathbf{k}_{+j}}\rangle_j\right]_{ij}=
        \displaystyle{\begin{pmatrix}
        \langle a_{1|\mathbf{k}_{+1}}\rangle_1 & \langle a_{1|\mathbf{k}_{+2}}\rangle_2 & \cdots & \langle a_{1|\mathbf{k}_{+n}}\rangle_n \\
        \langle a_{2|\mathbf{k}_{+1}}\rangle_1 & \langle a_{2|\mathbf{k}_{+2}}\rangle_2 & \cdots & \langle a_{2|\mathbf{k}_{+n}}\rangle_n \\
        \vdots & \vdots & \ddots & \vdots \\
        \langle a_{n|\mathbf{k}_{+1}}\rangle_1 & \langle a_{n|\mathbf{k}_{+2}}\rangle_2 & \cdots & \langle a_{n|\mathbf{k}_{+n}}\rangle_n
    \end{pmatrix}}.
    \end{equation}
\end{itemize}

Therefore, we attribute the problem to computing each elements in the aforementioned two types before substituting them to Eq.~(\ref{sieq_alevel2_PC}) or Eq.~(\ref{sieq_alevel3_PC}). The result of each element should be a function of $x_1, x_2, \dots, x_n$ (transformed from $q_{j|i}$ manually) and game parameters.

There are $n^2$ elements in each matrix. Their diagonals are equal, meaning we can compute $n$ fewer elements. Therefore, the total computation is $n^2+n^2-n=(2n-1)n$ elements. The computational complexity is $\mathrm{O}(n^2)$, which is feasible within polynomial time.

\subsubsection{Special linear system}\label{sec_linear}
We can further simplify the calculation when faced with special payoff structures. Although the payoff function $a_{i|\mathbf{k}}$ in a multiplayer game can be arbitrary, one of the most common cases is the linear payoff function. For example, the public goods game is a 2-strategy game with linear payoff function.

Given a co-player configuration $\mathbf{k}$, the linear payoff function is defined as containing only primary terms for $k_1,k_2,\dots,k_n$ and a constant term. Let us denote the coefficient matrix $\mathbf{b}$ and the constant vector $\mathbf{c}$,
\begin{equation}\label{sieq_b_c}
    \mathbf{b}=\begin{pmatrix}
        b_{11} & b_{12} & \cdots & b_{1n} \\
        b_{21} & b_{22} & \cdots & b_{2n} \\
        \vdots & \vdots & \ddots & \vdots \\
        b_{n1} & b_{n2} & \cdots & b_{nn}
    \end{pmatrix},~
    \mathbf{c}=\begin{pmatrix}
        c_1 \\
        c_2 \\
        \vdots \\
        c_n
    \end{pmatrix}.
\end{equation}
Then, the payoff for different strategies in a single game with co-player configuration $\mathbf{k}$ can be expressed as 
\begin{equation}
    \begin{pmatrix}
        a_{1|\mathbf{k}} \\
        a_{2|\mathbf{k}} \\
        \vdots \\
        a_{n|\mathbf{k}}
    \end{pmatrix}
    =\mathbf{b}\cdot\mathbf{k}^\top+\mathbf{c}
    =\begin{pmatrix}
        b_{11} & b_{12} & \cdots & b_{1n} \\
        b_{21} & b_{22} & \cdots & b_{2n} \\
        \vdots & \vdots & \ddots & \vdots \\
        b_{n1} & b_{n2} & \cdots & b_{nn}
    \end{pmatrix}
    \begin{pmatrix}
        k_1 \\
        k_2 \\
        \vdots \\
        k_n
    \end{pmatrix}
    +\begin{pmatrix}
        c_1 \\
        c_2 \\
        \vdots \\
        c_n
    \end{pmatrix},
\end{equation}
or
\begin{equation}\label{sieq_linear_a_PC}
    a_{i|\mathbf{k}}=\sum_{l=1}^n b_{il}k_l+c_i.
\end{equation}

Our goal is to obtain the simplified replicator equation for any linear payoff function depicted by $\mathbf{b}$ and $\mathbf{c}$. Making a small transformation to Eq.~(\ref{sieq_linear_a_PC}), we have
\begin{equation}
    a_{i|\mathbf{k}_{+j}}=\sum_{l=1}^n b_{il}k_l+b_{ij}+c_i.
\end{equation}

Importantly, we recall that a simple special case of Eq.~(\ref{sieq_simplify}) (Theorem~\ref{1} in \ref{sec_theorem}) is
\begin{equation}\label{sieq_simplify2}
    \sum_{\sum_{j=1}^n k_j=k}
    \frac{k!}{\prod_{j=1}^n k_j!} \left(\prod_{j=1}^n {z_j}^{k_j}\right)
    k_i
    =
    kz_i.
\end{equation}
In this way, we have
\begin{subequations}
\begin{align}
    &\langle a_{i|\mathbf{k}_{+j}}\rangle_i
    =(k-1)\sum_{l=1}^n b_{il}q_{l|i}+b_{ij}+c_i
    =(k-2)\sum_{l=1}^n b_{il}x_l+b_{ii}+b_{ij}+c_i, \label{sieq_aiji_bc} \\
    &\langle a_{i|\mathbf{k}_{+j}}\rangle_j
    =(k-1)\sum_{l=1}^n b_{il}q_{l|j}+b_{ij}+c_i
    =(k-2)\sum_{l=1}^n b_{il}x_l+2b_{ij}+c_i,
    \label{sieq_aijj_bc} 
\end{align}
\end{subequations}
which gives the elements of the $\langle a_{i|\mathbf{k}_{+j}}\rangle_i$ type and the $\langle a_{i|\mathbf{k}_{+j}}\rangle_j$ type. We can directly utilize the replicator equation divided into expected single-game payoffs given by Eq.~(\ref{sieq_alevel3_PC}). After calculating and organizing, we obtain
\begin{equation}\label{sieq_linear_PC}
    \dot{x}_1=\frac{\delta(k-2)}{2(k-1)}x_i\left(
    (k+1)(\Bar{\pi}_i-\Bar{\pi})
    +3\sum_{j=1}^n x_j (b_{ii}-b_{ij}-b_{ji}-b_{jj})
    +6\sum_{j=1}^n \sum_{l=1}^n x_j x_l b_{jl}
    \right),
\end{equation}
where $\Bar{\pi}_i$ is the mean payoff of $i$-players in a well-mixed population, 
\begin{equation}
    \Bar{\pi}_i=k\sum_{l=1}^n x_l b_{il}+c_i,
\end{equation}
and $\Bar{\pi}$ is the mean payoff of all individuals in a well-mixed population,
\begin{equation}
    \Bar{\pi}=\sum_{i=1}^n x_i\Bar{\pi}_i=k\sum_{i=1}^n \sum_{l=1}^n x_i x_l b_{il}+\sum_{i=1}^n x_i c_i.
\end{equation}

We know that the replicator equation in a well-mixed population is $\dot{x}_i=x_i(\Bar{\pi}_i-\Bar{\pi})$. In this way, Eq.~(\ref{sieq_linear_PC}) clearly shows the additional terms brought by pairwise comparison in a structured population compared to the well-mixed population.

\subsection{Comparison with two-strategy dynamics}\label{sec_PCn=2}
Let us discuss on how our replicator equations for $n$-strategy systems reduce to the 2-strategy system under pairwise comparison. When $n=2$, the configuration $\mathbf{k}=(k_1,k_2)=(k_1,k-k_1)$ can be represented by only $k_1$, where $k_1=0,1,\dots,k$. Similarly, we have $\mathbf{k}_{+1}=(k_1+1,k-k_1-1)$, $\mathbf{k}_{+2}=(k_1,k-k_1+1)$, where $k_1=0,1,\dots,k-1$.

For $n=2$, the accumulated payoff of a $j$-player with neighbor configuration $\mathbf{k}$ can be expressed as
\begin{align}
    \pi_j^\mathbf{k}=a_{j|\mathbf{k}}+
     k_1 \sum_{k'_1=0}^{k-1}\frac{(k-1)!}{k'_1!(k-k'_1-1)!}
     {q_{1|1}}^{k'_1}{q_{2|1}}^{k-k'_1-1}
     a_{j|\mathbf{k}'_{+1}}
     +
     (k-k_1) \sum_{k'_1=0}^{k-1}\frac{(k-1)!}{k'_1!(k-k'_1-1)!}
     {q_{1|2}}^{k'_1}{q_{2|2}}^{k-k'_1-1}
     a_{j|\mathbf{k}'_{+2}},
\end{align}
whereas the accumulated payoff of an $i$-player, neighboring a focal $j$-player with neighbor configuration $\mathbf{k}$, can be expressed as
\begin{align}
    \pi_{i|j}^\mathbf{k}=&~a_{i|\mathbf{k}_{-i,+j}}+
     \sum_{k'_1=0}^{k-1}\frac{(k-1)!}{k'_1!(k-k'_1-1)!}
     {q_{1|i}}^{k'_1}{q_{2|i}}^{k-k'_1-1}
     \Bigg(a_{i|\mathbf{k}'_{+j}}+
     k'_1 \sum_{k''_1=0}^{k-1}\frac{(k-1)!}{k''_1!(k-k''_1-1)!}
     {q_{1|1}}^{k''_1}{q_{2|1}}^{k-k''_1-1}
     a_{i|\mathbf{k}''_{+1}} \nonumber\\
     &+(k-k'_1-1) \sum_{k''_1=0}^{k-1}\frac{(k-1)!}{k''_1!(k-k''_1-1)!}
     {q_{1|2}}^{k''_1}{q_{2|2}}^{k-k''_1-1}
     a_{i|\mathbf{k}''_{+2}}
     \Bigg).
\end{align}

Comparing Eqs.~(\ref{sieq_replicator_PC}), (\ref{sieq_pilevel1_PC}), (\ref{sieq_pilevel1.5_PC}), and (\ref{sieq_pilevel2_PC}), we first write the primitive $n$-strategy replicator equation of Eq.~(\ref{sieq_replicator_PC}) as follows:
\begin{equation}\label{sieq_PCforn=2}
    \dot{x}_i=\frac{\delta}{2} \sum_{j=1,j\neq i}^n x_j \sum_{\sum_{i'=1}^n k_{i'}=k}\frac{k!}{\prod_{i'=1}^n k_{i'}!} \left(\prod_{i'=1}^n {q_{{i'}|j}}^{k_{i'}}\right) \frac{k_i}{k}\left(
    \pi_{i|j}^\mathbf{k}-\pi_j^\mathbf{k}
    \right).
\end{equation}

Then, we discuss the corresponding $n=2$ case. For $n=2$, the system state can be described by only one of $x_1$ and $x_2$, because $x_1+x_2=1$, $\dot{x}_1+\dot{x}_2=0$. Let us choose $x_1$ to describe the system. According to Eq.~(\ref{sieq_PCforn=2}), the replicator equation for $n=2$ can be written as
\begin{align}\label{sieq_replicator_PC_n2}
    \dot{x}_1=\frac{\delta}{2}
    (1-x_1) \sum_{k_1=0}^{k}\frac{k!}{k_1!(k-k_1)!}
    {q_{1|2}}^{k_1}{q_{2|2}}^{k-k_1} \frac{k_1}{k} (\pi_{1|2}^\mathbf{k}-\pi_2^\mathbf{k}),
\end{align}
which is complete for application. If we want to analyze Eq.~(\ref{sieq_replicator_PC_n2}) further, we can do the following calculation.
\begin{align}
    \pi_{1|2}^\mathbf{k}-\pi_2^\mathbf{k}
    =&~a_{1|\mathbf{k}_{-1,+2}}+
    \sum_{k'_1=0}^{k-1}\frac{(k-1)!}{k'_1!(k-k'_1-1)!}
    {q_{1|1}}^{k'_1}{q_{2|1}}^{k-k'_1-1}
    \Bigg(a_{1|\mathbf{k}'_{+2}}+
    k'_1 \sum_{k''_1=0}^{k-1}\frac{(k-1)!}{k''_1!(k-k''_1-1)!}
    {q_{1|1}}^{k''_1}{q_{2|1}}^{k-k''_1-1}
    a_{1|\mathbf{k}''_{+1}} \nonumber\\
    &+(k-k'_1-1) \sum_{k''_1=0}^{k-1}\frac{(k-1)!}{k''_1!(k-k''_1-1)!}
    {q_{1|2}}^{k''_1}{q_{2|2}}^{k-k''_1-1}
    a_{1|\mathbf{k}''_{+2}}
    \Bigg)
    -a_{2|\mathbf{k}} \nonumber\\
    &-
    k_1 \sum_{k'_1=0}^{k-1}\frac{(k-1)!}{k'_1!(k-k'_1-1)!}
    {q_{1|1}}^{k'_1}{q_{2|1}}^{k-k'_1-1}
    a_{2|\mathbf{k}'_{+1}}
    -
    (k-k_1) \sum_{k'_1=0}^{k-1}\frac{(k-1)!}{k'_1!(k-k'_1-1)!}
    {q_{1|2}}^{k'_1}{q_{2|2}}^{k-k'_1-1}
    a_{2|\mathbf{k}'_{+2}} \nonumber\\
    =&~a_{1|\mathbf{k}_{-1,+2}}+
    \sum_{k'_1=0}^{k-1}\frac{(k-1)!}{k'_1!(k-k'_1-1)!}
    {q_{1|1}}^{k'_1}{q_{2|1}}^{k-k'_1-1}
    a_{1|\mathbf{k}'_{+2}}+
    (k-1)q_{1|1} \sum_{k'_1=0}^{k-1}\frac{(k-1)!}{k'_1!(k-k'_1-1)!}
    {q_{1|1}}^{k'_1}{q_{2|1}}^{k-k'_1-1}
    a_{1|\mathbf{k}'_{+1}} \nonumber\\
    &+(k-1)q_{2|1} \sum_{k'_1=0}^{k-1}\frac{(k-1)!}{k'_1!(k-k'_1-1)!}
    {q_{1|2}}^{k'_1}{q_{2|2}}^{k-k'_1-1}
    a_{1|\mathbf{k}'_{+2}}
    -a_{2|\mathbf{k}} \nonumber\\
    &-
    k_1 \sum_{k'_1=0}^{k-1}\frac{(k-1)!}{k'_1!(k-k'_1-1)!}
    {q_{1|1}}^{k'_1}{q_{2|1}}^{k-k'_1-1}
    a_{2|\mathbf{k}'_{+1}}
    -
    (k-k_1) \sum_{k'_1=0}^{k-1}\frac{(k-1)!}{k'_1!(k-k'_1-1)!}
    {q_{1|2}}^{k'_1}{q_{2|2}}^{k-k'_1-1}
    a_{2|\mathbf{k}'_{+2}},
\end{align}
and
\begin{align}\label{sieq_replin2part2}
    &~\sum_{k_1=0}^{k}\frac{k!}{k_1!(k-k_1)!}
    {q_{1|2}}^{k_1}{q_{2|2}}^{k-k_1} \frac{k_1}{k} (\pi_{1|2}^\mathbf{k}-\pi_2^\mathbf{k}) \nonumber\\
    =&~q_{1|2}\sum_{k_1=0}^{k-1}\frac{(k-1)!}{k_1!(k-k_1-1)!}
    {q_{1|2}}^{k_1}{q_{2|2}}^{k-k_1-1}
    (a_{1|\mathbf{k}_{+2}}-a_{2|\mathbf{k}_{+1}})
    +q_{1|2}\sum_{k'_1=0}^{k-1}\frac{(k-1)!}{k'_1!(k-k'_1-1)!}
    {q_{1|1}}^{k'_1}{q_{2|1}}^{k-k'_1-1}
    a_{1|\mathbf{k}'_{+2}} \nonumber\\
    &+q_{1|1}(k-1)q_{1|2}\sum_{k'_1=0}^{k-1}\frac{(k-1)!}{k'_1!(k-k'_1-1)!}
    {q_{1|1}}^{k'_1}{q_{2|1}}^{k-k'_1-1}
    a_{1|\mathbf{k}'_{+1}}
    +q_{2|1}(k-1)q_{1|2}\sum_{k'_1=0}^{k-1}\frac{(k-1)!}{k'_1!(k-k'_1-1)!}
    {q_{1|2}}^{k'_1}{q_{2|2}}^{k-k'_1-1}
    a_{1|\mathbf{k}'_{+2}} \nonumber\\
    &-q_{1|2}[1+(k-1)q_{1|2}]\sum_{k'_1=0}^{k-1}\frac{(k-1)!}{k'_1!(k-k'_1-1)!}
    {q_{1|1}}^{k'_1}{q_{2|1}}^{k-k'_1-1}
    a_{2|\mathbf{k}'_{+1}} \nonumber\\
    &-(k-1)q_{1|2}q_{2|2}\sum_{k'_1=0}^{k-1}\frac{(k-1)!}{k'_1!(k-k'_1-1)!}
    {q_{1|2}}^{k'_1}{q_{2|2}}^{k-k'_1-1}
    a_{2|\mathbf{k}'_{+2}} \nonumber\\
    =&~q_{1|2}\Big\{
    \langle a_{1|\mathbf{k}_{+2}}\rangle_1-\langle a_{2|\mathbf{k}_{+1}}\rangle_2
    +[(k-2)x_1+1]\left(\langle a_{1|\mathbf{k}_{+1}}\rangle_1-\langle a_{2|\mathbf{k}_{+1}}\rangle_1\right) \nonumber\\
    &+[(k-2)(1-x_1)+1]\left(\langle a_{1|\mathbf{k}_{+2}}\rangle_2-\langle a_{2|\mathbf{k}_{+2}}\rangle_2\right)
    \Big\}.
\end{align}

Applying Eq.~(\ref{sieq_replin2part2}) to Eq.~(\ref{sieq_replicator_PC_n2}), we can obtain Eq.~(\ref{sieq_PC_single_n2}), which allows us to compute the replicator dynamics through expected single-game payoffs.
\begin{align}\label{sieq_PC_single_n2}
    \dot{x}_1=&~\frac{\delta(k-2)}{2(k-1)}x_1(1-x_1)
    \Big\{\langle a_{1|\mathbf{k}_{+2}}\rangle_1-\langle a_{2|\mathbf{k}_{+1}}\rangle_2
    +[(k-2)x_1+1]\left(\langle a_{1|\mathbf{k}_{+1}}\rangle_1-\langle a_{2|\mathbf{k}_{+1}}\rangle_1\right) \nonumber\\
    &+[(k-2)(1-x_1)+1]\left(\langle a_{1|\mathbf{k}_{+2}}\rangle_2-\langle a_{2|\mathbf{k}_{+2}}\rangle_2\right)
    \Big\}.
\end{align}
To be sure, Eq.~(\ref{sieq_PC_single_n2}) can also be obtained by applying $n=2$ to Eq.~(\ref{sieq_alevel3_PC}) directly.

Next, we present the case of $n=2$ for special linear systems. According to Eq.~(\ref{sieq_linear_PC}), when it is sufficient for $\mathbf{b}$ and $\mathbf{c}$ to describe the payoff structure, the replicator dynamics for $n=2$ can be simplified as
\begin{equation}\label{sieq_n2_PC_linear}
    \dot{x}_1=\frac{\delta(k-2)(k+1)}{2(k-1)}
    \left[x_1(\Bar{\pi}_1-\Bar{\pi})
    +\frac{3}{k+1}x_1(1-x_1)(1-2x_1)(b_{11}-b_{12}-b_{21}+b_{22})\right],
\end{equation}
where $x_1(\Bar{\pi}_1-\Bar{\pi})$ corresponds to the replicator dynamics in a well-mixed population, which can be obtained by the knowledge for well-mixed populations. Using $\mathbf{b}$ and $\mathbf{c}$ to express the payoff, we can also write them as
\begin{equation}
    x_1(\Bar{\pi}_1-\Bar{\pi})=x_1(1-x_1)[k(b_{11}-b_{21})x_1+k(b_{12}-b_{22})(1-x_1)+c_1-c_2].
\end{equation}

\section{: Applications}\label{sec_appli}
We give five examples here to demonstrate how to apply our general replicator equations to specific models. We include the traditional public goods game ($n=2$), public goods games with peer punishment ($n=3$, linear) \& pool punishment ($n=3$, nonlinear), public goods games with the reward mechanism ($n=3$, linear), and the multi-stage public goods game ($n=4$, linear). It is worth noting that all applications are done under the pairwise comparison rule for strategy evolution.

\subsection{The traditional public goods game ($n=2$)}\label{sec_PGG}
The traditional public goods game in structured populations has been studied under both pairwise comparison~\cite{wang2023inertia,zhang2023cooperation} and death-birth~\cite{li2014cooperation,li2016evolutionary}. Here, we present the results under pairwise comparison within our framework.

As mentioned in the main text, there are $n=2$ strategies in the traditional public goods game:

$1=\text{Cooperation}$ ($C$);

$2=\text{Defection}$ ($D$).

Cooperation means investing $c$ in the common pool containing the co-players in $\mathbf{k}$ and oneself. Defection means investing nothing. The investment of all these $k+1$ players ($k_1 c$ for a focal defector and $(k_1+1)c$ for a focal cooperator) is enlarged by a synergy factor $r$ ($r>1$). The resultant public goods ($rk_1 c$ for a focal defector and $r(k_1+1)c$ for a focal cooperator) are equally distributed to all $k+1$ players. Therefore, for co-player configuration $\mathbf{k}=(k_1,k_2)$, we have the following payoff calculation in a single game,
\begin{subequations}\label{sieq_a_pgg}
\begin{align}
a_{1|\mathbf{k}}&=\frac{r(k_1+1)c}{k+1}-c
=\frac{rc}{k+1}k_1+\frac{rc}{k+1}-c, \\
a_{2|\mathbf{k}}&=\frac{rk_1c}{k+1}
=\frac{rc}{k+1}k_1.
\end{align}
\end{subequations}

\subsubsection{The well-mixed population}
The frequencies of strategies 1 and 2 are denoted by $x_1$ and $x_2=1-x_1$. In a well-mixed population, the mean payoffs of the two strategies are calculated as follows.
\begin{subequations}
\begin{align}
\Bar{\pi}_1&
=\sum_{k_1=0}^{k}\frac{k!}{k_1! (k-k_1)!} {x_1}^{k_1}(1-x_1)^{k-k_1} a_{1|\mathbf{k}}
=\frac{rc}{k+1}kx_1+\frac{rc}{k+1}-c, \\
\Bar{\pi}_2&
=\sum_{k_1=0}^{k}\frac{k!}{k_1! (k-k_1)!} {x_1}^{k_1}(1-x_1)^{k-k_1} a_{2|\mathbf{k}}
=\frac{rc}{k+1}kx_1.
\end{align}
\end{subequations}
The mean payoff of the total population is
\begin{equation}
    \Bar{\pi}
=x_1\Bar{\pi}_1+(1-x_1)\Bar{\pi}_2
=(r-1)c x_1.
\end{equation}
According to the traditional replicator dynamics in well-mixed populations, we have
\begin{equation}
\dot{x}_1
=x_1(\Bar{\pi}_1-\Bar{\pi})
=x_1(1-x_1)\left(\frac{rc}{k+1}-c\right).
\end{equation}

It is clear that the system has two equilibrium points, $x_1=0$ and $x_1=1$. When $r<k+1$, the system is stable at $x_1=0$. When $r>k+1$, the system is stable at $x_1=1$.

\subsubsection{The structured population}\label{sec_PGG_st}
We observe that the payoff structure in Eq.~(\ref{sieq_a_pgg}) is a linear function of $\mathbf{k}$. That is, the payoff in a single game can be expressed as $(a_{1|\mathbf{k}},a_{2|\mathbf{k}})^\top=\mathbf{b}\cdot\mathbf{k}^\top+\mathbf{c}$, where
\begin{equation}
    \mathbf{b}=
    \begin{pmatrix}
        \dfrac{rc}{k+1} & 0 \\[1em]
        \dfrac{rc}{k+1} & 0 
    \end{pmatrix},~
    \mathbf{c}=
    \begin{pmatrix}
        \dfrac{rc}{k+1}-c \\[1em]
        0
    \end{pmatrix}.
\end{equation}
Therefore, we can utilize the conclusion of special linear systems given by Eq.~(\ref{sieq_n2_PC_linear}). The replicator equation for the traditional public goods game in a structured population is
\begin{equation}
    \dot{x}_1=\frac{\delta(k-2)(k+1)}{2(k-1)}
    x_1(\Bar{\pi}_1-\Bar{\pi})
    =\frac{\delta(k-2)(k+1)}{2(k-1)}x_1(1-x_1)\left(\frac{rc}{k+1}-c\right).
\end{equation}
The coefficient in front of $x_1(\Bar{\pi}_1-\Bar{\pi})$ does not affect the equilibrium points and stability. Therefore, we can say that evolutionary dynamics of the public goods game in structured populations under pairwise comparison has no difference from the well-mixed population in the weak selection limit. We can attribute this to $b_{11}-b_{12}-b_{21}+b_{22}=0$.

\subsubsection{When does pairwise comparison have no effect on evolution?}\label{sec_PCequateWMandST}
It is an existing conclusion in previous literature that the equilibrium points and stability of public goods games under pairwise comparison in structured populations are equivalent to the well-mixed population~\cite{wang2023inertia}. This is also the reason why many works prefer death-birth when studying structured populations~\cite{li2014cooperation,li2016evolutionary,su2019spatial,wang2023inertia,wang2023greediness}, whose evolutionary dynamics can make difference from the well-mixed population.

To our knowledge, there is no previous findings of an equivalence between the PC rule in well-mixed and structured populations for nonlinear payoff functions. The accidental equivalence happens for linear payoff functions, such as the public goods game. Nonetheless, the linearity does not necessarily lead to the equivalence between structured and well-mixed populations. For example, all two-player games are linear. While the prisoner's dilemma game is equivalent in structured and well-mixed populations under the PC rule, the snowdrift is not~\cite{ohtsuki2006replicator}.

Here, we conclude the general condition for linear multiplayer games, that pairwise comparison in a structured population has no effect on evolution compared to well-mixed populations. This is to let the terms other than $x_i(\Bar{\pi}_i-\Bar{\pi})$ equal zero in Eq.~(\ref{sieq_linear_PC}), that is,
\begin{equation}\label{sieq_condition_PCnoeffect}
    b_{ii}-\sum_{j=1}^n x_j (b_{ij}+b_{ji}+b_{jj})
    +2\sum_{j=1}^n \sum_{l=1}^n x_j x_l b_{jl}=0,
\end{equation}
for $i=1,2,\dots,n$. Once Eq.~(\ref{sieq_condition_PCnoeffect}) is satisfied, we have $\dot{x}_i\propto x_i(\Bar{\pi}_i-\Bar{\pi})$ and pairwise comparison in structured populations makes no difference compared to the well-mixed population. We can see that the condition is only dependent of $\mathbf{b}$ but independent of $\mathbf{c}$. For 2-strategy linear multiplayer games, the condition reduces to $b_{11}-b_{12}-b_{21}+b_{22}=0$, which is accidentally the case of the public goods game.

Using the judgment of Eq.~(\ref{sieq_condition_PCnoeffect}), we know that the public goods game is not the only multiplayer game where the PC rule plays no role. The multi-stage public goods game, which is a 4-strategy system~\cite{szolnoki2022tactical}, is also an example equivalent in structured and well-mixed populations under weak selection and the PC rule, which we will show in Supplementary Note~\ref{sec_multipgg}.

One may say that using other update rules, such as death-birth, can avoid the accidental equivalence between structured and well-mixed populations under weak selection and can make new findings. However, we would like to stress an advantage of the equivalence under pairwise comparison. We have known that the traditional public goods game with the PC rule generates the same results as in the well-mixed population. That is to say, we can avoid blurring the effect of structured populations on the traditional 2-strategy model when we introduce a third strategy. When introducing the third strategy to the traditional public goods game and studying the coevolutionary games of more than two strategies, we can compare the structured and well-mixed populations to obtain the difference created by the additional strategies only. This cannot be realized by the death-birth rule. In other words, the PC rule focuses on the independent role of additional strategies in structured populations, while the death-birth rule would mix the effect of the traditional two strategies in structured populations, reducing the scientific validity of the conclusions on the additional strategies.

\subsection{Public goods games with peer punishment ($n=3$)}\label{sec_peer}
As mentioned in the main text, there are $n=3$ strategies in the public goods game with peer punishment~\cite{helbing2010evolutionary}:

$1=\text{Cooperation}$ ($C$);

$2=\text{Defection}$ ($D$);

$3=\text{Peer punishment}$ ($E$).

Based on the traditional public goods game, the peer punishment strategy is introduced as an additional strategy. A punishing player pays a cost $\alpha$ to punish a defective co-player. The punished defective player is charged with a fine $\beta$. As a result, given $k_2$ defective co-players, a punishing player has $\alpha k_2$ paid. Similarly, given $k_3$ punishing co-players, a defective player has $\beta k_3$ charged. Meanwhile, we assume punishing players also perform the cooperative behavior, investing $c$ to the common pool. This makes the cooperative players second-order free-riders.

Given the co-player configuration $\mathbf{k}=(k_1,k_2,k_3)$, we have the following payoff calculation in a single game.
\begin{subequations}\label{sieq_a_peer}
\begin{align}
a_{1|\mathbf{k}}&
=\frac{r(k_1 +1+k_3)c}{k+1}-c
=\frac{rc}{k+1}k_1+\frac{rc}{k+1}k_3+\frac{rc}{k+1}-c, \\
a_{2|\mathbf{k}}&
=\frac{r(k_1 +k_3)c}{k+1}-\beta k_3
=\frac{rc}{k+1}k_1+\left(\frac{rc}{k+1}-\beta\right) k_3, \\
a_{3|\mathbf{k}}&
=\frac{r(k_1+k_3 +1)c}{k+1}-c-\alpha k_2
=\frac{rc}{k+1}k_1-\alpha k_2+\frac{rc}{k+1}k_3+\frac{rc}{k+1}-c.
\end{align}
\end{subequations}

\subsubsection{The well-mixed population}\label{sec_peer_wellmixed}
The frequencies of strategies 1, 2, 3 are denoted by $x_1$, $x_2$, and $x_3$ (or $x_C$, $x_D$, and $x_E$ in the main text for straightforward understanding), respectively. In a well-mixed population, the mean payoffs of the three strategies are calculated as follows.
\begin{subequations}
\begin{align}
\Bar{\pi}_1&
=\sum_{k_1+k_2+k_3=k}\frac{k!}{k_1! k_2! k_3!} {x_1}^{k_1}{x_2}^{k_2}{x_3}^{k_3} a_{1|\mathbf{k}}
=\frac{rc}{k+1}kx_1+\frac{rc}{k+1}kx_3+\frac{rc}{k+1}-c, \\
\Bar{\pi}_2&
=\sum_{k_1+k_2+k_3=k}\frac{k!}{k_1! k_2! k_3!} {x_1}^{k_1}{x_2}^{k_2}{x_3}^{k_3} a_{2|\mathbf{k}}
=\frac{rc}{k+1}kx_1+\left(\frac{rc}{k+1}-\beta\right) kx_3, \\
\Bar{\pi}_3&
=\sum_{k_1+k_2+k_3=k}\frac{k!}{k_1! k_2! k_3!} {x_1}^{k_1}{x_2}^{k_2}{x_3}^{k_3} a_{3|\mathbf{k}}
=\frac{rc}{k+1}kx_1-\alpha kx_2+\frac{rc}{k+1}kx_3+\frac{rc}{k+1}-c.
\end{align}
\end{subequations}
The mean payoff of the total population is then calculated by 
\begin{equation}
    \Bar{\pi}
=x_1\Bar{\pi}_1+x_2\Bar{\pi}_2+x_3\Bar{\pi}_3
=\frac{rc}{k+1}kx_1+\frac{rc}{k+1}kx_3+(x_1+x_3)\left(\frac{rc}{k+1}-c\right)-kx_2x_3(\alpha+\beta).
\end{equation}

On this basis, we can write the replicator equations of the well-mixed population $\dot{x}_i=x_i(\Bar{\pi}_i-\Bar{\pi})$ as follows.
\begin{subequations}
\begin{align}
\dot{x}_1&
=x_1(\Bar{\pi}_1-\Bar{\pi})
=x_1\left[(1-x_1-x_3)\left(\frac{rc}{k+1}-c\right)+kx_2x_3(\alpha+\beta)\right], \\
\dot{x}_2&
=x_2(\Bar{\pi}_2-\Bar{\pi})
=x_2\left[-(x_1+x_3)\left(\frac{rc}{k+1}-c\right)+k\left(x_2 x_3 (\alpha+\beta)-x_3 \beta\right)\right], \\
\dot{x}_3&
=x_3(\Bar{\pi}_3-\Bar{\pi})
=x_3\left[(1-x_1-x_3)\left(\frac{rc}{k+1}-c\right)+k\left(x_2 x_3 (\alpha+\beta)-x_2 \alpha\right)\right].
\end{align}
\end{subequations}

We denote the system state $\mathbf{x}=(x_1,x_2,x_3)$. Solving $\dot{\mathbf{x}}=\mathbf{0}$, we obtain equilibrium points, which can be divided into three categories. The first and second categories are single equilibrium points: a point on the $D$-vertex, $\mathbf{x}^{(D)}=(0,1,0)$, and a point on the $DE$-edge, $\mathbf{x}^{(DE)}=(0,x_2^{(DE)},x_3^{(DE)})$, where 
\begin{subequations}
\begin{align}
x_2^{(DE)}&=\frac{1}{k(\alpha+\beta)}\left(\frac{rc}{k+1}-c+k\beta \right), \label{sieq_peer_x2DE_wm}\\
x_3^{(DE)}&=1-x_1^{(DE)}=\frac{1}{k(\alpha+\beta)}\left(-\frac{rc}{k+1}+c+k\alpha \right).
\end{align}
\end{subequations}

The third category contains infinite equilibrium points on the $CE$-edge, denoted by $\mathbf{x}^{(CE)}=(x_1^{(CE)},0,x_3^{(CE)})$, where $0\leq x_1^{(CE)}\leq 1$, $0\leq x_3^{(CE)}\leq 1$, $x_1^{(CE)}+x_3^{(CE)}=1$. This category covers other two vertex points $(1,0,0)$ and $(0,0,1)$.

The stability of $\mathbf{x}^{(D)}$ and $\mathbf{x}^{(DE)}$ can be studied by the regular method. We cancel $x_2=1-x_1-x_3$ and study the dynamics depicted by $\dot{x}_1$ and $\dot{x}_3$,
\begin{subequations}\label{sieq_peer_x1x3}
\begin{align}
\dot{x}_1&
=x_1(1-x_1-x_3)\left(\frac{rc}{k+1}-c+kx_3(\alpha+\beta)\right), \\
\dot{x}_3&
=x_3(1-x_1-x_3)\left(\frac{rc}{k+1}-c+kx_3(\alpha+\beta)-k\alpha \right).
\end{align}
\end{subequations}

The Jacobian matrix of system (\ref{sieq_peer_x1x3}) is 
\begin{align}\label{sieq_J_peer_wm}
    J&=\begin{pmatrix}
    \displaystyle{\frac{\partial \dot{x}_1}{\partial x_1}} & 
    \displaystyle{\frac{\partial \dot{x}_1}{\partial x_3}} \\[1em]
    \displaystyle{\frac{\partial \dot{x}_3}{\partial x_1}} & 
    \displaystyle{\frac{\partial \dot{x}_3}{\partial x_3}}
    \end{pmatrix} \nonumber\\
    &=\begin{pmatrix}
    \displaystyle{(1-2x_1-x_3)\left(\frac{rc}{k+1}-c+kx_3 (\alpha+\beta)\right)} & 
    \displaystyle{-x_1\left(\frac{rc}{k+1}-c+k(1-x_1) (\alpha+\beta)\right)} \\[1em]
    \displaystyle{-x_3\left(\frac{rc}{k+1}-c+kx_3 (\alpha+\beta)-k\alpha \right)} & 
    \displaystyle{(1-x_1-2x_3)\left(\frac{rc}{k+1}-c-k\alpha \right)+(2-2x_1-3x_3)kx_3 (\alpha+\beta)}
    \end{pmatrix}.
\end{align}

Substituting the value of $\mathbf{x}^{(D)}$ into Eq.~(\ref{sieq_J_peer_wm}), we have
\begin{equation}\label{sieq_J_xD_wm}
    \left.J\right|_{\mathbf{x}=\mathbf{x}^{(D)}}=
    \begin{pmatrix}
    \displaystyle{\frac{rc}{k+1}-c} & 
    \displaystyle{0} \\[1em]
    \displaystyle{0} & 
    \displaystyle{\frac{rc}{k+1}-c-k\alpha}
    \end{pmatrix}.
\end{equation}
The conditions ensuring $\left.J\right|_{\mathbf{x}=\mathbf{x}^{(D)}}$ negative-definite are $r<k+1$ and $r<(1+k\alpha/c)(k+1)$. Considering $k+1<(1+k\alpha/c)(k+1)$, we know that $\mathbf{x}^{(D)}$ is stable if and only if $r<k+1$.

Substituting the value of $\mathbf{x}^{(DE)}$ into Eq.~(\ref{sieq_J_peer_wm}), we have
\begin{equation}
    \left.J\right|_{\mathbf{x}=\mathbf{x}^{(DE)}}=
    \begin{pmatrix}
    \displaystyle{(1-x_3^{(DE)})\left(\frac{rc}{k+1}-c+kx_3^{(DE)} (\alpha+\beta)\right)} & 
    \displaystyle{0} \\[1em]
    \displaystyle{-x_3^{(DE)}\left(\frac{rc}{k+1}-c+kx_3^{(DE)} (\alpha+\beta)-k\alpha \right)} & 
    \displaystyle{(1-2x_3^{(DE)})\left(\frac{rc}{k+1}-c-k\alpha \right)+(2-3x_3^{(DE)})kx_3^{(DE)} (\alpha+\beta)}
    \end{pmatrix}.
\end{equation}
The first-order sequential principal subformula is
\begin{equation}
    (1-x_3^{(DE)})\left(\frac{rc}{k+1}-c+kx_3^{(DE)} (\alpha+\beta)\right)=
    (1-x_3^{(DE)})\left(\frac{rc}{k+1}-c+\left(-\frac{rc}{k+1}+c+k\alpha \right)\right)=
    (1-x_3^{(DE)})k\alpha>0.
\end{equation}
Therefore, $\mathbf{x}^{(DE)}$ is unstable.

Next, we study the stability of $\mathbf{x}^{(CE)}=(x_1^{(CE)},0,x_3^{(CE)})$, which contains infinite equilibrium points satisfying $x_1^{(CE)}+x_3^{(CE)}=1$. This is because when $x_2=0$, we have $\dot{x}_1=\dot{x}_3$ and strategies 1 and 3 are indistinguishable. Concerning this, we can treat $x_1+x_3$ as a whole and study the system depicted by $\dot{x}_2$ and $\dot{x}_1+\dot{x}_3$. We cancel $x_1+x_3=1-x_2$ and have 
\begin{equation}
    \dot{x}_2=-x_2(1-x_2)\left(\frac{rc}{k+1}-c\right)+kx_2x_3\left(x_2 (\alpha+\beta)-\beta\right).
\end{equation}
The element in the single-order Jacobian matrix is 
\begin{equation}
    \frac{\mathrm{d}\dot{x}_2}{\mathrm{d} x_2}=-(1-2x_2)\left(\frac{rc}{k+1}-c\right)+kx_3\left(2x_2 (\alpha+\beta)-\beta\right).
\end{equation}
Substituting the value of $\mathbf{x}^{(CE)}$ into $\mathrm{d}\dot{x}_2/\mathrm{d} x_2$, we have
\begin{equation}
    \left.\frac{\mathrm{d}\dot{x}_2}{\mathrm{d} x_2}\right|_{\mathbf{x}=\mathbf{x}^{(CE)}}
    =-\left(\frac{rc}{k+1}-c\right)-kx_3^{(CE)}\beta.
\end{equation}
Therefore, $\mathbf{x}^{(CE)}$ is stable if $x_1^{(CE)}<x_{1,\star}^{(CE)}$ (or $x_3^{(CE)}>x_{3,\star}^{(CE)}$), where 
\begin{equation}\label{sieq_peer_x1CE_wm}
    x_{1,\star}^{(CE)}=1+\left(\frac{r}{k+1}-1\right)\frac{c}{k\beta}\equiv 
    \left(1+\frac{\alpha}{\beta}\right) x_2^{(DE)}, 
\end{equation}
or $x_{3,\star}^{(CE)}=\left(-\frac{r}{k+1}+1\right)\frac{c}{k\beta}$. That is, while every point on the $CE$-edge is equilibrium, only the points on the side of $x_1^{(CE)}<x_{1,\star}^{(CE)}$ are stable. Moreover, we note that $x_{1,\star}^{(CE)}>1$ when $r>k+1$. That is, $x_1^{(CE)}<x_{1,\star}^{(CE)}$ always holds and the points on the $CE$-edge are stable everywhere for $r>k+1$.

\subsubsection{The structured population}
We notice that the payoff structure given by Eq.~(\ref{sieq_a_peer}) is linear, which means that we can utilize the simplified method for special linear systems given by Supplementary Note~\ref{sec_linear} for convenience.

Comparing the payoff structure $a_{i|\mathbf{k}}$ in Eq.~(\ref{sieq_a_peer}) with Eqs.~(\ref{sieq_b_c})--(\ref{sieq_linear_a_PC}), we extract matrices $\mathbf{b}$ and $\mathbf{c}$, 
\begin{equation}
    \mathbf{b}=
    \begin{pmatrix}
        \dfrac{rc}{k+1} & 0 & \dfrac{rc}{k+1} \\[1em]
        \dfrac{rc}{k+1} & 0 & \dfrac{rc}{k+1}-\beta \\[1em]
        \dfrac{rc}{k+1} & -\alpha & \dfrac{rc}{k+1}
    \end{pmatrix},~
    \mathbf{c}=
    \begin{pmatrix}
        \dfrac{rc}{k+1}-c \\[1em]
        0 \\
        \dfrac{rc}{k+1}-c
    \end{pmatrix}.
\end{equation}
According to Eq.~(\ref{sieq_linear_PC}), let us calculate
\begin{subequations}
\begin{align}
3\sum_{j=1}^3 x_j (b_{11}-b_{1j}-b_{j1}-b_{jj})&
=-6(x_1+x_3)\frac{rc}{k+1}, \\
3\sum_{j=1}^3 x_j (b_{22}-b_{2j}-b_{j2}-b_{jj})&
=-6(x_1+x_3)\frac{rc}{k+1}+3x_3 (\alpha+\beta), \\
3\sum_{j=1}^3 x_j (b_{33}-b_{3j}-b_{j3}-b_{jj})&
=-6(x_1+x_3)\frac{rc}{k+1}+3x_2 (\alpha+\beta), \\
6\sum_{j=1}^3 \sum_{l=1}^3 x_j x_l b_{jl}&
=6(x_1+x_3)\frac{rc}{k+1}-6x_2 x_3 (\alpha+\beta).
\end{align}
\end{subequations}
Then, we can substitute them into Eq.~(\ref{sieq_linear_PC}). Meanwhile, the expression of $\Bar{\pi}_i$ and $\Bar{\pi}$ is the same as we give in Supplementary Note~\ref{sec_peer_wellmixed} for well-mixed populations. In this way, we obtain the replicator equations in structured populations as follows.
\begin{subequations}
\begin{align}
\dot{x}_1&
=\frac{\delta(k-2)}{2(k-1)} x_1\left\{(k+1)\left[(1-x_1-x_3)\left(\frac{rc}{k+1}-c\right)+kx_2x_3(\alpha+\beta)\right]-6x_2 x_3 (\alpha+\beta)\right\}, \\
\dot{x}_2&
=\frac{\delta(k-2)}{2(k-1)}x_2\left\{(k+1)\left[-(x_1+x_3)\left(\frac{rc}{k+1}-c\right)+k\left(x_2 x_3 (\alpha+\beta)-x_3 \beta\right)\right]-6x_2 x_3 (\alpha+\beta)+3x_3 (\alpha+\beta)\right\}, \\
\dot{x}_3&
=\frac{\delta(k-2)}{2(k-1)}x_3\left\{(k+1)\left[(1-x_1-x_3)\left(\frac{rc}{k+1}-c\right)+k\left(x_2 x_3 (\alpha+\beta)-x_2 \alpha\right)\right]-6x_2 x_3 (\alpha+\beta)+3x_2 (\alpha+\beta)\right\}.
\end{align}
\end{subequations}

Similarly, solving $\dot{\mathbf{x}}=\mathbf{0}$, we obtain equilibrium points, which can be divided into three categories. The first and second categories are single equilibrium points: a point on the $D$-vertex, $\mathbf{x}^{(D)}=(0,1,0)$, and a point on the $DE$-edge, $\mathbf{x}^{(DE)}=(0,x_2^{(DE)},x_3^{(DE)})$, where 
\begin{subequations}
\begin{align}
    x_2^{(DE)}&=
    \frac{1}{(k-2)(k+3)(\alpha+\beta)}\left[(k+1)\left(\frac{rc}{k+1}-c+k\beta\right)-3(\alpha+\beta)\right], \label{sieq_peer_x2DE_st}\\
    x_3^{(DE)}&=1-x_2^{(DE)}=
    \frac{1}{(k-2)(k+3)(\alpha+\beta)}\left[(k+1)\left(-\frac{rc}{k+1}+c+k\alpha\right)-3(\alpha+\beta)\right].
\end{align}
\end{subequations}

The third category contains infinite equilibrium points on the $CE$-edge, denoted by $\mathbf{x}^{(CE)}=(x_1^{(CE)},0,x_3^{(CE)})$, where $0\leq x_1^{(CE)}\leq 1$, $0\leq x_3^{(CE)}\leq 1$, $x_1^{(CE)}+x_3^{(CE)}=1$. This category covers other two vertex points $(1,0,0)$ and $(0,0,1)$.

Again, the stability of $\mathbf{x}^{(D)}$ and $\mathbf{x}^{(DE)}$ can be studied by the regular method. We cancel $x_2=1-x_1-x_3$ and study the dynamics depicted by $\dot{x}_1$ and $\dot{x}_3$,
\begin{subequations}\label{sieq_PC_peer_x1x3}
\begin{align}
\dot{x}_1&
=\frac{\delta(k-2)}{2(k-1)} x_1(1-x_1-x_3)\left\{(k+1)\left(\frac{rc}{k+1}-c\right)+x_3(k-2)(k+3)(\alpha+\beta)\right\}, \\
\dot{x}_3&
=\frac{\delta(k-2)}{2(k-1)} x_3(1-x_1-x_3)\left\{(k+1)\left(\frac{rc}{k+1}-c-k\alpha\right)+x_3(k-2)(k+3)(\alpha+\beta)+3(\alpha+\beta)\right\}.
\end{align}
\end{subequations}

The Jacobian matrix of system (\ref{sieq_PC_peer_x1x3}) is 
\begin{equation}\label{sieq_J_peerPC}
    J=\begin{pmatrix}
    \displaystyle{\frac{\partial \dot{x}_1}{\partial x_1}} & 
    \displaystyle{\frac{\partial \dot{x}_1}{\partial x_3}} \\[1em]
    \displaystyle{\frac{\partial \dot{x}_3}{\partial x_1}} & 
    \displaystyle{\frac{\partial \dot{x}_3}{\partial x_3}}
    \end{pmatrix},
\end{equation}
where
\begin{subequations}
\begin{align}
    \frac{\partial \dot{x}_1}{\partial x_1}&
    =\frac{\delta(k-2)}{2(k-1)} (1-2x_1-x_3)\left\{(k+1)\left(\frac{rc}{k+1}-c\right)+x_3(k-2)(k+3)(\alpha+\beta)\right\}, \\
    \frac{\partial \dot{x}_1}{\partial x_3}&
    =-\frac{\delta(k-2)}{2(k-1)} x_1\left\{(k+1)\left(\frac{rc}{k+1}-c\right)-(1-x_1-2x_3)(k-2)(k+3)(\alpha+\beta)\right\}, \\
    \frac{\partial \dot{x}_3}{\partial x_1}&
    =-\frac{\delta(k-2)}{2(k-1)} x_3\left\{(k+1)\left(\frac{rc}{k+1}-c-k\alpha\right)+x_3(k-2)(k+3)(\alpha+\beta)+3(\alpha+\beta)\right\}, \\
    \frac{\partial \dot{x}_3}{\partial x_3}&
    =\frac{\delta(k-2)}{2(k-1)}\left\{(1-x_1-2x_3)\left[(k+1)\left(\frac{rc}{k+1}-c-k\alpha\right)+3(\alpha+\beta)\right]+x_3(2-2x_1-3x_3)(k-2)(k+3)(\alpha+\beta)
    \right\}.
\end{align}
\end{subequations}

Substituting the value of $\mathbf{x}^{(D)}$ into Eq.~(\ref{sieq_J_peerPC}), we have
\begin{equation}\label{sieq_J_xD_st}
    \left.J\right|_{\mathbf{x}=\mathbf{x}^{(D)}}=\frac{\delta(k-2)}{2(k-1)}
    \begin{pmatrix}
    \displaystyle{(k+1)\left(\frac{rc}{k+1}-c\right)} & 
    \displaystyle{0} \\[1em]
    \displaystyle{0} & 
    \displaystyle{(k+1)\left(\frac{rc}{k+1}-c-k\alpha\right)+3(\alpha+\beta)}
    \end{pmatrix}.
\end{equation}
The condition ensuring $\left.J\right|_{\mathbf{x}=\mathbf{x}^{(D)}}$ negative-definite is $r<k+1$ and $r<(1+k\alpha/c)(k+1)-3(\alpha+\beta)/c$. It is hard to compare them in size, but we can express the second condition as $x_3^{(DE)}>0$, which holds if $\mathbf{x}^{(DE)}$ exists. Therefore, if $r<k+1$ and $\mathbf{x}^{(DE)}$ exists, the equilibrium $\mathbf{x}^{(D)}$ is stable.

Substitute the value of $\mathbf{x}^{(DE)}$ into Eq.~(\ref{sieq_J_peerPC}), we have $\left.\partial \dot{x}_1/\partial x_3\right|_{\mathbf{x}=\mathbf{x}^{(DE)}}=0$. Therefore, the first- and second-order sequential principal subformulas' negativity is equivalent to $\left.\partial \dot{x}_1/\partial x_1\right|_{\mathbf{x}=\mathbf{x}^{(DE)}}<0$ and $\left.\partial \dot{x}_3/\partial x_3\right|_{\mathbf{x}=\mathbf{x}^{(DE)}}<0$. We have
\begin{equation}
    \left.\frac{\partial \dot{x}_1}{\partial x_1}\right|_{\mathbf{x}=\mathbf{x}^{(DE)}}
    =\frac{\delta(k-2)}{2(k-1)} (1-x_3^{(DE)})\left[(k+1)k\alpha-3(\alpha+\beta)\right],
\end{equation}
whose sign is hard to judge. Instead, let us calculate
\begin{equation}
    \left.\frac{\partial \dot{x}_3}{\partial x_3}\right|_{\mathbf{x}=\mathbf{x}^{(DE)}}
    =\frac{\delta(k-2)}{2(k-1)}(1-x_3^{(DE)})\left[(k+1)\left(-\frac{rc}{k+1}+c+k\alpha\right)-3(\alpha+\beta)\right].
\end{equation}
We can see that as long as $0<x_3^{(DE)}<1$, we have $\left.\partial \dot{x}_3/\partial x_3\right|_{\mathbf{x}=\mathbf{x}^{(DE)}}>0$. Therefore, if $\mathbf{x}^{(DE)}$ exists, $\mathbf{x}^{(DE)}$ cannot be stable.

Next, we study the stability of $\mathbf{x}^{(CE)}=(x_1^{(CE)},0,x_3^{(CE)})$, which contains infinite equilibrium points satisfying $x_1^{(CE)}+x_3^{(CE)}=1$. Similar to well-mixed populations, this is because when $x_2=0$, we have $\dot{x}_1=\dot{x}_3$ and strategies 1 and 3 are indistinguishable. Concerning this, we can treat $x_1+x_3$ as a whole and study the system depicted by $\dot{x}_2$ and $\dot{x}_1+\dot{x}_3$. We can cancel $x_1+x_3=1-x_2$ and have 
\begin{equation}
    \dot{x}_2=\frac{\delta(k-2)(k+1)}{2(k-1)}\left\{
    -x_2(1-x_2)\left(\frac{rc}{k+1}-c\right)+kx_2x_3
    \left(\frac{3-(k-2)(k+3)x_2}{k(k+1)} (\alpha+\beta)-\beta\right)\right\}.
\end{equation}
The element in the single-order Jacobian matrix is 
\begin{equation}
    \frac{\mathrm{d}\dot{x}_2}{\mathrm{d} x_2}=
    \frac{\delta(k-2)(k+1)}{2(k-1)}\left\{
    -(1-2x_2)\left(\frac{rc}{k+1}-c\right)+kx_3\left(\frac{3-2(k-2)(k+3)x_2}{k(k+1)} (\alpha+\beta)-\beta\right)\right\}.
\end{equation}
Substituting the value of $\mathbf{x}^{(CE)}$ into $\mathrm{d}\dot{x}_2/\mathrm{d} x_2$, we have
\begin{equation}
    \left.\frac{\mathrm{d}\dot{x}_2}{\mathrm{d} x_2}\right|_{\mathbf{x}=\mathbf{x}^{(CE)}}
    =\frac{\delta(k-2)(k+1)}{2(k-1)}\left\{
    -\left(\frac{rc}{k+1}-c\right)-kx_3^{(CE)}\left(\frac{3}{k(k+1)} (\alpha+\beta)-\beta\right)\right\}.
\end{equation}
Therefore, $\mathbf{x}^{(CE)}$ is stable if $x_1^{(CE)}<x_{1,\star}^{(CE)}$ (or $x_3^{(CE)}>x_{3,\star}^{(CE)}$), where 
\begin{equation}\label{sieq_peer_x1CE_st}
    x_{1,\star}^{(CE)}=1+\left(\frac{r}{k+1}-1\right)\frac{c}{k\beta-3(\alpha+\beta)/(k+1)}\equiv
    \frac{(k-2)(k+3)(\alpha+\beta)}{k(k+1)\beta-3(\alpha+\beta)}x_2^{(DE)}, 
\end{equation}
or $x_{3,\star}^{(CE)}=\left(-\frac{r}{k+1}+1\right)\frac{c}{k\beta-3(\alpha+\beta)/(k+1)}$. That is, while every point on the $CE$-edge is equilibrium, only the points on the side of $x_1^{(CE)}<x_{1,\star}^{(CE)}$ are stable. Moreover, we note that $x_{1,\star}^{(CE)}>1$ when $r>k+1$. That is, $x_1^{(CE)}<x_{1,\star}^{(CE)}$ always holds and the points on the $CE$-edge are stable everywhere for $r>k+1$. The conclusions here are based on a regular size of $\beta$. If the value of $\beta$ is very small, then the conclusion may be reversed, which is a different phenomenon compared with the well-mixed population.

\subsubsection{Discussion}\label{sec_peer_disucss}
According to stability analysis, we can divide the results into two cases. First, when $r>k+1$, the fixation of defection at $\mathbf{x}^{(D)}$ is unstable and the points on the $CE$-edge are stable everywhere. The dilemma is overcome through the principles known in the traditional public goods game. 

Second, when $r<k+1$, the fixation of defection occurs in the traditional public goods game, but in this 3-strategy system with peer punishment, the situations can be different. Here, we discuss the effect of peer punishment when $r<k+1$ and show how the results in Fig.~\ref{fig_peer_table} in the main text are obtained.

As mentioned in the main text, when the punishment strength is intermediate ($\beta_0<\beta<\beta^\star$ for structured populations and $\beta>\beta_0^\text{WM}$ for well-mixed populations, as revealed later), there are a stable vertex equilibrium $\mathbf{x}^{(D)}$, an unstable edge equilibrium $\mathbf{x}^{(DE)}$, and a stable equilibrium line $\mathbf{x}^{(CE)}$. The vertex equilibrium $\mathbf{x}^{(D)}$ and the equilibrium line $\mathbf{x}^{(CE)}$ are bi-stable, depending on the initial state space divided by $\mathbf{x}^{(DE)}$ and $\mathbf{x}_\star^{(CE)}=(x_{1,\star}^{(CE)},0,x_{3,\star}^{(CE)})$. In well-mixed populations, $x_2^{(DE)}\to 1$ as $\beta\to \infty$ according to Eq.~(\ref{sieq_peer_x2DE_wm}), and $\mathbf{x}^{(D)}$ is always stable regardless of the value of $\beta$ according to Eq.~(\ref{sieq_J_xD_wm}). This is the analytical illustration that peer punishment cannot truly resolve social dilemmas in a well-mixed population. However, in structured populations, $x_2^{(DE)}\to k(k+1)/[(k-2)(k+3)]>1$ as $\beta\to \infty$ according to Eq.~(\ref{sieq_peer_x2DE_st}). In particular, $x_2^{(DE)} >1$ happens (i.e., $\mathbf{x}^{(DE)}$ no longer exists) when 
\begin{align}\label{sieq_peer_star}
    &~\frac{1}{(k-2)(k+3)(\alpha+\beta)}\left[(k+1)\left(\frac{rc}{k+1}-c+k\beta\right)-3(\alpha+\beta)\right]>1 \nonumber\\
    \Leftrightarrow&~\beta>\beta^\star\equiv \frac{k+1}{3}\left(-\frac{rc}{k+1}+c+k\alpha\right)-\alpha.
\end{align}
Meanwhile, according to Eq.~(\ref{sieq_J_xD_st}), $\mathbf{x}^{(D)}$ becomes unstable at the same time when Eq.~(\ref{sieq_peer_star}) is satisfied. The bi-stable system state becomes mono-stable on the $CE$-edge. This is the analytical interpretation that peer punishment can resolve social dilemmas in structured populations.

On the other hand, structured populations do not always facilitate the advantage of peer punishment compared to well-mixed populations. This property can be observed from the critical punishment strength over which peer punishment starts to play a role. In a structured population, according to Eq.~(\ref{sieq_peer_x2DE_st}), $x_2^{(DE)}>0$ means 
\begin{align}\label{sieq_peer_beta0st}
    &~\frac{1}{(k-2)(k+3)(\alpha+\beta)}\left[(k+1)\left(\frac{rc}{k+1}-c+k\beta\right)-3(\alpha+\beta)\right]>0 \nonumber\\
    \Leftrightarrow&~\beta>\beta_0\equiv \frac{k+1}{k^2+k-3}\left(-\frac{rc}{k+1}+c\right)+\frac{3\alpha}{k^2+k-3}.
\end{align}
And, according to Eq.~(\ref{sieq_peer_x1CE_st}), $x_{1,\star}^{(CE)}>0 \Leftrightarrow x_2^{(DE)}>0$. When $\beta<\beta_0$, the system is mono-stable at the vertex equilibrium $\mathbf{x}^{(D)}$, as it is in the traditional public goods game. Peer punishment starts functioning and creates bi-stability in structured populations when $\beta>\beta_0$.

In a well-mixed population, according to Eq.~(\ref{sieq_peer_x2DE_wm}), $x_2^{(DE)}>0$ means 
\begin{equation}\label{sieq_peer_beta0wm}
    \frac{1}{k(\alpha+\beta)}\left(\frac{rc}{k+1}-c+k\beta \right)>0 
    \Leftrightarrow \beta> \beta_0^\text{WM} \equiv \frac{1}{k}\left(-\frac{rc}{k+1}+c\right),
\end{equation}
and $x_{1,\star}^{(CE)}>0 \Leftrightarrow x_2^{(DE)}>0$ according to Eq.~(\ref{sieq_peer_x1CE_wm}). Peer punishment starts working and creates bi-stability in well-mixed populations when $\beta>\beta_0^\text{WM}$.

Comparing Eqs.~(\ref{sieq_peer_beta0st}) and (\ref{sieq_peer_beta0wm}), we see that $\beta_0^\text{WM}<\beta_0$ always holds. That is, peer punishment first starts to play a role in well-mixed populations when increasing the punishment strength.

Actually, there is a considerable interval of $\beta$ that $x_2^{(DE)}$ in structured populations are smaller than the ones in well-mixed populations. That is, a structured population enlarges the initial state space leading to full defection, thus less effective to utilize peer punishment. A structured population becomes more advantageous only when its $x_2^{(DE)}$ are greater than the ones in well-mixed populations. According to Eqs.~(\ref{sieq_peer_x2DE_wm}) and (\ref{sieq_peer_x2DE_st}), this means
\begin{align}
    &~\frac{1}{(k-2)(k+3)(\alpha+\beta)}\left[(k+1)\left(\frac{rc}{k+1}-c+k\beta\right)-3(\alpha+\beta)\right]
    >\frac{1}{k(\alpha+\beta)}\left(\frac{rc}{k+1}-c+k\beta \right) \nonumber\\
    \Leftrightarrow&~\beta>\beta_= \equiv \frac{2}{k}\left(-\frac{rc}{k+1}+c\right)+\alpha.
\end{align}
Only when $\beta>\beta_=$, a structured population has a smaller initial state space leading to full defection, thus more effective to utilize peer punishment than well-mixed populations. In particular, $\beta>\beta^\star$ completely eliminates the initial state space leading to full defection as shown in Eq.~(\ref{sieq_peer_star}), while a well-mixed population cannot realize this.

\subsection{Public goods games with pool punishment ($n=3$)}\label{sec_pool}
As mentioned in the main text, there are $n=3$ strategies in the public goods game with pool punishment~\cite{szolnoki2011phase}.

$1=\text{Cooperation}$ ($C$);

$2=\text{Defection}$ ($D$);

$3=\text{Pool punishment}$ ($O$).

Based on the traditional public goods game, the pool punishment strategy is introduced as an additional strategy. A punishing player always pays a cost $\alpha$ to establish the institution for pool punishment. A defective player is charged with a fine $\beta$ if there is at least one punishing co-player. That is, given $k_3$ punishing co-players, a defective player has $\beta f(k_3)$ charged, where $f(k_3)=1$ if $k_3>0$ and $f(k_3)=0$ if $k_3=0$. Meanwhile, we assume punishing players also perform the cooperative behavior, investing $c$ to the common pool. Again, this makes the cooperative players second-order free-riders.

Therefore, given the co-player configuration $\mathbf{k}=(k_1,k_2,k_3)$, we have the following payoff calculation in a single game.
\begin{subequations}\label{sieq_a_pool}
\begin{align}
a_{1|\mathbf{k}}&
=\frac{r(k_1 +1+k_3)c}{k+1}-c
=\frac{rc}{k+1}k_1+\frac{rc}{k+1}k_3+\frac{rc}{k+1}-c, \\
a_{2|\mathbf{k}}&
=\frac{r(k_1 +k_3)c}{k+1}-\beta f(k_3)
=\frac{rc}{k+1}k_1+\frac{rc}{k+1}k_3-\beta f(k_3), \\
a_{3|\mathbf{k}}&
=\frac{r(k_1+k_3 +1)c}{k+1}-c-\alpha
=\frac{rc}{k+1}k_1+\frac{rc}{k+1}k_3+\frac{rc}{k+1}-c-\alpha.
\end{align}
\end{subequations}

\subsubsection{The well-mixed population}
The frequencies of strategies 1, 2, 3 are denoted by $x_1$, $x_2$, and $x_3$ (or $x_C$, $x_D$, and $x_O$ in the main text), respectively. In a well-mixed population, the mean payoffs of the three strategies are calculated as follows.
\begin{subequations}
\begin{align}
\Bar{\pi}_1&
=\sum_{k_1+k_2+k_3=k}\frac{k!}{k_1! k_2! k_3!} {x_1}^{k_1}{x_2}^{k_2}{x_3}^{k_3} a_{1|\mathbf{k}}
=\frac{rc}{k+1}kx_1+\frac{rc}{k+1}kx_3+\frac{rc}{k+1}-c, \\
\Bar{\pi}_2&
=\sum_{k_1+k_2+k_3=k}\frac{k!}{k_1! k_2! k_3!} {x_1}^{k_1}{x_2}^{k_2}{x_3}^{k_3} a_{2|\mathbf{k}}
=\frac{rc}{k+1}kx_1+\frac{rc}{k+1}kx_3-\beta\left[1-(1-x_3)^k\right] ,\\
\Bar{\pi}_3&
=\sum_{k_1+k_2+k_3=k}\frac{k!}{k_1! k_2! k_3!} {x_1}^{k_1}{x_2}^{k_2}{x_3}^{k_3} a_{3|\mathbf{k}}
=\frac{rc}{k+1}kx_1+\frac{rc}{k+1}kx_3+\frac{rc}{k+1}-c-\alpha.
\end{align}
\end{subequations}
The mean payoff of the total population is then calculated by 
\begin{equation}
    \Bar{\pi}
=x_1\Bar{\pi}_1+x_2\Bar{\pi}_2+x_3\Bar{\pi}_3
=\frac{rc}{k+1}kx_1+\frac{rc}{k+1}kx_3+(x_1+x_3)\left(\frac{rc}{k+1}-c\right)-x_3\alpha-x_2\left[1-(1-x_3)^k\right].
\end{equation}

On this basis, we can write the replicator equations of the well-mixed population $\dot{x}_i=x_i(\Bar{\pi}_i-\Bar{\pi})$ as follows.
\begin{subequations}
\begin{align}
\dot{x}_1&
=x_1(\Bar{\pi}_1-\Bar{\pi})
=x_1\left\{(1-x_1-x_3)\left(\frac{rc}{k+1}-c\right)+x_3\alpha+x_2\beta\left[1-(1-x_3)^k\right]\right\}, \\
\dot{x}_2&
=x_2(\Bar{\pi}_2-\Bar{\pi})
=x_2\left\{-(x_1+x_3)\left(\frac{rc}{k+1}-c\right)+x_3\alpha-(1-x_2)\beta\left[1-(1-x_3)^k\right]\right\}, \\
\dot{x}_3&
=x_3(\Bar{\pi}_3-\Bar{\pi})
=x_3\left\{(1-x_1-x_3)\left(\frac{rc}{k+1}-c\right)-(1-x_3)\alpha+x_2\beta\left[1-(1-x_3)^k\right]\right\}.
\end{align}
\end{subequations}

We denote the system state $\mathbf{x}=(x_1,x_2,x_3)$. Solving $\dot{\mathbf{x}}=\mathbf{0}$, we obtain four possible equilibrium points: a point on the $C$-vertex, $\mathbf{x}^{(C)}=(1,0,0)$, a point on the $D$-vertex, $\mathbf{x}^{(D)}=(0,1,0)$, a point on the $O$-vertex, $\mathbf{x}^{(O)}=(0,0,1)$, and a point on the $DO$-edge, $\mathbf{x}^{(DO)}=(0,x_2^{(DO)},x_3^{(DO)})$, where
\begin{subequations}
\begin{align}
x_2^{(DO)}&=\sqrt[k]{1+\frac{1}{\beta}\left(\frac{rc}{k+1}-c-\alpha \right)}, \label{sieq_pool_x2DO_wm}\\
x_3^{(DO)}&=1-x_2^{(DO)}=1-\sqrt[k]{1+\frac{1}{\beta}\left(\frac{rc}{k+1}-c-\alpha \right)}.
\end{align}
\end{subequations}

The stability of these four equilibrium points can be studied by the regular method. We cancel $x_2=1-x_1-x_3$ and study the dynamics depicted by $\dot{x}_1$ and $\dot{x}_3$,
\begin{subequations}\label{sieq_pool_x1x3}
\begin{align}
\dot{x}_1&
=x_1\left\{(1-x_1-x_3)\left(\frac{rc}{k+1}-c\right)+x_3\alpha+(1-x_1-x_3)\left[1-(1-x_3)^k\right]\right\}, \\
\dot{x}_3&
=x_3\left\{(1-x_1-x_3)\left(\frac{rc}{k+1}-c\right)-(1-x_3)\alpha+(1-x_1-x_3)\left[1-(1-x_3)^k\right]\right\}.
\end{align}
\end{subequations}

The Jacobian matrix of system (\ref{sieq_pool_x1x3}) is 
\begin{equation}\label{sieq_J_pool_PC}
    J=\begin{pmatrix}
    \displaystyle{\frac{\partial \dot{x}_1}{\partial x_1}} & 
    \displaystyle{\frac{\partial \dot{x}_1}{\partial x_3}} \\[1em]
    \displaystyle{\frac{\partial \dot{x}_3}{\partial x_1}} & 
    \displaystyle{\frac{\partial \dot{x}_3}{\partial x_3}}
    \end{pmatrix},
\end{equation}
where
\begin{subequations}
\begin{align}
    \frac{\partial \dot{x}_1}{\partial x_1}
    =&~(1-2x_1-x_3)\left(\frac{rc}{k+1}-c\right)+x_3\alpha+(1-2x_1-x_3)\left[1-(1-x_3)^k\right], \\
    \frac{\partial \dot{x}_1}{\partial x_3}
    =&~x_1\left\{-\frac{rc}{k+1}+c+\alpha-\beta+(1-x_3)^{k-1}\beta\left[1-x_3+k(1-x_1-x_3)\right]\right\}, \\
    \frac{\partial \dot{x}_3}{\partial x_1}
    =&-x_3\left\{\frac{rc}{k+1}-c+\beta\left[1-(1-x_3)^k\right]\right\}, \\
    \frac{\partial \dot{x}_3}{\partial x_3}
    =&~(1-x_1-2x_3)\left(\frac{rc}{k+1}-c\right)-(1-2x_3)\alpha \nonumber\\
    &+\beta\left\{1-x_1-2x_3+(1-x_3)^{k-1}\left[-(1-x_1-x_3)+(k+2)(1-x_3)x_3-(k+1)x_1x_3\right]\right\}.
\end{align}
\end{subequations}

Substituting the value of $\mathbf{x}^{(C)}$ into Eq.~(\ref{sieq_J_pool_PC}), we have
\begin{equation}
    \left.J\right|_{\mathbf{x}=\mathbf{x}^{(C)}}=
    \begin{pmatrix}
    \displaystyle{-\frac{rc}{k+1}+c} & 
    \displaystyle{-\frac{rc}{k+1}+c+\alpha-\beta} \\[1em]
    \displaystyle{0} & 
    \displaystyle{-\alpha}
    \end{pmatrix}.
\end{equation}
The condition ensuring $\left.J\right|_{\mathbf{x}=\mathbf{x}^{(C)}}$ negative-definite is $r>k+1$. Therefore, the $C$-vertex equilibrium $\mathbf{x}^{(C)}$ is stable if and only if $r>k+1$.

Substituting the value of $\mathbf{x}^{(D)}$ into Eq.~(\ref{sieq_J_pool_PC}), we have
\begin{equation}\label{sieq_pool_J_xD_wm}
    \left.J\right|_{\mathbf{x}=\mathbf{x}^{(D)}}=
    \begin{pmatrix}
    \displaystyle{\frac{rc}{k+1}-c} & 
    \displaystyle{0} \\[1em]
    \displaystyle{0} & 
    \displaystyle{\frac{rc}{k+1}-c-\alpha}
    \end{pmatrix}.
\end{equation}
The condition ensuring $\left.J\right|_{\mathbf{x}=\mathbf{x}^{(D)}}$ negative-definite is $r<k+1$. Therefore, the $D$-vertex equilibrium $\mathbf{x}^{(D)}$ is stable if and only if $r<k+1$.

Substituting the value of $\mathbf{x}^{(O)}$ into Eq.~(\ref{sieq_J_pool_PC}), we have $\left.\partial \dot{x}_1/\partial x_1\right|_{\mathbf{x}=\mathbf{x}^{(O)}}=\alpha>0$. Therefore, the Jacobian matrix at $\mathbf{x}^{(O)}$ is not negative-definite and the $O$-vertex equilibrium $\mathbf{x}^{(O)}$ is not stable.

Substituting the value of $\mathbf{x}^{(DO)}$ into Eq.~(\ref{sieq_J_pool_PC}), we note that
\begin{equation}
    \left.\frac{\partial\dot{x}_1}{\partial x_1}\right|_{\mathbf{x}=\mathbf{x}^{(DO)}}
    =\sqrt[k]{1+\frac{1}{\beta}\left(\frac{rc}{k+1}-c-\alpha \right)}\left\{\left(\frac{rc}{k+1}-c-\alpha \right)+\beta\left[-\frac{1}{\beta}\left(\frac{rc}{k+1}-c-\alpha \right)\right]\right\}+\alpha=\alpha>0.
\end{equation}
Therefore, the Jacobian matrix at $\mathbf{x}^{(DO)}$ is not negative-definite and the equilibrium $\mathbf{x}^{(DO)}$ on the $DO$-edge is not stable.

\subsubsection{The structured population}
Next, we study pool punishment in structured populations. We notice that the payoff structure given by Eq.~(\ref{sieq_a_pool}) is nonlinear. Therefore, to obtain the replicator dynamics in a structured population, we need to utilize Eq.~(\ref{sieq_alevel3_PC}) given by Supplementary Note~\ref{sec_PCsingle}. The process is to calculate all elements of the `$\langle a_{i|\mathbf{k}_{+j}}\rangle_i$ type' and the `$\langle a_{i|\mathbf{k}_{+j}}\rangle_j$ type' manually, which is tedious. For this $n=3$ system, the $\langle a_{i|\mathbf{k}_{+j}}\rangle_i$ type is 
\begin{equation}
        \left[\langle a_{i|\mathbf{k}_{+j}}\rangle_i\right]_{ij}=
        \displaystyle{\begin{pmatrix}
        \langle a_{1|\mathbf{k}_{+1}}\rangle_1 & \langle a_{1|\mathbf{k}_{+2}}\rangle_1 & \langle a_{1|\mathbf{k}_{+3}}\rangle_1 \\
        \langle a_{2|\mathbf{k}_{+1}}\rangle_2 & \langle a_{2|\mathbf{k}_{+2}}\rangle_2 & \langle a_{2|\mathbf{k}_{+3}}\rangle_2 \\
        \langle a_{3|\mathbf{k}_{+1}}\rangle_3 & \langle a_{3|\mathbf{k}_{+2}}\rangle_3 & \langle a_{3|\mathbf{k}_{+3}}\rangle_3
    \end{pmatrix}},
\end{equation}
where, by applying Eq.~(\ref{sieq_a_pool}) to Eq.~(\ref{sieq_<a>}), we have
\begin{align}\label{sieq_poolatype1}
    \langle a_{1|\mathbf{k}_{+1}}\rangle_1
    &=\sum_{k_1+k_2+k_3=k-1}\frac{(k-1)!}{k_1!k_2!k_3!} q_{1|1}^{k_1}q_{2|1}^{k_2}q_{3|1}^{k_3}
    a_{1|\mathbf{k}_{+1}} \nonumber\\
    &=\sum_{k_1+k_2+k_3=k-1}\frac{(k-1)!}{k_1!k_2!k_3!} q_{1|1}^{k_1}q_{2|1}^{k_2}q_{3|1}^{k_3}
    \left(\frac{rc}{k+1}(k_1+1)+\frac{rc}{k+1}k_3+\frac{rc}{k+1}-c\right) \nonumber\\
    &=\frac{rc}{k+1}(k-1)q_{1|1}+\frac{rc}{k+1}(k-1)q_{3|1}+\frac{2rc}{k+1}-c \nonumber\\
    &=\frac{rc}{k+1}(k-2)x_1+\frac{rc}{k+1}(k-2)x_3+\frac{3rc}{k+1}-c,
\end{align}
and similarly, 
\begin{subequations}
    \begin{align}
        \langle a_{1|\mathbf{k}_{+2}}\rangle_1
        &=\frac{rc}{k+1}(k-2)x_1+\frac{rc}{k+1}(k-2)x_3+\frac{2rc}{k+1}-c,\label{sieq_poolatype21} \\
        \langle a_{1|\mathbf{k}_{+3}}\rangle_1
        &=\frac{rc}{k+1}(k-2)x_1+\frac{rc}{k+1}(k-2)x_3+\frac{3rc}{k+1}-c;
        \\
        \langle a_{2|\mathbf{k}_{+1}}\rangle_2
        &=\frac{rc}{k+1}(k-2)x_1+\frac{rc}{k+1}(k-2)x_3+\frac{rc}{k+1}-\beta\left[1-\left(1-\frac{k-2}{k-1}x_3\right)^{k-1}\right], \\
        \langle a_{2|\mathbf{k}_{+2}}\rangle_2
        &=\frac{rc}{k+1}(k-2)x_1+\frac{rc}{k+1}(k-2)x_3-\beta\left[1-\left(1-\frac{k-2}{k-1}x_3\right)^{k-1}\right], \\
        \langle a_{2|\mathbf{k}_{+3}}\rangle_2
        &=\frac{rc}{k+1}(k-2)x_1+\frac{rc}{k+1}(k-2)x_3+\frac{rc}{k+1}-\beta; 
        \\
        \langle a_{3|\mathbf{k}_{+1}}\rangle_3
        &=\frac{rc}{k+1}(k-2)x_1+\frac{rc}{k+1}(k-2)x_3+\frac{3rc}{k+1}-c-\alpha, \\
        \langle a_{3|\mathbf{k}_{+2}}\rangle_3
        &=\frac{rc}{k+1}(k-2)x_1+\frac{rc}{k+1}(k-2)x_3+\frac{2rc}{k+1}-c-\alpha, \\
        \langle a_{3|\mathbf{k}_{+3}}\rangle_3
        &=\frac{rc}{k+1}(k-2)x_1+\frac{rc}{k+1}(k-2)x_3+\frac{3rc}{k+1}-c-\alpha. \label{sieq_poolatype22}
    \end{align}
\end{subequations}

Next, the $\langle a_{i|\mathbf{k}_{+j}}\rangle_j$ type is 
\begin{equation}
        \left[\langle a_{i|\mathbf{k}_{+j}}\rangle_j\right]_{ij}=
        \displaystyle{\begin{pmatrix}
        \langle a_{1|\mathbf{k}_{+1}}\rangle_1 & \langle a_{1|\mathbf{k}_{+2}}\rangle_2 & \langle a_{1|\mathbf{k}_{+3}}\rangle_3 \\
        \langle a_{2|\mathbf{k}_{+1}}\rangle_1 & \langle a_{2|\mathbf{k}_{+2}}\rangle_2 & \langle a_{2|\mathbf{k}_{+3}}\rangle_3 \\
        \langle a_{3|\mathbf{k}_{+1}}\rangle_1 & \langle a_{3|\mathbf{k}_{+2}}\rangle_2 & \langle a_{3|\mathbf{k}_{+3}}\rangle_3
    \end{pmatrix}},
\end{equation}
where the diagonal elements have been calculated previously, and the remaining elements are
\begin{subequations}
    \begin{align}
        \langle a_{1|\mathbf{k}_{+2}}\rangle_2
        &=\frac{rc}{k+1}(k-2)x_1+\frac{rc}{k+1}(k-2)x_3+\frac{2rc}{k+1}-c, \label{sieq_poolatype31}\\
        \langle a_{1|\mathbf{k}_{+3}}\rangle_3
        &=\frac{rc}{k+1}(k-2)x_1+\frac{rc}{k+1}(k-2)x_3+\frac{3rc}{k+1}-c;
        \\
        \langle a_{2|\mathbf{k}_{+1}}\rangle_1
        &=\frac{rc}{k+1}(k-2)x_1+\frac{rc}{k+1}(k-2)x_3+\frac{2rc}{k+1}-\beta\left[1-\left(1-\frac{k-2}{k-1}x_3\right)^{k-1}\right], \\
        \langle a_{2|\mathbf{k}_{+3}}\rangle_3
        &=\frac{rc}{k+1}(k-2)x_1+\frac{rc}{k+1}(k-2)x_3+\frac{2rc}{k+1}-\beta; 
        \\
        \langle a_{3|\mathbf{k}_{+1}}\rangle_1
        &=\frac{rc}{k+1}(k-2)x_1+\frac{rc}{k+1}(k-2)x_3+\frac{3rc}{k+1}-c-\alpha, \\
        \langle a_{3|\mathbf{k}_{+2}}\rangle_2
        &=\frac{rc}{k+1}(k-2)x_1+\frac{rc}{k+1}(k-2)x_3+\frac{rc}{k+1}-c-\alpha.\label{sieq_poolatype32}
    \end{align}
\end{subequations}
According to Eqs.~(\ref{sieq_poolatype1}), (\ref{sieq_poolatype21})--(\ref{sieq_poolatype22}), and (\ref{sieq_poolatype31})--(\ref{sieq_poolatype32}), there are totaling $(2n-1)n=15$ elements in these two `$\langle a_{i|\mathbf{k}_{+j}}\rangle_i$' and `$\langle a_{i|\mathbf{k}_{+j}}\rangle_j$' types.

We can apply these elements to Eq.~(\ref{sieq_alevel3_PC}), which is
\begin{equation}\label{sieq_alevel3_PC_pool}
    \dot{x}_i=\frac{\delta(k-2)}{2(k-1)}x_i \sum_{j=1}^3 x_j
    \Bigg(
    \langle a_{i|\mathbf{k}_{+j}}\rangle_i+
    (k-1) \langle a_{i|\mathbf{k}_{+j}}\rangle_j
    +\langle a_{i|\mathbf{k}_{+i}}\rangle_i
    -\langle a_{j|\mathbf{k}_{+i}}\rangle_j
    -\langle a_{j|\mathbf{k}_{+i}}\rangle_i
    -(k-2)\sum_{l=1}^3 x_l 
    \langle a_{j|\mathbf{k}_{+l}}\rangle_l
    -\langle a_{j|\mathbf{k}_{+j}}\rangle_j
    \Bigg).
\end{equation}

Let us do this step by step. For $i=1$, we have
\begin{subequations}
    \begin{align}
        \sum_{j=1}^3 x_j\langle a_{1|\mathbf{k}_{+j}}\rangle_1
        =&~\frac{rc}{k+1}(k-1)x_1+\frac{rc}{k+1}(k-1)x_3+\frac{2rc}{k+1}-c, \label{sieq_poolasum11}\\
        \sum_{j=1}^3 x_j\langle a_{1|\mathbf{k}_{+j}}\rangle_j
        =&~\frac{rc}{k+1}kx_1+\frac{rc}{k+1}kx_3+\frac{rc}{k+1}-c, \\
        \sum_{j=1}^3 x_j\langle a_{1|\mathbf{k}_{+1}}\rangle_1
        =&~\frac{rc}{k+1}(k-2)x_1+\frac{rc}{k+1}(k-2)x_3+\frac{3rc}{k+1}-c, \\
        \sum_{j=1}^3 x_j\langle a_{j|\mathbf{k}_{+1}}\rangle_1
        =&~\frac{rc}{k+1}(k-1)x_1+\frac{rc}{k+1}(k-1)x_3+\frac{2rc}{k+1}-c(x_1+x_3)-\alpha x_3 \nonumber\\
        &-\beta(1-x_1-x_3)\left[1-\left(1-\frac{k-2}{k-1}x_3\right)^{k-1}\right], \\
        \sum_{j=1}^3 x_j\langle a_{j|\mathbf{k}_{+1}}\rangle_j
        =&~\frac{rc}{k+1}kx_1+\frac{rc}{k+1}kx_3+\frac{rc}{k+1}-c(x_1+x_3)-\alpha x_3-\beta(1-x_1-x_3)\left[1-\left(1-\frac{k-2}{k-1}x_3\right)^{k-1}\right], \\
        \sum_{j=1}^3 x_j\sum_{l=1}^3 x_l\langle a_{j|\mathbf{k}_{+l}}\rangle_l
        =&~(rc-c)(x_1+x_3)-\alpha x_3-\beta(1-x_1-x_3)(1-x_3)\left[1-\left(1-\frac{k-2}{k-1}x_3\right)^{k-1}\right]-\beta(1-x_1-x_3)x_3, \\
        \sum_{j=1}^3 x_j\langle a_{j|\mathbf{k}_{+j}}\rangle_j
        =&~(rc-c)(x_1+x_3)-\alpha x_3-\beta(1-x_1-x_3)\left[1-\left(1-\frac{k-2}{k-1}x_3\right)^{k-1}\right],\label{sieq_poolasum12}
    \end{align}
\end{subequations}
which can be used to obtain $\dot{x}_1$.

For $i=3$, $\sum_{j=1}^3 x_j\sum_{l=1}^3 x_l\langle a_{j|\mathbf{k}_{+l}}\rangle_l$ and $\sum_{j=1}^3 x_j\langle a_{j|\mathbf{k}_{+j}}\rangle_j$ have been obtained previously, and we have the remaining terms:
\begin{subequations}
    \begin{align}
        \sum_{j=1}^3 x_j\langle a_{3|\mathbf{k}_{+j}}\rangle_3
        =&~\frac{rc}{k+1}(k-1)x_1+\frac{rc}{k+1}(k-1)x_3+\frac{2rc}{k+1}-c-\alpha, \label{sieq_poolasum21}\\
        \sum_{j=1}^3 x_j\langle a_{3|\mathbf{k}_{+j}}\rangle_j
        =&~\frac{rc}{k+1}kx_1+\frac{rc}{k+1}kx_3+\frac{rc}{k+1}-c-\alpha, \\
        \sum_{j=1}^3 x_j\langle a_{3|\mathbf{k}_{+3}}\rangle_3
        =&~\frac{rc}{k+1}(k-2)x_1+\frac{rc}{k+1}(k-2)x_3+\frac{3rc}{k+1}-c-\alpha, \\
        \sum_{j=1}^3 x_j\langle a_{j|\mathbf{k}_{+3}}\rangle_3
        =&~\frac{rc}{k+1}(k-1)x_1+\frac{rc}{k+1}(k-1)x_3+\frac{2rc}{k+1}-c(x_1+x_3)-\alpha x_3-\beta(1-x_1-x_3), \\
        \sum_{j=1}^3 x_j\langle a_{j|\mathbf{k}_{+3}}\rangle_j
        =&~\frac{rc}{k+1}kx_1+\frac{rc}{k+1}kx_3+\frac{rc}{k+1}-c(x_1+x_3)-\alpha x_3-\beta(1-x_1-x_3),\label{sieq_poolasum22}
    \end{align}
\end{subequations}
which can be used to obtain $\dot{x}_3$.

Applying Eqs.~(\ref{sieq_poolasum11})--(\ref{sieq_poolasum12}) and (\ref{sieq_poolasum21})--(\ref{sieq_poolasum22}) to Eq.~(\ref{sieq_alevel3_PC_pool}), we obtain the following simple replicator equation:
\begin{subequations}\label{sieq_PC_pool_x1x3}
\begin{align}
\dot{x}_1&
=\frac{\delta(k-2)(k+1)}{2(k-1)} x_1\left\{(1-x_1-x_3)\left(\frac{rc}{k+1}-c\right)+x_3\alpha+(1-x_1-x_3)\beta
\left[1-\left(1-\frac{k-2}{k+1}x_3\right)\left(1-\frac{k-2}{k-1}x_3\right)^{k-1}\right]\right\}, \\
\dot{x}_3&
=\frac{\delta(k-2)(k+1)}{2(k-1)} x_3\left\{(1-x_1-x_3)\left(\frac{rc}{k+1}-c\right)-(1-x_3)\alpha+(1-x_1-x_3)\beta
\left[1-\frac{k-1}{k+1}\left(1-\frac{k-2}{k-1}x_3\right)^k\right]\right\}.
\end{align}
\end{subequations}
Please note that $\dot{x}_2=-\dot{x}_1-\dot{x}_3$ has been canceled.

Similarly, solving $\dot{\mathbf{x}}=\mathbf{0}$, we obtain four possible equilibrium points: a point on the $C$-vertex, $\mathbf{x}^{(C)}=(1,0,0)$, a point on the $D$-vertex, $\mathbf{x}^{(D)}=(0,1,0)$, a point on the $O$-vertex, $\mathbf{x}^{(O)}=(0,0,1)$, and a point on the $DO$-edge, $\mathbf{x}^{(DO)}=(0,x_2^{(DO)},x_3^{(DO)})$, where
\begin{subequations}
\begin{align}
x_2^{(DO)}&=\frac{k-1}{k-2}\left(-\frac{1}{k-1}+\sqrt[k]{\frac{k+1}{k-1}\left[1+\frac{1}{\beta}\left(\frac{rc}{k+1}-c-\alpha \right)\right]}\right), \label{sieq_pool_x2DO_st}\\
x_3^{(DO)}&=1-x_2^{(DO)}=\frac{k-1}{k-2}\left(1-\sqrt[k]{\frac{k+1}{k-1}\left[1+\frac{1}{\beta}\left(\frac{rc}{k+1}-c-\alpha \right)\right]}\right).
\end{align}
\end{subequations}

The Jacobian matrix of system (\ref{sieq_PC_pool_x1x3}) is 
\begin{equation}\label{sieq_J_PC_pool}
    J=\begin{pmatrix}
    \displaystyle{\frac{\partial \dot{x}_1}{\partial x_1}} & 
    \displaystyle{\frac{\partial \dot{x}_1}{\partial x_3}} \\[1em]
    \displaystyle{\frac{\partial \dot{x}_3}{\partial x_1}} & 
    \displaystyle{\frac{\partial \dot{x}_3}{\partial x_3}}
    \end{pmatrix},
\end{equation}
where
\begin{align}
    \frac{\partial \dot{x}_1}{\partial x_1}
    =&~\frac{\delta(k-2)(k+1)}{2(k-1)}\Bigg\{(1-2x_1-x_3)\left(\frac{rc}{k+1}-c\right)+x_3\alpha \nonumber\\
    &+(1-2x_1-x_3)\beta \left[1-\left(1-\frac{k-2}{k+1}x_3\right)\left(1-\frac{k-2}{k-1}x_3\right)^{k-1}\right]\Bigg\}, \\
    \frac{\partial \dot{x}_1}{\partial x_3}
    =&~\frac{\delta(k-2)(k+1)}{2(k-1)} x_1\Bigg\{-\frac{rc}{k+1}+c+\alpha-\beta\left[1-\left(1-\frac{k-2}{k+1}x_3\right)\left(1-\frac{k-2}{k-1}x_3\right)^{k-1}\right] \nonumber\\
    &+(1-x_1-x_3)\beta\left(1-\frac{k-2}{k-1}x_3\right)^{k-2}\frac{k-2}{k+1}\left(k+2-\frac{k(k-2)}{k-1}x_3\right)\Bigg\}, \\
    \frac{\partial \dot{x}_3}{\partial x_1}
    =&-\frac{\delta(k-2)(k+1)}{2(k-1)} x_3\left\{\frac{rc}{k+1}-c+\beta\left[1-\frac{k-1}{k+1}\left(1-\frac{k-2}{k-1}x_3\right)^k \right]\right\}, \\
    \frac{\partial \dot{x}_3}{\partial x_3}
    =&~\frac{\delta(k-2)(k+1)}{2(k-1)}\Bigg\{(1-x_1-2x_3)\left(\frac{rc}{k+1}-c\right)-(1-2x_3)\alpha+(1-x_1-2x_3)\beta\left[1-\frac{k-1}{k+1}\left(1-\frac{k-2}{k-1}x_3\right)^k \right] \nonumber\\
    &+x_3(1-x_1-x_3)\beta\frac{k(k-2)}{k+1}\left(1-\frac{k-2}{k-1}x_3\right)^{k-1}
    \Bigg\}.
\end{align}

Substituting the value of $\mathbf{x}^{(C)}$ into Eq.~(\ref{sieq_J_PC_pool}), we have
\begin{equation}
    \left.J\right|_{\mathbf{x}=\mathbf{x}^{(C)}}=
    \frac{\delta(k-2)(k+1)}{2(k-1)}
    \begin{pmatrix}
    \displaystyle{-\frac{rc}{k+1}+c} & 
    \displaystyle{-\frac{rc}{k+1}+c+\alpha} \\[1em]
    \displaystyle{0} & 
    \displaystyle{-\frac{rc}{k+1}+c-\alpha}
    \end{pmatrix}.
\end{equation}
The condition ensuring $\left.J\right|_{\mathbf{x}=\mathbf{x}^{(C)}}$ negative-definite is $r>k+1$. Therefore, the $C$-vertex equilibrium $\mathbf{x}^{(D)}$ is stable if and only if $r>k+1$.

Substituting the value of $\mathbf{x}^{(D)}$ into Eq.~(\ref{sieq_J_PC_pool}), we have
\begin{equation}\label{sieq_pool_J_xD_st}
    \left.J\right|_{\mathbf{x}=\mathbf{x}^{(D)}}=
    \frac{\delta(k-2)(k+1)}{2(k-1)}
    \begin{pmatrix}
    \displaystyle{\frac{rc}{k+1}-c} & 
    \displaystyle{0} \\[1em]
    \displaystyle{0} & 
    \displaystyle{\frac{rc}{k+1}-c-\alpha+\frac{2\beta}{k+1}}
    \end{pmatrix}.
\end{equation}
The condition ensuring $\left.J\right|_{\mathbf{x}=\mathbf{x}^{(D)}}$ negative-definite and the $D$-vertex equilibrium $\mathbf{x}^{(D)}$ stable is $r<k+1$ and $\beta<\beta^\star$, where
\begin{equation}
    \beta^\star=\frac{k+1}{2}\left(-\frac{rc}{k+1}+c+\alpha\right).
\end{equation}
This indicates that when $\beta>\beta^\star$, the $D$-vertex equilibrium cannot be stable even if $r<k+1$.

Substituting the value of $\mathbf{x}^{(O)}$ into Eq.~(\ref{sieq_J_PC_pool}), we have $\left.\partial \dot{x}_1/\partial x_1\right|_{\mathbf{x}=\mathbf{x}^{(O)}}\propto\alpha>0$. Therefore, the Jacobian matrix at $\mathbf{x}^{(O)}$ is not negative-definite and the $O$-vertex equilibrium $\mathbf{x}^{(O)}$ is not stable.

Substituting the value of $\mathbf{x}^{(DO)}$ into Eq.~(\ref{sieq_J_PC_pool}), we note that $\left.\partial \dot{x}_3/\partial x_1\right|_{\mathbf{x}=\mathbf{x}^{(DO)}}=0$. Therefore, the first- and second-order sequential principal subformulas' negativity is equivalent to $\left.\partial \dot{x}_1/\partial x_1\right|_{\mathbf{x}=\mathbf{x}^{(DO)}}<0$ and $\left.\partial \dot{x}_3/\partial x_3\right|_{\mathbf{x}=\mathbf{x}^{(DO)}}<0$. Let us calculate
\begin{align}
    \left.\frac{\partial \dot{x}_3}{\partial x_3}\right|_{\mathbf{x}=\mathbf{x}^{(DO)}}
    =&~\frac{\delta(k-2)(k+1)}{2(k-1)}\Bigg\{(1-2x_3^{(DO)})\left(\frac{rc}{k+1}-c-\alpha+\beta\left[1-\frac{k-1}{k+1}\left(1-\frac{k-2}{k-1}x_3^{(DO)}\right)^k \right]\right) \nonumber\\
    &+x_3^{(DO)}(1-x_3^{(DO)})\beta\frac{k(k-2)}{k+1}\left(1-\frac{k-2}{k-1}x_3^{(DO)}\right)^{k-1}
    \Bigg\} \nonumber\\
    =&~\frac{\delta(k-2)(k+1)}{2(k-1)}
    x_3^{(DO)}(1-x_3^{(DO)})\beta\frac{k(k-2)}{k+1}\left(1-\frac{k-2}{k-1}x_3^{(DO)}\right)^{k-1}>0.
\end{align}
Therefore, the equilibrium $\mathbf{x}^{(DO)}$ on the $DO$-edge is not stable.

\subsubsection{Discussion}
According to stability analysis, we can divide the results into two cases. First, when $r>k+1$, the fixation of defection at $\mathbf{x}^{(D)}$ is unstable and the fixation of cooperation at $\mathbf{x}^{(C)}$ is stable. The dilemma is overcome through known principles known in the traditional public goods game. 

Second, when $r<k+1$, the fixation of defection occurs in the traditional public goods game, but in this 3-strategy system with pool punishment, the situations can be different. Here, we discuss the effect of pool punishment when $r<k+1$ and show how the results in Fig.~\ref{fig_pool_table} in the main text are obtained.

As mentioned in the main text, when the punishment strength is intermediate ($\beta_0<\beta<\beta^\star$ for structured populations and $\beta>\beta_0^\text{WM}$ for well-mixed populations, as revealed later), there are a stable vertex equilibrium $\mathbf{x}^{(D)}$, an unstable vertex equilibrium $\mathbf{x}^{(O)}$, an unstable vertex equilibrium $\mathbf{x}^{(C)}$, and an unstable edge equilibrium $\mathbf{x}^{(DO)}$. In the 3-strategy system state space, $\mathbf{x}^{(D)}$ is the only stable equilibrium point. However, on the $DO$ edge where $x_1=0$, the vertex equilibrium points $\mathbf{x}^{(D)}$ and $\mathbf{x}^{(O)}$ are bi-stable, depending on the initial state divided by $\mathbf{x}^{(DO)}$. In well-mixed populations, $x_2^{(DO)}\to 1$ as $\beta\to \infty$ according to Eq.~(\ref{sieq_pool_x2DO_wm}), and $\mathbf{x}^{(D)}$ is always stable regardless of the value of $\beta$ according to Eq.~(\ref{sieq_pool_J_xD_wm}). This is the analytical illustration that pool punishment cannot truly resolve social dilemmas in a well-mixed population. However, in structured populations, $x_2^{(DO)}\to [(k+1)^{1/k}(k-1)^{1-1/k}-1]/(k-2)>1$ as $\beta\to \infty$ according to Eq.~(\ref{sieq_pool_x2DO_st}). In particular, $x_2^{(DO)} >1$ happens (i.e., $\mathbf{x}^{(DO)}$ no longer exists) when 
\begin{equation}\label{sieq_pool_star}
    \frac{k-1}{k-2}\left(-\frac{1}{k-1}+\sqrt[k]{\frac{k+1}{k-1}\left[1+\frac{1}{\beta}\left(\frac{rc}{k+1}-c-\alpha \right)\right]}\right)>1
    \Leftrightarrow
    \beta>\beta^\star\equiv \frac{k+1}{2}\left(-\frac{rc}{k+1}+c+\alpha\right).
\end{equation}
Meanwhile, according to Eq.~(\ref{sieq_pool_J_xD_st}), $\mathbf{x}^{(D)}$ becomes unstable at the same time when Eq.~(\ref{sieq_pool_star}) is satisfied, while $\mathbf{x}^{(O)}$ and $\mathbf{x}^{(C)}$ remain unstable. In this case, the system enters a rock-paper-scissors cycle, where defection conquers cooperation ($\partial \dot{x}_1/\partial x_1<0$ when $x_3=0$), cooperation conquers pool punishment ($\partial \dot{x}_1/\partial x_1>0$ when $x_2=0$), and pool punishment conquers defection ($\partial \dot{x}_3/\partial x_3>0$ when $x_1=0$). This is the analytical interpretation that pool punishment can resolve social dilemmas in structured populations.

On the other hand, we also find that structured populations do not always facilitate the advantage of pool punishment compared to well-mixed populations. This property can be observed from the critical punishment strength over which pool punishment starts to play a role. In a structured population, according to Eq.~(\ref{sieq_pool_x2DO_st}), $x_2^{(DO)}>0$ means 
\begin{align}\label{sieq_pool_beta0st}
    &~\frac{k-1}{k-2}\left(-\frac{1}{k-1}+\sqrt[k]{\frac{k+1}{k-1}\left[1+\frac{1}{\beta}\left(\frac{rc}{k+1}-c-\alpha \right)\right]}\right)>0 \nonumber\\
    \Leftrightarrow&~\beta>\beta_0\equiv \frac{(k+1)(k-1)^{k-1}}{1-(k+1)(k-1)^{k-1}}\left(\frac{rc}{k+1}-c-\alpha \right).
\end{align}
When $\beta<\beta_0$, $\mathbf{x}^{(DO)}$ is not a real vector and the system is stable at the vertex equilibrium $\mathbf{x}^{(D)}$, even along the $DO$-edge. Pool punishment starts working and creates bi-stability along the $DO$-edge in structured populations when $\beta>\beta_0$.

In a well-mixed population, according to Eq.~(\ref{sieq_pool_x2DO_wm}), $x_2^{(DO)}>0$ means 
\begin{equation}\label{sieq_pool_beta0wm}
    \sqrt[k]{1+\frac{1}{\beta}\left(\frac{rc}{k+1}-c-\alpha \right)}>0 
    \Leftrightarrow \beta> \beta_0^\text{WM} \equiv -\frac{rc}{k+1}+c+\alpha.
\end{equation}
Pool punishment starts functioning and creates bi-stability along the $DO$-edge in well-mixed populations when $\beta>\beta_0^\text{WM}$.

Comparing Eqs.~(\ref{sieq_pool_beta0st}) and (\ref{sieq_pool_beta0wm}), we see that $\beta_0^\text{WM}<\beta_0$ always holds. That is, pool punishment first becomes efficient in well-mixed populations when increasing the punishment strength.

Actually, there is a considerable interval of $\beta$ that $x_2^{(DO)}$ in structured populations are smaller than the ones in well-mixed populations. That is, a structured population enlarges the initial state interval leading to full defection along the $DO$-edge, thus less effective to utilize pool punishment. A structured population becomes more advantageous only when its $x_2^{(DO)}$ are greater than the ones in well-mixed populations. According to Eqs.~(\ref{sieq_pool_x2DO_wm}) and (\ref{sieq_pool_x2DO_st}), this means
\begin{align}
    &~\frac{k-1}{k-2}\left(-\frac{1}{k-1}+\sqrt[k]{\frac{k+1}{k-1}\left[1+\frac{1}{\beta}\left(\frac{rc}{k+1}-c-\alpha \right)\right]}\right)
    >\sqrt[k]{1+\frac{1}{\beta}\left(\frac{rc}{k+1}-c-\alpha \right)} \nonumber\\
    \Leftrightarrow&~\beta>\beta_= \equiv \frac{\left[(k+1)^{\frac{1}{k}}(k-1)^{1-\frac{1}{k}}-k+2\right]^k}{1-\left[(k+1)^{\frac{1}{k}}(k-1)^{1-\frac{1}{k}}-k+2\right]^k}\left(\frac{rc}{k+1}-c-\alpha \right).
\end{align}
Only when $\beta>\beta_=$, a structured population has a smaller initial state interval leading to full defection along the $DO$-edge, thus more effective to utilize pool punishment than well-mixed populations. In particular, $\beta>\beta^\star$ can even destabilize the full defection equilibrium as shown in Eq.~(\ref{sieq_pool_star}) and transform the system to a rock-paper-scissors-like cyclic dominance, while a well-mixed population cannot realize this.

\subsection{Public goods games with the reward mechanism ($n=3$)}\label{sec_reward}
As a different mechanism from punishment, we study public goods games with the rewarding mechanism. In this model, there are $n=3$ strategies~\cite{szolnoki2010reward}: 

$1=\text{Cooperation ($C$)}$; 

$2=\text{Defection ($D$)}$; 

$3=\text{Reward ($R$)}$. 

On the basis of the public goods game, a rewarding player, as the third strategy, pays a normalized cost $\alpha/k$ to reward a co-player who cooperates. An individual with cooperative behavior receives a normalized reward $\gamma/k$. Here, the cost $\alpha$ and reward $\gamma$ are normalized to keep comparable with the previous work~\cite{szolnoki2010reward}. We assume that rewarding players also perform the cooperative behavior (i.e., investing $c$ in the common pool). Therefore, there are $k_1+k_3$ co-players who perform the cooperative behavior, which incurs the cost $\alpha(k_1+k_3)/k$ to each rewarding player. Given $k_3$ rewarding co-players, each cooperator receives a reward $\gamma k_3/k$. Each rewarding player, as performing the cooperative behavior, also receives a reward $\gamma k_3/k$.

To sum up, given the co-player configuration $\mathbf{k}=(k_1,k_2,k_3)$, we have the following payoff calculation in a single game.
\begin{subequations}\label{sieq_a_reward}
\begin{align}
a_{1|\mathbf{k}}&
=\frac{r(k_1 +1+k_3)c}{k+1}-c+\frac{\gamma k_3}{k}
=\frac{rc}{k+1}k_1+\left(\frac{rc}{k+1}+\frac{\gamma}{k}\right)k_3+\frac{rc}{k+1}-c, \\
a_{2|\mathbf{k}}&
=\frac{r(k_1 +k_3)c}{k+1}
=\frac{rc}{k+1}k_1+\frac{rc}{k+1}k_3, \\
a_{3|\mathbf{k}}&
=\frac{r(k_1+k_3 +1)c}{k+1}-c+\frac{\gamma k_3}{k}-\frac{\alpha (k_1+k_3)}{k}
=\left(\frac{rc}{k+1}-\frac{\alpha}{k}\right)k_1+\left(\frac{rc}{k+1}+\frac{\gamma-\alpha}{k}\right)k_3+\frac{rc}{k+1}-c.
\end{align}
\end{subequations}

\subsubsection{The well-mixed population}\label{sec_reward_wellmixed}
The frequencies of strategies 1, 2, 3 are denoted by $x_1$, $x_2$, and $x_3$ (or $x_C$, $x_D$, and $x_R$ for straightforward understanding), respectively. In a well-mixed population, the mean payoffs of the three strategies are calculated as follows.
\begin{subequations}
\begin{align}
\Bar{\pi}_1&
=\sum_{k_1+k_2+k_3=k}\frac{k!}{k_1! k_2! k_3!} {x_1}^{k_1}{x_2}^{k_2}{x_3}^{k_3} a_{1|\mathbf{k}}
=\frac{rc}{k+1}kx_1+\left(\frac{rc}{k+1}+\frac{\gamma}{k}\right)kx_3+\frac{rc}{k+1}-c, \\
\Bar{\pi}_2&
=\sum_{k_1+k_2+k_3=k}\frac{k!}{k_1! k_2! k_3!} {x_1}^{k_1}{x_2}^{k_2}{x_3}^{k_3} a_{2|\mathbf{k}}
=\frac{rc}{k+1}kx_1+\frac{rc}{k+1}kx_3, \\
\Bar{\pi}_3&
=\sum_{k_1+k_2+k_3=k}\frac{k!}{k_1! k_2! k_3!} {x_1}^{k_1}{x_2}^{k_2}{x_3}^{k_3} a_{3|\mathbf{k}}
=\left(\frac{rc}{k+1}-\frac{\alpha}{k}\right)kx_1+\left(\frac{rc}{k+1}+\frac{\gamma-\alpha}{k}\right)kx_3+\frac{rc}{k+1}-c.
\end{align}
\end{subequations}
The mean payoff of the total population is then calculated by 
\begin{equation}
    \Bar{\pi}
=x_1\Bar{\pi}_1+x_2\Bar{\pi}_2+x_3\Bar{\pi}_3
=\frac{rc}{k+1}kx_1+\frac{rc}{k+1}kx_3+(x_1+x_3)\left(\frac{rc}{k+1}-c\right)+x_3 (x_1+x_3)(\gamma-\alpha).
\end{equation}

On this basis, we can write the replicator equations of the well-mixed population $\dot{x}_i=x_i(\Bar{\pi}_i-\Bar{\pi})$ as follows.
\begin{subequations}
\begin{align}
\dot{x}_1&
=x_1(\Bar{\pi}_1-\Bar{\pi})
=x_1\left[(1-x_1-x_3)\left(\frac{rc}{k+1}-c\right)+x_3(1-x_1-x_3)\gamma+x_3(x_1+x_3)\alpha\right], \\
\dot{x}_2&
=x_2(\Bar{\pi}_2-\Bar{\pi})
=x_2\left[-(x_1+x_3)\left(\frac{rc}{k+1}-c\right)-x_3(x_1+x_3)\gamma+x_3(x_1+x_3)\alpha\right], \\
\dot{x}_3&
=x_3(\Bar{\pi}_3-\Bar{\pi})
=x_3\left[(1-x_1-x_3)\left(\frac{rc}{k+1}-c\right)+x_3(1-x_1-x_3)\gamma-(1-x_3)(x_1+x_3)\alpha\right].
\end{align}
\end{subequations}

We denote the system state $\mathbf{x}=(x_1,x_2,x_3)$. Solving $\dot{\mathbf{x}}=\mathbf{0}$, we obtain four equilibrium points: a point on the $C$-vertex, $\mathbf{x}^{(C)}=(1,0,0)$, a point on the $D$-vertex, $\mathbf{x}^{(D)}=(0,1,0)$, a point on the $R$-vertex, $\mathbf{x}^{(R)}=(0,0,1)$, and a point on the $DR$-edge, $\mathbf{x}^{(DR)}=(0,x_2^{(DR)},x_3^{(DR)})$, where 
\begin{subequations}
\begin{align}
x_2^{(DR)}&=1-\frac{1}{\gamma-\alpha}\left(-\frac{rc}{k+1}+c\right), \label{sieq_reward_x2DR_wm}\\
x_3^{(DR)}&=\frac{1}{\gamma-\alpha}\left(-\frac{rc}{k+1}+c\right).\label{sieq_reward_x3DR_wm}
\end{align}
\end{subequations}

The stability of these four equilibrium points can be studied using the regular method. We cancel $x_2=1-x_1-x_3$ and study the dynamics depicted by $\dot{x}_1$ and $\dot{x}_3$,
\begin{subequations}\label{sieq_reward_x1x3_WM}
\begin{align}
\dot{x}_1&
=x_1\left[(1-x_1-x_3)\left(\frac{rc}{k+1}-c\right)+x_3(1-x_1-x_3)\gamma+x_3(x_1+x_3)\alpha\right], \\
\dot{x}_3&
=x_3\left[(1-x_1-x_3)\left(\frac{rc}{k+1}-c\right)+x_3(1-x_1-x_3)\gamma-(1-x_3)(x_1+x_3)\alpha\right].
\end{align}
\end{subequations}

The Jacobian matrix of system (\ref{sieq_reward_x1x3_WM}) is 
\begin{equation}\label{sieq_J_reward_wm}
    J=\begin{pmatrix}
    \displaystyle{\frac{\partial \dot{x}_1}{\partial x_1}} & 
    \displaystyle{\frac{\partial \dot{x}_1}{\partial x_3}} \\[1em]
    \displaystyle{\frac{\partial \dot{x}_3}{\partial x_1}} & 
    \displaystyle{\frac{\partial \dot{x}_3}{\partial x_3}}
    \end{pmatrix},
\end{equation}
where
\begin{subequations}
    \begin{align}
    \frac{\partial \dot{x}_1}{\partial x_1}&
    =(1-2x_1-x_3)\left(\frac{rc}{k+1}-c+x_3\gamma\right)+x_3(2x_1+x_3)\alpha, \\
    \frac{\partial \dot{x}_1}{\partial x_3}&
    =x_1\left(-\frac{rc}{k+1}+c+(1-x_1-2x_3)\gamma+(x_1+2x_3)\alpha\right), \\
    \frac{\partial \dot{x}_3}{\partial x_1}&
    =x_3\left(-\frac{rc}{k+1}+c-x_3\gamma-(1-x_3)\alpha\right), \\
    \frac{\partial \dot{x}_3}{\partial x_3}&
    =(1-x_1-2x_3)\left(\frac{rc}{k+1}-c\right)+x_3(2-2x_1-3x_3)\gamma-(x_1+2x_3)\alpha+x_3(2x_1+3x_3)\alpha.
    \end{align}
\end{subequations}

Substituting the value of $\mathbf{x}^{(C)}$ into Eq.~(\ref{sieq_J_reward_wm}), we have
\begin{equation}
    \left.J\right|_{\mathbf{x}=\mathbf{x}^{(C)}}=
    \begin{pmatrix}
    \displaystyle{-\frac{rc}{k+1}+c} & 
    \displaystyle{-\frac{rc}{k+1}+c+\alpha} \\[1em]
    \displaystyle{0} & 
    \displaystyle{-\alpha}
    \end{pmatrix}.
\end{equation}
The conditions ensuring $\left.J\right|_{\mathbf{x}=\mathbf{x}^{(C)}}$ negative-definite are $r>k+1$ and $\alpha>0$. The second condition holds clearly. Therefore, we know that $\mathbf{x}^{(C)}$ is stable if and only if $r>k+1$.

Substituting the value of $\mathbf{x}^{(D)}$ into Eq.~(\ref{sieq_J_reward_wm}), we have
\begin{equation}\label{sieq_reward_J_xD_wm}
    \left.J\right|_{\mathbf{x}=\mathbf{x}^{(D)}}=
    \begin{pmatrix}
    \displaystyle{\frac{rc}{k+1}-c} & 
    \displaystyle{0} \\[1em]
    \displaystyle{0} & 
    \displaystyle{\frac{rc}{k+1}-c}
    \end{pmatrix}.
\end{equation}
The condition ensuring $\left.J\right|_{\mathbf{x}=\mathbf{x}^{(D)}}$ negative-definite is $r<k+1$. Therefore, the equilibrium $\mathbf{x}^{(D)}$ is stable when $r<k+1$.

Substituting the value of $\mathbf{x}^{(R)}$ into Eq.~(\ref{sieq_J_reward_wm}), we have
\begin{equation}
    \left.J\right|_{\mathbf{x}=\mathbf{x}^{(R)}}=
    \begin{pmatrix}
    \displaystyle{\alpha} & 
    \displaystyle{0} \\[1em]
    \displaystyle{-\frac{rc}{k+1}+c-\gamma} & 
    \displaystyle{-\frac{rc}{k+1}+c-\gamma+\alpha}
    \end{pmatrix}.
\end{equation}
Since $\alpha>0$, $\left.J\right|_{\mathbf{x}=\mathbf{x}^{(R)}}$ is not negative-definite. Therefore, the equilibrium $\mathbf{x}^{(R)}$ is not stable.

Substituting the value of $\mathbf{x}^{(DR)}$ into Eq.~(\ref{sieq_J_reward_wm}), we obtain $\left.J\right|_{\mathbf{x}=\mathbf{x}^{(DR)}}$, in which
\begin{align}\label{sieq_J_reward_DR_WM}
    \left.\frac{\partial\dot{x}_1}{\partial x_1}\right|_{\mathbf{x}=\mathbf{x}^{(DR)}}
    &=(1-x_3^{(DR)})\left(\frac{rc}{k+1}-c\right)+x_3^{(DR)}\left(\gamma+x_3^{(DR)} (\gamma-\alpha)\right) \nonumber\\
    &=\left[\alpha+\frac{2}{\gamma-\alpha}\left(-\frac{rc}{k+1}+c\right)\right]\left(-\frac{rc}{k+1}+c\right)\frac{1}{\gamma-\alpha}>0.
\end{align}
Eq.~(\ref{sieq_J_reward_DR_WM}) holds because $\gamma-\alpha>0$ as long as $x_3^{(DR)}$ exists according to Eq.~(\ref{sieq_reward_x3DR_wm}), and the case that we focus on is $r<k+1$. Therefore, the equilibrium point $\mathbf{x}^{(DR)}$ is not stable.

\subsubsection{The structured population}
We notice that the payoff structure given by Eq.~(\ref{sieq_a_reward}) is linear, which means that we can utilize the simplified method for special linear systems given by Supplementary Note~\ref{sec_linear} for convenience.

Comparing the payoff structure $a_{i|\mathbf{k}}$ in Eq.~(\ref{sieq_a_reward}) with Eqs.~(\ref{sieq_b_c})--(\ref{sieq_linear_a_PC}), we extract matrices $\mathbf{b}$ and $\mathbf{c}$, 
\begin{equation}
    \mathbf{b}=
    \begin{pmatrix}
        \dfrac{rc}{k+1} & 0 & \dfrac{rc}{k+1}+\dfrac{\gamma}{k} \\[1em]
        \dfrac{rc}{k+1} & 0 & \dfrac{rc}{k+1} \\[1em]
        \dfrac{rc}{k+1}-\dfrac{\alpha}{k} & 0 & \dfrac{rc}{k+1}+\dfrac{\gamma-\alpha}{k}
    \end{pmatrix},~
    \mathbf{c}=
    \begin{pmatrix}
        \dfrac{rc}{k+1}-c \\[1em]
        0 \\
        \dfrac{rc}{k+1}-c
    \end{pmatrix}.
\end{equation}
According to Eq.~(\ref{sieq_linear_PC}), let us calculate
\begin{subequations}
\begin{align}
3\sum_{j=1}^3 x_j (b_{11}-b_{1j}-b_{j1}-b_{jj})&
=-6(x_1+x_3)\frac{rc}{k+1}-6x_3 \frac{\gamma-\alpha}{k}, \\
3\sum_{j=1}^3 x_j (b_{22}-b_{2j}-b_{j2}-b_{jj})&
=-6(x_1+x_3)\frac{rc}{k+1}-3x_3 \frac{\gamma-\alpha}{k}, \\
3\sum_{j=1}^3 x_j (b_{33}-b_{3j}-b_{j3}-b_{jj})&
=-6(x_1+x_3)\frac{rc}{k+1}+(3x_2-6x_3)\frac{\gamma-\alpha}{k}, \\
6\sum_{j=1}^3 \sum_{l=1}^3 x_j x_l b_{jl}&
=6(x_1+x_3)\frac{rc}{k+1}+6x_3 (x_1+x_3)\frac{\gamma-\alpha}{k}.
\end{align}
\end{subequations}
Then, we insert them into Eq.~(\ref{sieq_linear_PC}). Meanwhile, the expressions of $\Bar{\pi}_i$ and $\Bar{\pi}$ are the same as we give in Supplementary Note~\ref{sec_reward_wellmixed} for well-mixed populations. In this way, we obtain the replicator equations in structured populations as follows.
\begin{subequations}
\begin{align}
\dot{x}_1&
=\frac{\delta(k-2)}{2(k-1)} x_1\left\{(k+1)\left[(1-x_1-x_3)\left(\frac{rc}{k+1}-c\right)+x_3(1-x_1-x_3)\gamma+x_3(x_1+x_3)\alpha\right]-6x_2 x_3 \frac{\gamma-\alpha}{k}\right\}, \\
\dot{x}_2&
=\frac{\delta(k-2)}{2(k-1)}x_2\left\{(k+1)\left[-(x_1+x_3)\left(\frac{rc}{k+1}-c\right)-x_3(x_1+x_3)\gamma+x_3(x_1+x_3)\alpha\right]-3x_3(1-2x_1-2x_3)\frac{\gamma-\alpha}{k}\right\}, \\
\dot{x}_3&
=\frac{\delta(k-2)}{2(k-1)}x_3\left\{(k+1)\left[(1-x_1-x_3)\left(\frac{rc}{k+1}-c\right)+x_3(1-x_1-x_3)\gamma-(1-x_3)(x_1+x_3)\alpha\right]+3x_2(1-2x_3)\frac{\gamma-\alpha}{k}\right\}.
\end{align}
\end{subequations}

Similarly, solving $\dot{\mathbf{x}}=\mathbf{0}$, we can obtain five possible equilibrium points: a point on the $C$-vertex, $\mathbf{x}^{(C)}=(1,0,0)$, a point on the $D$-vertex, $\mathbf{x}^{(D)}=(0,1,0)$, a point on the $R$-vertex, $\mathbf{x}^{(R)}=(0,0,1)$, a point on the $DR$-edge, $\mathbf{x}^{(DR)}=(0,x_2^{(DR)},x_3^{(DR)})$, where 
\begin{subequations}
\begin{align}
x_2^{(DR)}&=\frac{k(k+1)}{(k^2+k-6)(\gamma-\alpha)}\left(\frac{rc}{k+1}-c\right)+\frac{k^2+k-3}{k^2+k-6}, \label{sieq_reward_x2DR_PC}\\
x_3^{(DR)}&=\frac{k(k+1)}{(k^2+k-6)(\gamma-\alpha)}\left(-\frac{rc}{k+1}+c\right)-\frac{3}{k^2+k-6},
\end{align}
\end{subequations}
and finally, an interior point $\mathbf{x}^{(CDR)}=(x_1^{(CDR)},x_2^{(CDR)},x_3^{(CDR)})$, where
\begin{subequations}
\begin{align}
x_1^{(CDR)}&=1-\frac{k(k+1)\alpha}{k(k+1)\alpha+3(\gamma-\alpha)}-\frac{k(k+1)}{k(k+1)\gamma-3(\gamma-\alpha)}\left(-\frac{rc}{k+1}+c\right), \\
x_2^{(CDR)}&=\frac{k(k+1)\alpha}{k(k+1)\alpha+3(\gamma-\alpha)}, \\
x_3^{(CDR)}&=\frac{k(k+1)}{k(k+1)\gamma-3(\gamma-\alpha)}\left(-\frac{rc}{k+1}+c\right).
\end{align}
\end{subequations}
We note that in public goods games with the reward mechanism, structured populations create a new equilibrium point $\mathbf{x}^{(CDR)}$ that does not exist in well-mixed populations. This is a new phenomenon that we did not observe in punishment mechanisms.

We cancel $x_2=1-x_1-x_3$ and study the dynamics depicted by $\dot{x}_1$ and $\dot{x}_3$,
\begin{subequations}\label{sieq_reward_x1x3_PC}
\begin{align}
\dot{x}_1&
=\frac{\delta(k-2)(k+1)}{2(k-1)} x_1\left\{
(1-x_1-x_3)\left[\frac{rc}{k+1}-c+x_3 \left(\gamma-6\frac{\gamma-\alpha}{k(k+1)}\right)\right]+x_3 (x_1+x_3)\alpha\right\}, \\
\dot{x}_3&
=\frac{\delta(k-2)(k+1)}{2(k-1)} x_3\left\{
x_3(1-x_1-x_3)\left[\frac{rc}{k+1}-c+x_3 \left(\gamma-6\frac{\gamma-\alpha}{k(k+1)}\right)+3\frac{\gamma-\alpha}{k(k+1)}\right]-(1-x_3) (x_1+x_3)\alpha\right\}.
\end{align}
\end{subequations}

The Jacobian matrix of system (\ref{sieq_reward_x1x3_PC}) is 
\begin{equation}\label{sieq_J_reward_PC}
    J=\begin{pmatrix}
    \displaystyle{\frac{\partial \dot{x}_1}{\partial x_1}} & 
    \displaystyle{\frac{\partial \dot{x}_1}{\partial x_3}} \\[1em]
    \displaystyle{\frac{\partial \dot{x}_3}{\partial x_1}} & 
    \displaystyle{\frac{\partial \dot{x}_3}{\partial x_3}}
    \end{pmatrix},
\end{equation}
where
\begin{subequations}
    \begin{align}
    \frac{\partial \dot{x}_1}{\partial x_1}
    =&~\frac{\delta(k-2)(k+1)}{2(k-1)}\left\{
    (1-2x_1-x_3)\left[\frac{rc}{k+1}-c+x_3 \left(\gamma-6\frac{\gamma-\alpha}{k(k+1)}\right)\right]+x_3(2x_1+x_3)\alpha\right\}, \\
    \frac{\partial \dot{x}_1}{\partial x_3}
    =&~\frac{\delta(k-2)(k+1)}{2(k-1)}x_1 \left[
    -\frac{rc}{k+1}+c-2x_3 \left(\gamma-6\frac{\gamma-\alpha}{k(k+1)}\right)+(x_1+2x_3)\alpha\right], \\
    \frac{\partial \dot{x}_3}{\partial x_1}
    =&~\frac{\delta(k-2)(k+1)}{2(k-1)}x_3 \left[
    -\frac{rc}{k+1}+c-x_3 \left(\gamma-6\frac{\gamma-\alpha}{k(k+1)}\right)-3\frac{\gamma-\alpha}{k(k+1)}-(1-x_3)\alpha\right], \\
    \frac{\partial \dot{x}_3}{\partial x_3}
    =&~\frac{\delta(k-2)(k+1)}{2(k-1)}\Bigg\{
    (1-x_1-2x_3)\left(\frac{rc}{k+1}-c+3\frac{\gamma-\alpha}{k(k+1)}\right)+(2-2x_1-3x_3)x_3 \left(\gamma-6\frac{\gamma-\alpha}{k(k+1)}\right) \nonumber\\
    &-(1-2x_3)x_1 \alpha-(2-3x_3)x_3 \alpha\Bigg\}. \label{sieq_reward_Jx3x3_PC}
    \end{align}
\end{subequations}

Substituting the value of $\mathbf{x}^{(C)}$ into Eq.~(\ref{sieq_J_reward_PC}), we have
\begin{equation}
    \left.J\right|_{\mathbf{x}=\mathbf{x}^{(C)}}=
    \frac{\delta(k-2)(k+1)}{2(k-1)}
    \begin{pmatrix}
    \displaystyle{-\frac{rc}{k+1}+c} & 
    \displaystyle{-\frac{rc}{k+1}+c+\alpha} \\[1em]
    \displaystyle{0} & 
    \displaystyle{-\alpha}
    \end{pmatrix}.
\end{equation}
The conditions ensuring $\left.J\right|_{\mathbf{x}=\mathbf{x}^{(C)}}$ negative-definite are $r>k+1$ and $\alpha>0$. The second condition holds clearly. Therefore, we know that $\mathbf{x}^{(C)}$ is stable if and only if $r>k+1$.

Substituting the value of $\mathbf{x}^{(D)}$ into Eq.~(\ref{sieq_J_reward_PC}), we have
\begin{equation}\label{sieq_reward_J_xD_st}
    \left.J\right|_{\mathbf{x}=\mathbf{x}^{(D)}}=
    \frac{\delta(k-2)(k+1)}{2(k-1)}
    \begin{pmatrix}
    \displaystyle{\frac{rc}{k+1}-c} & 
    \displaystyle{0} \\[1em]
    \displaystyle{0} & 
    \displaystyle{\frac{rc}{k+1}-c+3\frac{\gamma-\alpha}{k(k+1)}}
    \end{pmatrix}.
\end{equation}
The conditions ensuring $\left.J\right|_{\mathbf{x}=\mathbf{x}^{(D)}}$ negative-definite and the $D$-vertex equilibrium $\mathbf{x}^{(D)}$ stable are $r<k+1$ and $\gamma<\gamma^\star$, where
\begin{equation}
    \gamma^\star=\frac{k(k+1)}{3}\left(-\frac{rc}{k+1}+c\right)+\alpha.
\end{equation}
This indicates that when $\gamma>\gamma^\star$, the $D$-vertex equilibrium cannot be stable even if $r<k+1$.

Substituting the value of $\mathbf{x}^{(R)}$ into Eq.~(\ref{sieq_J_reward_PC}), we have
\begin{equation}
    \left.J\right|_{\mathbf{x}=\mathbf{x}^{(R)}}=
    \frac{\delta(k-2)(k+1)}{2(k-1)}
    \begin{pmatrix}
    \displaystyle{\alpha} & 
    \displaystyle{0} \\[1em]
    \displaystyle{-\frac{rc}{k+1}+c-\gamma-3\frac{\gamma-\alpha}{k(k+1)}} & 
    \displaystyle{-\frac{rc}{k+1}+c-\frac{k^2+k-3}{k(k+1)}(\gamma-\alpha)}
    \end{pmatrix}.
\end{equation}
Since $\alpha>0$, $\left.J\right|_{\mathbf{x}=\mathbf{x}^{(R)}}$ is not negative-definite. Therefore, the $R$-vertex equilibrium point $\mathbf{x}^{(R)}$ is not stable.

Substituting the value of $\mathbf{x}^{(DR)}$ into Eq.~(\ref{sieq_J_reward_PC}), we obtain $\left.J\right|_{\mathbf{x}=\mathbf{x}^{(DR)}}$, where we note that $\left.\partial \dot{x}_1/\partial x_3\right|_{\mathbf{x}=\mathbf{x}^{(DR)}}=0$. Therefore, the first- and second-order sequential principal subformulas' negativity is equivalent to $\left.\partial \dot{x}_1/\partial x_1\right|_{\mathbf{x}=\mathbf{x}^{(DR)}}<0$ and $\left.\partial \dot{x}_3/\partial x_3\right|_{\mathbf{x}=\mathbf{x}^{(DR)}}<0$. Let us study $\left.\partial \dot{x}_3/\partial x_3\right|_{\mathbf{x}=\mathbf{x}^{(DR)}}$. Reorganizing Eq.~(\ref{sieq_reward_Jx3x3_PC}) but keeping $x_3^{(DR)}$, we obtain
\begin{equation}\label{sieq_reward_Jx3x3_DR_PC}
    \left.\frac{\partial \dot{x}_3}{\partial x_3}\right|_{\mathbf{x}=\mathbf{x}^{(DR)}}
    =\frac{\delta(k-2)(k+1)}{2(k-1)}\left(\mathcal{A}(x_3^{(DR)})^2+\mathcal{B}x_3^{(DR)}+\mathcal{C}\right),
\end{equation}
where $\mathcal{A}$, $\mathcal{B}$, and $\mathcal{C}$ are constants, 
\begin{subequations}
    \begin{align}
    \mathcal{A}&=-3\frac{k^2+k-6}{k(k+1)}(\gamma-\alpha), \\
    \mathcal{B}&=2\left(-\frac{rc}{k+1}+c+\frac{k^2+k-9}{k(k+1)}(\gamma-\alpha)\right), \\
    \mathcal{C}&=\frac{rc}{k+1}-c+3\frac{\gamma-\alpha}{k(k+1)}.
    \end{align}
\end{subequations}
Importantly, we notice that
\begin{equation}
    x_3^{(DR)}<1 \Leftrightarrow \gamma-\alpha>\frac{k(k+1)}{k^2+k-3}\left(-\frac{rc}{k+1}+c\right)>0,
\end{equation}
which indicates $\mathcal{A}<0$.

According to Eq.~(\ref{sieq_reward_Jx3x3_DR_PC}), $\left.\partial \dot{x}_3/\partial x_3\right|_{\mathbf{x}=\mathbf{x}^{(DR)}}$ is a quadratic function of $x_3^{(DR)}$, with $\mathcal{A}<0$ as we found. The properties of quadratic functions indicate that $\left.\partial \dot{x}_3/\partial x_3\right|_{\mathbf{x}=\mathbf{x}^{(DR)}}\geq 0$ in the interval $0\leq x_3^{(DR)}\leq 1$ as long as (1) $\left.\partial \dot{x}_3/\partial x_3\right|_{\mathbf{x}=\mathbf{x}^{(DR)}}\geq 0$ at $x_3^{(DR)}=0$ and (2) $\left.\partial \dot{x}_3/\partial x_3\right|_{\mathbf{x}=\mathbf{x}^{(DR)}}\geq 0$ at $x_3^{(DR)}=1$. Let us examine these two conditions. First, when $x_3^{(DR)}=0$, we have 
\begin{equation}
    \left.\frac{\partial \dot{x}_3}{\partial x_3}\right|_{\mathbf{x}=\mathbf{x}^{(DR)}}
    =\frac{rc}{k+1}-c+3\frac{\gamma-\alpha}{k(k+1)}=-\frac{k(k+1)}{(k^2+k-6)(\gamma-\alpha)}x_3^{(DR)}=0.
\end{equation}
Second, when $x_3^{(DR)}=1$, we have 
\begin{equation}
    \left.\frac{\partial \dot{x}_3}{\partial x_3}\right|_{\mathbf{x}=\mathbf{x}^{(DR)}}
    =-\frac{rc}{k+1}+c-\frac{k^2+k-3}{k(k+1)}(\gamma-\alpha)=-\frac{k(k+1)}{(k^2+k-6)(\gamma-\alpha)}(1-x_3^{(DR)})=0.
\end{equation}

Therefore, $\left.\partial \dot{x}_3/\partial x_3\right|_{\mathbf{x}=\mathbf{x}^{(DR)}}\geq 0$ and $\left.J\right|_{\mathbf{x}=\mathbf{x}^{(DR)}}$ is not negative-definite in the interval $0\leq x_3^{(DR)}\leq 1$. The equilibrium point $\mathbf{x}^{(DR)}$ is not stable when it exists.

Substituting the value of $\mathbf{x}^{(CDR)}$ into Eq.~(\ref{sieq_J_reward_PC}), we obtain $\left.J\right|_{\mathbf{x}=\mathbf{x}^{(CDR)}}$, where
\begin{align}\label{sieq_J_x1x1_XCDR}
    \left.\frac{\partial \dot{x}_1}{\partial x_1}\right|_{\mathbf{x}=\mathbf{x}^{(CDR)}}
    =&~\frac{\delta(k-2)(k+1)}{2(k-1)}\Bigg\{
    (1-2x_1^{(CDR)}-x_3^{(CDR)})\left[\frac{rc}{k+1}-c+x_3^{(CDR)} \left(\gamma-6\frac{\gamma-\alpha}{k(k+1)}\right)\right]\nonumber\\
    &+x_3^{(CDR)}(2x_1^{(CDR)}+x_3^{(CDR)})\alpha\Bigg\}.
\end{align}

At $\mathbf{x}^{(CDR)}$, we have $\dot{x}_1=0$, that is,
\begin{equation}\label{sieq_dx10}
    \frac{\delta(k-2)(k+1)}{2(k-1)}\left\{(1-x_1^{(CDR)}-x_3^{(CDR)})\left[\frac{rc}{k+1}-c+x_3^{(CDR)} \left(\gamma-6\frac{\gamma-\alpha}{k(k+1)}\right)\right]+x_3^{(CDR)} (x_1^{(CDR)}+x_3^{(CDR)})\alpha\right\}=0.
\end{equation}
We calculate Eq.~(\ref{sieq_J_x1x1_XCDR}) minus Eq.~(\ref{sieq_dx10}), (i.e., $\left.\partial \dot{x}_3/\partial x_3\right|_{\mathbf{x}=\mathbf{x}^{(CDR)}}-0=\left.\partial \dot{x}_3/\partial x_3\right|_{\mathbf{x}=\mathbf{x}^{(CDR)}}$) and further organize the result, which leads to 
\begin{align}
    \left.\frac{\partial \dot{x}_1}{\partial x_1}\right|_{\mathbf{x}=\mathbf{x}^{(CDR)}}
    &=\frac{\delta(k-2)(k+1)}{2(k-1)}x_1^{(CDR)}\left(-\frac{rc}{k+1}+c-x_3^{(CDR)} \frac{k^2+k-6}{k(k+1)}(\gamma-\alpha)\right) \nonumber\\
    &=\frac{\delta(k-2)(k+1)}{2(k-1)}x_1^{(CDR)}\frac{k(k+1)\alpha+3(\gamma-\alpha)}{k(k+1)\gamma-3(\gamma-\alpha)}\left(-\frac{rc}{k+1}+c\right) \nonumber\\
    &=\frac{\delta(k-2)(k+1)}{2(k-1)}x_1^{(CDR)}\frac{x_3^{(CDR)}}{x_2^{(CDR)}}\alpha>0.
\end{align}
Therefore, $\left.J\right|_{\mathbf{x}=\mathbf{x}^{(CDR)}}$ is not negative-definite and the interior equilibrium point $\mathbf{x}^{(CDR)}$ is not stable when it exists.

\subsubsection{Discussion}
Similar to the case in punishment, according to stability analysis, we can divide the results into two cases. First, when $r>k+1$, the fixation of defection at $\mathbf{x}^{(D)}$ is unstable, and the fixation of cooperation at $\mathbf{x}^{(C)}$ is stable. The dilemma is overcome through known principles known in the traditional public goods game. 

Second, when $r<k+1$, the fixation of defection occurs in the traditional public goods game, but in this 3-strategy system with the reward mechanism, the situations can be different. Here, we discuss the effect of reward when $r<k+1$.

\begin{figure}
	\centering
		\includegraphics[width=\textwidth]{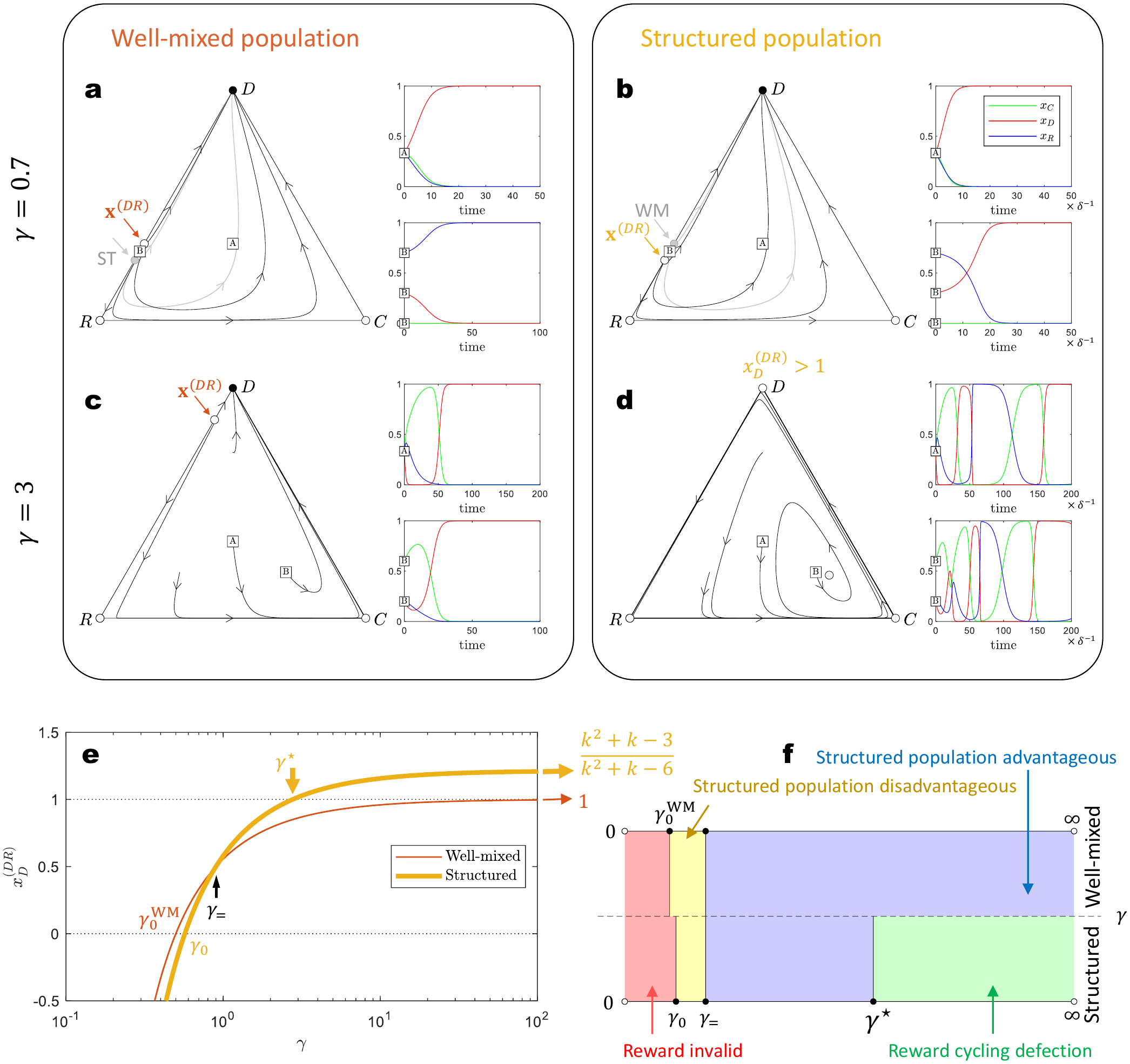}
	\caption{\textbf{The reward mechanism can resolve the social dilemma of public goods game in structured populations.} 
    \textbf{a} and \textbf{b}, In the three-strategy system space, the state consistently converges to full $D$. However, along the $DR$ edge, an unstable equilibrium point, $\mathbf{x}^{(DR)}$, creates a bi-stable space. In the $D$ versus $R$ dynamics, the final state, either $D$ or $R$, is determined by the initial conditions. Under mild reward ($\gamma=0.7$), a structured population tends to favor defection, reducing the basin leading to the $R$ outcome. 
    \textbf{c} and \textbf{d}, Conversely, with strong reward ($\gamma=3$), structured populations result in the cyclic dominance of the three strategies around the interior equilibrium $\mathbf{x}^{(CDR)}$, thereby preventing the full $D$ state in public goods games. In contrast, the state space in well-mixed populations remains two distinct basins on the $DR$ edge, preventing cyclic dominance. 
    \textbf{e}, As the reward strength $\gamma$ increases, $x_D^{(DR)}$ increases, expanding the initial space leading to the $R$ outcome. In well-mixed populations, $x_D^{(DR)}\to 1$ as $\gamma\to \infty$, and the basin leading to defection cannot be completely eliminated. However, in structured populations, $x_D^{(DR)}\to (k^2+k-3)/(k^2+k-6)>1$ when $\gamma>\gamma^\star$, invariably resulting in the cyclic dominance of the three strategies. 
    \textbf{f}, The diagram of the different effects of reward in well-mixed versus structured populations. Structured populations are advantageous in promoting cooperation under strong reward but are a bit less effective when the reward is mild. 
    \textbf{Input parameters}: $r=3$, $c=1$, $\alpha=0.1$, $k=4$.} 
	\label{fig_reward}
\end{figure}

When the reward strength is intermediate ($\gamma_0<\gamma<\gamma^\star$ for structured populations and $\gamma>\gamma_0^\text{WM}$ for well-mixed populations, as revealed later), there are a stable vertex equilibrium $\mathbf{x}^{(D)}$, an unstable vertex equilibrium $\mathbf{x}^{(R)}$, an unstable vertex equilibrium $\mathbf{x}^{(C)}$, an unstable edge equilibrium $\mathbf{x}^{(DR)}$, and for structured populations, a possible unstable interior equilibrium $\mathbf{x}^{(CDR)}$. In the 3-strategy system state space, $\mathbf{x}^{(D)}$ is the only stable equilibrium point. However, on the $DR$ edge where $x_1=0$, the vertex equilibrium points $\mathbf{x}^{(D)}$ and $\mathbf{x}^{(R)}$ are bi-stable, depending on the initial state divided by $\mathbf{x}^{(DR)}$. In well-mixed populations, $x_2^{(DR)}\to 1$ as $\gamma\to \infty$ according to Eq.~(\ref{sieq_reward_x2DR_wm}), and $\mathbf{x}^{(D)}$ is always stable regardless of the value of $\gamma$ according to Eq.~(\ref{sieq_reward_J_xD_wm}). In this way, we analytically illustrate that reward cannot truly resolve social dilemmas in a well-mixed population, similar to punishment. However, in structured populations, $x_2^{(DR)}\to (k^2+k-3)/(k^2+k-6)>1$ as $\gamma\to \infty$ according to Eq.~(\ref{sieq_reward_x2DR_PC}). In particular, $x_2^{(DR)} >1$ happens (i.e., $\mathbf{x}^{(DR)}$ no longer exists) when 
\begin{equation}\label{sieq_reward_star}
    \frac{k(k+1)}{(k^2+k-6)(\gamma-\alpha)}\left(\frac{rc}{k+1}-c\right)+\frac{k^2+k-3}{k^2+k-6}>1 
    \Leftrightarrow\gamma>\gamma^\star\equiv \frac{k(k+1)}{3}\left(-\frac{rc}{k+1}+c\right)+\alpha.
\end{equation}
Meanwhile, according to Eq.~(\ref{sieq_reward_J_xD_st}), $\mathbf{x}^{(D)}$ becomes unstable at the same time when Eq.~(\ref{sieq_reward_star}) is satisfied, while $\mathbf{x}^{(R)}$ and $\mathbf{x}^{(C)}$ remain unstable. In this case, the system enters a rock-paper-scissors cycle (Fig.~\ref{fig_reward}\textbf{c}, \textbf{d}), where defection conquers cooperation ($\partial \dot{x}_1/\partial x_1<0$ when $x_3=0$), cooperation conquers reward ($\partial \dot{x}_1/\partial x_1>0$ when $x_2=0$), and reward conquers defection ($\partial \dot{x}_3/\partial x_3>0$ when $x_1=0$). In this way, we analytically confirm that reward can resolve social dilemmas in structured populations (Fig.~\ref{fig_reward}\textbf{e}). 

On the other hand, we also find that structured populations do not always facilitate the advantage of reward compared to well-mixed populations. This property can be observed from the critical reward strength over which the reward mechanism starts to play a role. In a structured population, according to Eq.~(\ref{sieq_reward_x2DR_PC}), $x_2^{(DR)}>0$ means 
\begin{equation}\label{sieq_reward_beta0st}
    \frac{k(k+1)}{(k^2+k-6)(\gamma-\alpha)}\left(\frac{rc}{k+1}-c\right)+\frac{k^2+k-3}{k^2+k-6}>0 
    \Leftrightarrow\gamma>\gamma_0\equiv \frac{k(k+1)}{k^2+k-3}\left(-\frac{rc}{k+1}+c\right)+\alpha.
\end{equation}
When $\gamma<\gamma$, we have $x_2^{(DR)}<0$ and the system is stable at the vertex equilibrium $\mathbf{x}^{(D)}$ even along the $DR$-edge. The reward mechanism starts working and creates bi-stability along the $DR$-edge in structured populations when $\gamma>\gamma_0$.

In a well-mixed population, according to Eq.~(\ref{sieq_reward_x2DR_wm}), $x_2^{(DR)}>0$ means 
\begin{equation}\label{sieq_reward_beta0wm}
    1-\frac{1}{\gamma-\alpha}\left(-\frac{rc}{k+1}+c\right)>0 
    \Leftrightarrow \gamma> \gamma_0^\text{WM} \equiv -\frac{rc}{k+1}+c+\alpha.
\end{equation}
The reward mechanism starts functioning and creates bi-stability along the $DR$-edge in well-mixed populations when $\gamma>\gamma_0^\text{WM}$.

Comparing Eqs.~(\ref{sieq_reward_beta0st}) and (\ref{sieq_reward_beta0wm}), we see that $\gamma_0^\text{WM}<\gamma_0$ always holds. That is, the reward mechanism first becomes efficient in well-mixed populations when increasing the reward strength.

There is a considerable interval of $\gamma$ that $x_2^{(DR)}$ in structured populations are smaller than the ones in well-mixed populations. That is, a structured population enlarges the initial state interval leading to full defection along the $DR$-edge, thus less effective in utilizing the reward mechanism (Fig.~\ref{fig_reward}\textbf{a}, \textbf{b}). A structured population becomes more advantageous only when its $x_2^{(DR)}$ are greater than the ones in well-mixed populations. According to Eqs.~(\ref{sieq_reward_x2DR_wm}) and (\ref{sieq_reward_x2DR_PC}), this means 
\begin{align}
    &~\frac{k(k+1)}{(k^2+k-6)(\gamma-\alpha)}\left(\frac{rc}{k+1}-c\right)+\frac{k^2+k-3}{k^2+k-6}
    >1-\frac{1}{\gamma-\alpha}\left(-\frac{rc}{k+1}+c\right) \nonumber\\
    \Leftrightarrow&~\gamma>\gamma_= \equiv \frac{1}{2}\left(-\frac{rc}{k+1}+c\right)+\alpha.
\end{align}
Only when $\gamma>\gamma_=$, a structured population has a smaller initial state interval leading to full defection along the $DR$-edge, thus more effective to utilize the reward mechanism than well-mixed populations. In particular, $\gamma>\gamma^\star$ can even destabilize the full defection equilibrium as shown in Eq.~(\ref{sieq_reward_star}) and transform the system to a rock-paper-scissors-like cyclic dominance, while a well-mixed population cannot realize this (Fig.~\ref{fig_reward}\textbf{f}). 

Similar principles were also found in pool punishment (Supplementary Note~\ref{sec_pool}), whose payoff function is nonlinear. However, through the reward mechanism, we reveal that rock-paper-scissors cycles could emerge even if the payoff function is linear.

\begin{figure}[t]
	\centering
		\includegraphics[width=.95\textwidth]{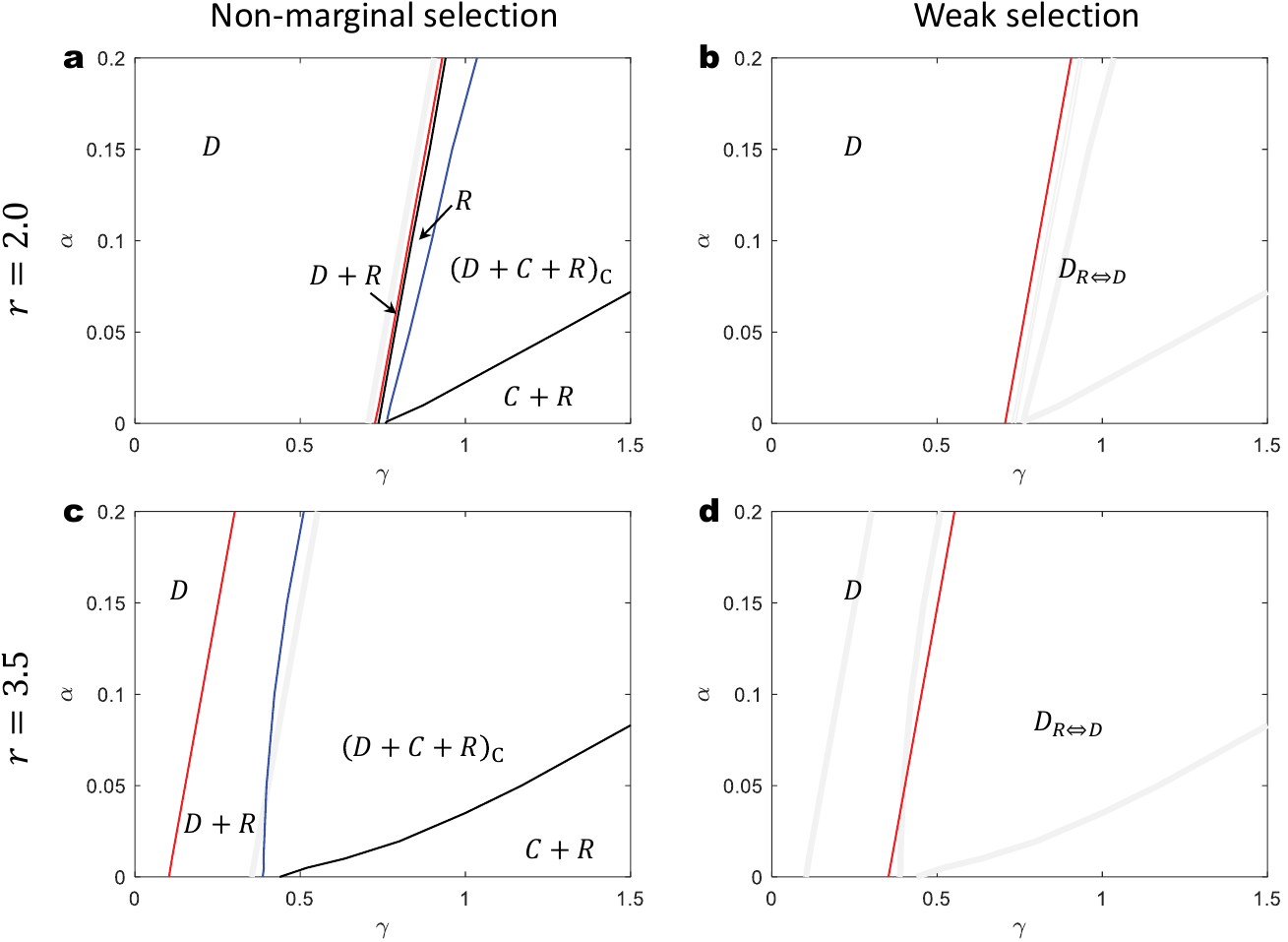}
	\caption{\textbf{Phase diagrams of the system behavior with the reward mechanism are qualitatively similar under non-marginal and weak selection strength.} 
    \textbf{a} and \textbf{c} (the data are in agreement with those published in Figs.~1(a) and 3(a) from ref.~\cite{szolnoki2010reward}), Numerical simulations under non-marginal selection ($\delta=2$). The phases are defined as follows: $D$---only $D$ exists; $R$---only $R$ exists; $(D+C+R)_\text{C}$---cyclic dominance among $D$, $C$, and $R$; $D+R$---$D$ and $R$ coexist; $C+R$---$C$ and $R$ coexist. 
    \textbf{b} and \textbf{d}, The phase diagram is divided by analytical $\gamma_0$ and $\gamma^\star$ (not visible in the presented parameter space) under weak selection ($\delta\to 0^+$). Here, $\gamma_0$ divides the $D$ and $D_{R\Leftrightarrow D}$ phases, while $\gamma^\star$ separates the $D_{R\Leftrightarrow D}$ and $(D+C+R)_\text{C}$ phases. Specifically, in \textbf{b}, $\gamma_0=12/17+\alpha$ (red), $\gamma^\star=4+\alpha$ (invisible); in \textbf{d}, $\gamma_0=6/17+\alpha$, $\gamma^\star=2+\alpha$ (invisible). The definition of the $D_{R\Leftrightarrow D}$ phase---the system finally evolves to full $D$ if cooperation is initially present, or to the fixation of either $R$ or $D$ in the absence of initial cooperators. \textbf{Other parameters:} $c=1$, $k=4$.} 
	\label{fig_phase_reward}
\end{figure}

These analytical results are in qualitative agreement with the $\alpha$-$\gamma$ phase diagrams from previous research~\cite{szolnoki2010reward}, as shown in Fig.~\ref{fig_phase_reward}. The phase diagrams under non-marginal selection indicate the existence of a cyclic dominance phase $(D+C+R)_\text{C}$ (Fig.~\ref{fig_phase_reward}\textbf{a}, \textbf{c}), where strategy $D$ invades $C$, strategy $C$ invades $R$, and strategy $R$ invades $D$. This phase is also predicted under weak selection in the previous analysis. However, the parameter space we show does not include this phase, with the analytical expressions of their boundaries listed in the caption of Fig.~\ref{fig_phase_reward}. Due to the role of a non-marginal selection strength, the effect of reward works in advance and induces the rock-paper-scissors cycling phase as we increase $\gamma$. In particular, there is even a $C+R$ phase under non-marginal selection, which does not exist under weak selection. This can be explained by analogizing the reward mechanism to a donation game---$R$-players donate $\alpha$ to each $C$- or $R$-co-player, who receives $\gamma$. Under weak selection, evolution cannot favor donors if we apply the pairwise comparison update rule~\cite{allen2014games}. However, under non-marginal selection, donors may form spatial clusters (which are more remarkable than the ones under weak selection) and thus survive. Again, while there are differences between outcomes derived from non-marginal selection through numerical simulation (Fig.~\ref{fig_phase_reward}\textbf{a}, \textbf{c}) and those obtained under weak selection with analytical methods (Fig.~\ref{fig_phase_reward}\textbf{b}, \textbf{d}), both approaches indicate some distinct behavioral patterns in structured populations that are not observed in well-mixed populations.

\subsection{The multi-stage public goods game ($n=4$)}\label{sec_multipgg}
Last, we involve another example, the multi-stage public goods game~\cite{szolnoki2022tactical}. In the simplest multi-stage public goods game, we consider two stages. Players can invest (i.e., cooperate) in the first and second stages. Namely, the number of available strategies is $n=4$: 

$1=\text{Cooperation in both stages ($CC$)}$; 

$2=\text{Cooperation only in the first stage ($CD$)}$; 

$3=\text{Cooperation only in the second stage ($DC$)}$; 

$4=\text{Defection in both stages ($DD$)}$. 

The traditional public goods game is equivalent to a one-stage public goods game, where players invest in the only stage and receive the produced public goods. In the extended two-stage public goods game, however, the produced public goods from the first stage (contributed by $CC$- and $CD$-players) are not distributed immediately, but reinvested to the second stage. The second stage uses the inherited public goods from the first stage plus the independent investment (contributed by $CC$- and $DC$-players) to produce the final public goods. The final public goods are then distributed to all players in the game. We assume that the synergy factor of the first stage is $r_1$, and the synergy factor of the second stage is $r_2$.

Therefore, given the co-player configuration $\mathbf{k}=(k_1,k_2,k_3,k_4)$, we have the following payoff calculation in a single multi-stage public goods game.
\begin{subequations}\label{sieq_a_multi}
\begin{align}
a_{1|\mathbf{k}}&
=\frac{r_2 [r_1(k_1+1+k_2)c+(k_1+1+k_3)c]}{k+1}-2c
=\frac{r_2(r_1+1)c}{k+1}k_1+\frac{r_2 r_1 c}{k+1}k_2+\frac{r_2 c}{k+1}k_3+\frac{r_2 (r_1+1)c}{k+1}-2c, \\
a_{2|\mathbf{k}}&
=\frac{r_2 [r_1(k_1+k_2+1)c+(k_1+k_3)c]}{k+1}-c
=\frac{r_2(r_1+1)c}{k+1}k_1+\frac{r_2 r_1 c}{k+1}k_2+\frac{r_2 c}{k+1}k_3+\frac{r_2 r_1 c}{k+1}-c, \\
a_{3|\mathbf{k}}&
=\frac{r_2 [r_1(k_1+k_2)c+(k_1+k_3+1)c]}{k+1}-c
=\frac{r_2(r_1+1)c}{k+1}k_1+\frac{r_2 r_1 c}{k+1}k_2+\frac{r_2 c}{k+1}k_3+\frac{r_2 c}{k+1}-c, \\
a_{4|\mathbf{k}}&
=\frac{r_2 [r_1(k_1+k_2)c+(k_1+k_3)c]}{k+1}
=\frac{r_2(r_1+1)c}{k+1}k_1+\frac{r_2 r_1 c}{k+1}k_2+\frac{r_2 c}{k+1}k_3.
\end{align}
\end{subequations}

\subsubsection{The structured population}
For this multi-stage public goods game model, we find that structured populations are equivalent to well-mixed populations under pairwise comparison and in the weak selection limit. Let us explain why. Observing the payoff structure given by Eq.~(\ref{sieq_a_multi}), we find it linear. We can again utilize the simplified method for special linear systems given by Supplementary Note~\ref{sec_linear} for convenience. Comparing the payoff structure $a_{i|\mathbf{k}}$ with Eqs.~(\ref{sieq_b_c})--(\ref{sieq_linear_a_PC}), we extract matrices $\mathbf{b}$ and $\mathbf{c}$, 
\begin{equation}
    \mathbf{b}=
    \begin{pmatrix}
        \dfrac{r_2(r_1+1)c}{k+1} & \dfrac{r_2 r_1 c}{k+1} & \dfrac{r_2 c}{k+1} & 0 \\[1em]
        \dfrac{r_2(r_1+1)c}{k+1} & \dfrac{r_2 r_1 c}{k+1} & \dfrac{r_2 c}{k+1} & 0 \\[1em]
        \dfrac{r_2(r_1+1)c}{k+1} & \dfrac{r_2 r_1 c}{k+1} & \dfrac{r_2 c}{k+1} & 0 \\[1em]
        \dfrac{r_2(r_1+1)c}{k+1} & \dfrac{r_2 r_1 c}{k+1} & \dfrac{r_2 c}{k+1} & 0
    \end{pmatrix},~
    \mathbf{c}=
    \begin{pmatrix}
        \dfrac{r_2(r_1+1)c}{k+1}-2c \\[1em]
        \dfrac{r_2 r_1 c}{k+1}-c \\[1em]
        \dfrac{r_2 c}{k+1}-c \\[1em]
        0
    \end{pmatrix}.
\end{equation}
The frequencies of strategies 1, 2, 3, 4 are denoted by $x_1$, $x_2$, $x_3$, and $x_4$. We then calculate
\begin{subequations}
\begin{align}
3\sum_{j=1}^4 x_j (b_{ii}-b_{ij}-b_{ji}-b_{jj})&
=-\frac{6r_2(r_1+1)c}{k+1}x_1-\frac{6r_2 r_1 c}{k+1}x_2-\frac{6r_2 c}{k+1}x_3, \quad \mbox{for $i=1,2,3,4$,} \label{sieq_multi_add1}\\
6\sum_{j=1}^4 \sum_{l=1}^4 x_j x_l b_{jl}&
=\frac{6r_2(r_1+1)c}{k+1}x_1+\frac{6r_2 r_1 c}{k+1}x_2+\frac{6r_2 c}{k+1}x_3, \label{sieq_multi_add2}
\end{align}
\end{subequations}
which leads to 
\begin{equation}\label{sieq_condition_PCnoeffect_multi}
    b_{ii}-\sum_{j=1}^4 x_j (b_{ij}+b_{ji}+b_{jj})
    +2\sum_{j=1}^4 \sum_{l=1}^4 x_j x_l b_{jl}=0.
\end{equation}
This meets the general condition Eq.~(\ref{sieq_condition_PCnoeffect}) proposed by Supplementary Note~\ref{sec_PCequateWMandST} that pairwise comparison equates well-mixed and structured populations under weak selection. The additional effect brought by network structures is zero, and according to Eq.~(\ref{sieq_linear_PC}), the replicator equations reduce to $\dot{x}_i\propto x_i (\Bar{\pi}_i-\Bar{\pi})$, equivalent to the ones in well-mixed populations.

With this in mind, we only put the analysis in the context of structured populations here, using the methods within our framework.

Given the additional terms Eqs.~(\ref{sieq_multi_add1}) and (\ref{sieq_multi_add2}), we still need calculate the mean payoffs for different strategies in well-mixed populations, which are
\begin{subequations}
\begin{align}
\Bar{\pi}_1&
=\sum_{k_1+k_2+k_3+k_4=k}\frac{k!}{k_1! k_2! k_3! k_4!} {x_1}^{k_1}{x_2}^{k_2}{x_3}^{k_3}{x_4}^{k_4} a_{1|\mathbf{k}}
=\frac{r_2 (r_1+1)c}{k+1}kx_1+\frac{r_2 r_1 c}{k+1}kx_2+\frac{r_2 c}{k+1}kx_3+\frac{r_2 (r_1+1)c}{k+1}-c, \label{sieq_multi_avepi1}\\
\Bar{\pi}_2&
=\sum_{k_1+k_2+k_3+k_4=k}\frac{k!}{k_1! k_2! k_3! k_4!} {x_1}^{k_1}{x_2}^{k_2}{x_3}^{k_3}{x_4}^{k_4} a_{2|\mathbf{k}}
=\frac{r_2 (r_1+1)c}{k+1}kx_1+\frac{r_2 r_1 c}{k+1}kx_2+\frac{r_2 c}{k+1}kx_3+\frac{r_2 r_1 c}{k+1}-c, \\
\Bar{\pi}_3&
=\sum_{k_1+k_2+k_3+k_4=k}\frac{k!}{k_1! k_2! k_3! k_4!} {x_1}^{k_1}{x_2}^{k_2}{x_3}^{k_3}{x_4}^{k_4} a_{3|\mathbf{k}}
=\frac{r_2 (r_1+1)c}{k+1}kx_1+\frac{r_2 r_1 c}{k+1}kx_2+\frac{r_2 c}{k+1}kx_3+\frac{r_2 c}{k+1}-c, \\
\Bar{\pi}_4&
=\sum_{k_1+k_2+k_3+k_4=k}\frac{k!}{k_1! k_2! k_3! k_4!} {x_1}^{k_1}{x_2}^{k_2}{x_3}^{k_3}{x_4}^{k_4} a_{4|\mathbf{k}}
=\frac{r_2 (r_1+1)c}{k+1}kx_1+\frac{r_2 r_1 c}{k+1}kx_2+\frac{r_2 c}{k+1}kx_3. \label{sieq_multi_avepi4}
\end{align}
\end{subequations}
The mean payoff of the total population is then calculated by 
\begin{equation}\label{sieq_multi_avepi}
    \Bar{\pi}
=x_1\Bar{\pi}_1+x_2\Bar{\pi}_2+x_3\Bar{\pi}_3+x_4\Bar{\pi}_4
=(k+1)\left(\frac{r_2 (r_1+1)c}{k+1}x_1+\frac{r_2 r_1 c}{k+1}x_2+\frac{r_2 c}{k+1}x_3\right)-(2x_1+x_2+x_3)c.
\end{equation}

Inserting Eqs.~(\ref{sieq_multi_add1}), (\ref{sieq_multi_add2}), (\ref{sieq_multi_avepi1})--(\ref{sieq_multi_avepi4}), and (\ref{sieq_multi_avepi}) into Eq.~(\ref{sieq_linear_PC}), we obtain the replicator equations for multi-stage public goods game in structured populations:
\begin{subequations}
\begin{align}
\dot{x}_1&
=\frac{\delta(k-2)(k+1)}{2(k-1)} x_1\left[(1-x_1)\left(\frac{r_2 (r_1+1)c}{k+1}-2c\right)-x_2\left(\frac{r_2 r_1 c}{k+1}-c\right)-x_3\left(\frac{r_2 c}{k+1}-c\right)\right], \label{sieq_multi_dx1}\\
\dot{x}_2&
=\frac{\delta(k-2)(k+1)}{2(k-1)} x_2\left[-x_1\left(\frac{r_2 (r_1+1)c}{k+1}-2c\right)+(1-x_2)\left(\frac{r_2 r_1 c}{k+1}-c\right)-x_3\left(\frac{r_2 c}{k+1}-c\right)\right], \\
\dot{x}_3&
=\frac{\delta(k-2)(k+1)}{2(k-1)} x_3\left[-x_1\left(\frac{r_2 (r_1+1)c}{k+1}-2c\right)-x_2\left(\frac{r_2 r_1 c}{k+1}-c\right)+(1-x_3)\left(\frac{r_2 c}{k+1}-c\right)\right], \label{sieq_multi_dx3}\\
\dot{x}_4&=-\dot{x}_1-\dot{x}_2-\dot{x}_3.
\end{align}
\end{subequations}

We denote the system state $\mathbf{x}=(x_1,x_2,x_3,x_4)$. Solving $\dot{\mathbf{x}}=\mathbf{0}$, we obtain four equilibrium points, denoted by
$\mathbf{x}^{(CC)}=(1,0,0,0)$, $\mathbf{x}^{(CD)}=(0,1,0,0)$, $\mathbf{x}^{(DC)}=(0,0,1,0)$, and $\mathbf{x}^{(DD)}=(0,0,0,1)$. To analyze their stability, we study the Jacobian matrix of the system composed by Eqs.~(\ref{sieq_multi_dx1})--(\ref{sieq_multi_dx3}) (cancel $\dot{x}_4=-\dot{x}_1-\dot{x}_2-\dot{x}_3$), 
\begin{align}\label{sieq_J_multi_PC}
    J&=\begin{pmatrix}
    \displaystyle{\frac{\partial \dot{x}_1}{\partial x_1}} & 
    \displaystyle{\frac{\partial \dot{x}_1}{\partial x_2}} & 
    \displaystyle{\frac{\partial \dot{x}_1}{\partial x_3}}
    \\[1em]
    \displaystyle{\frac{\partial \dot{x}_2}{\partial x_1}} & 
    \displaystyle{\frac{\partial \dot{x}_2}{\partial x_2}} & 
    \displaystyle{\frac{\partial \dot{x}_2}{\partial x_3}}
    \\[1em]
    \displaystyle{\frac{\partial \dot{x}_3}{\partial x_1}} & 
    \displaystyle{\frac{\partial \dot{x}_3}{\partial x_2}} & 
    \displaystyle{\frac{\partial \dot{x}_3}{\partial x_3}}
    \end{pmatrix} \nonumber\\
    &=\frac{\delta(k-2)(k+1)}{2(k-1)}
    \begin{pmatrix}
    \displaystyle{J_{11}} & 
    \displaystyle{-x_1\left(\frac{r_2 r_1 c}{k+1}-c\right)} & 
    \displaystyle{-x_1\left(\frac{r_2 c}{k+1}-c\right)}
    \\[1em]
    \displaystyle{-x_2\left(\frac{r_2 (r_1+1)c}{k+1}-2c\right)} & 
    \displaystyle{J_{22}} & 
    \displaystyle{-x_2\left(\frac{r_2 c}{k+1}-c\right)}
    \\[1em]
    \displaystyle{-x_3\left(\frac{r_2 (r_1+1)c}{k+1}-2c\right)} & 
    \displaystyle{-x_3\left(\frac{r_2 r_1 c}{k+1}-c\right)} & 
    \displaystyle{J_{33}}
    \end{pmatrix}, 
\end{align}
where 
\begin{subequations}
    \begin{align}
        J_{11}&=(1-2x_1)\left(\frac{r_2 (r_1+1)c}{k+1}-2c\right)-x_2\left(\frac{r_2 r_1 c}{k+1}-c\right)-x_3\left(\frac{r_2 c}{k+1}-c\right), \\
        J_{22}&=-x_1\left(\frac{r_2 (r_1+1)c}{k+1}-2c\right)+(1-2x_2)\left(\frac{r_2 r_1 c}{k+1}-c\right)-x_3\left(\frac{r_2 c}{k+1}-c\right), \\
        J_{33}&=-x_1\left(\frac{r_2 (r_1+1)c}{k+1}-2c\right)-x_2\left(\frac{r_2 r_1 c}{k+1}-c\right)+(1-2x_3)\left(\frac{r_2 c}{k+1}-c\right).
    \end{align}
\end{subequations}

Substituting the value of $\mathbf{x}^{(CC)}$ into Eq.~(\ref{sieq_J_multi_PC}), we have
\begin{equation}
    \left.J\right|_{\mathbf{x}=\mathbf{x}^{(CC)}}
    =\frac{\delta(k-2)(k+1)}{2(k-1)}
    \begin{pmatrix}
    \displaystyle{-\frac{r_2 (r_1+1)c}{k+1}+2c} & 
    \displaystyle{-\frac{r_2 r_1 c}{k+1}+c} & 
    \displaystyle{-\frac{r_2 c}{k+1}+c}
    \\[1em]
    \displaystyle{0} & 
    \displaystyle{-\frac{r_2 c}{k+1}+c} & 
    \displaystyle{0}
    \\[1em]
    \displaystyle{0} & 
    \displaystyle{0} & 
    \displaystyle{-\frac{r_2 r_1 c}{k+1}+c}
    \end{pmatrix}.
\end{equation}
The conditions ensuring $\left.J\right|_{\mathbf{x}=\mathbf{x}^{(CC)}}$ negative-definite are $r_1 r_2>2(k+1)$, $r_2>k+1$, and $r_1 r_2>k+1$. The last one covers the first one. Therefore, the equilibrium point $\mathbf{x}^{(CC)}$ is stable if and only if $r_1 r_2>k+1$, $r_2>k+1$.

Substituting the value of $\mathbf{x}^{(CD)}$ into Eq.~(\ref{sieq_J_multi_PC}), we have
\begin{equation}
    \left.J\right|_{\mathbf{x}=\mathbf{x}^{(CD)}}
    =\frac{\delta(k-2)(k+1)}{2(k-1)}
    \begin{pmatrix}
    \displaystyle{\frac{r_2 c}{k+1}-c} & 
    \displaystyle{0} & 
    \displaystyle{0}
    \\[1em]
    \displaystyle{-\frac{r_2 (r_1+1)c}{k+1}+2c} & 
    \displaystyle{-\frac{r_2 r_1 c}{k+1}+c} & 
    \displaystyle{-\frac{r_2 c}{k+1}+c}
    \\[1em]
    \displaystyle{0} & 
    \displaystyle{0} & 
    \displaystyle{-\frac{r_2 (r_1-1)c}{k+1}}
    \end{pmatrix}.
\end{equation}
The conditions ensuring $\left.J\right|_{\mathbf{x}=\mathbf{x}^{(CD)}}$ negative-definite are $r_2<k+1$, $r_1 r_2>k+1$, and $r_1>1$. If the first two conditions hold, then the last one is naturally valid. Therefore, $\mathbf{x}^{(CD)}$ is stable if and only if $r_1 r_2>k+1$, $r_2<k+1$.

Substituting the value of $\mathbf{x}^{(DC)}$ into Eq.~(\ref{sieq_J_multi_PC}), we have
\begin{equation}\label{sieq_J_multi_PC_JDC}
    \left.J\right|_{\mathbf{x}=\mathbf{x}^{(DC)}}
    =\frac{\delta(k-2)(k+1)}{2(k-1)}
    \begin{pmatrix}
    \displaystyle{\frac{r_2 r_1 c}{k+1}-c} & 
    \displaystyle{0} & 
    \displaystyle{0}
    \\[1em]
    \displaystyle{0} & 
    \displaystyle{\frac{r_2 (r_1-1)c}{k+1}-c} & 
    \displaystyle{0}
    \\[1em]
    \displaystyle{-\frac{r_2 (r_1+1)c}{k+1}+2c} & 
    \displaystyle{-\frac{r_2 r_1 c}{k+1}+c} & 
    \displaystyle{-\frac{r_2 c}{k+1}+c}
    \end{pmatrix}.
\end{equation}
The conditions ensuring $\left.J\right|_{\mathbf{x}=\mathbf{x}^{(DC)}}$ negative-definite are $r_1 r_2<k+1$, $r_1<1$, and $r_2>k+1$. If the first and third conditions hold, then the second one is naturally valid. Therefore, $\mathbf{x}^{(DC)}$ is stable if and only if $r_1 r_2<k+1$, $r_2>k+1$.

Substituting the value of $\mathbf{x}^{(DD)}$ into Eq.~(\ref{sieq_J_multi_PC}), we have
\begin{equation}
    \left.J\right|_{\mathbf{x}=\mathbf{x}^{(DD)}}
    =\frac{\delta(k-2)(k+1)}{2(k-1)}
    \begin{pmatrix}
    \displaystyle{\frac{r_2 (r_1+1)c}{k+1}-2c} & 
    \displaystyle{0} & 
    \displaystyle{0}
    \\[1em]
    \displaystyle{0} & 
    \displaystyle{\frac{r_2 r_1 c}{k+1}-c} & 
    \displaystyle{0}
    \\[1em]
    \displaystyle{0} & 
    \displaystyle{0} & 
    \displaystyle{\frac{r_2 c}{k+1}-c}
    \end{pmatrix}.
\end{equation}
The conditions ensuring $\left.J\right|_{\mathbf{x}=\mathbf{x}^{(DD)}}$ negative-definite are $r_2 (r_1+1)<2(k+1)$, $r_1 r_2<k+1$, and $r_2<k+1$. If the second and third conditions hold, then the first one is naturally valid. Therefore, $\mathbf{x}^{(DD)}$ is stable if and only if $r_1 r_2<k+1$, $r_2<k+1$.

\subsubsection{Discussion}
\begin{figure}
	\centering
		\includegraphics[width=.9\textwidth]{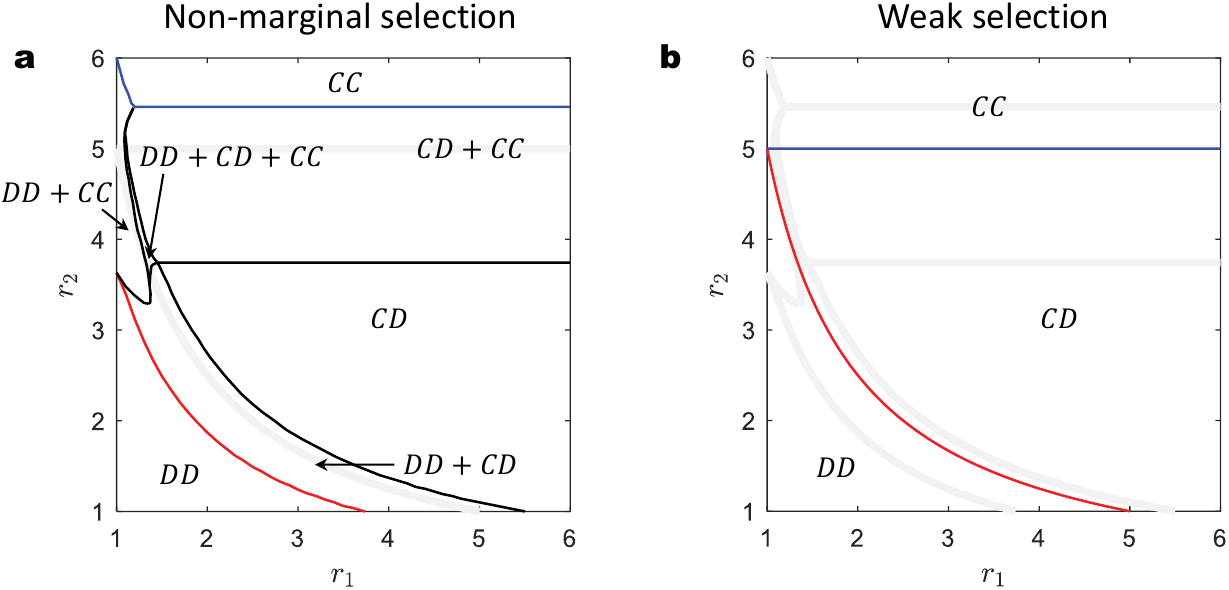}
	\caption{\textbf{Phase diagrams of the system behavior in multi-stage public goods games are qualitatively similar under non-marginal and weak selection strength.} 
    \textbf{a} (the data are in agreement with those published in Fig.~2 from ref.~\cite{szolnoki2022tactical}), Numerical simulations under non-marginal selection ($\delta=2$). The phases are defined directly by their name. For example, $CC$ represents a phase where only $CC$ exists, and  $CD+CC$ represents a phase where $CC$ and $CD$ coexist.  
    \textbf{b}, The phase diagram is divided by analytical $r_1 r_2=k+1$ (red) and $r_2=k+1$ (blue) under weak selection ($\delta\to 0^+$). Here, $r_1 r_2=k+1$ divides the $DD$ and $CD$ phases, while $r_2=k+1$ separates the $CD$ and $CC$ phases. \textbf{Other parameters:} $c=1$, $k=4$.} 
	\label{fig_phase_multi}
\end{figure}

To compare our analytical results under weak selection and the numerical ones from previous work under non-marginal selection~\cite{szolnoki2022tactical}, we refer to their phase diagrams on the $r_1$-$r_2$ plane (Fig.~\ref{fig_phase_multi}). Under a non-marginal selection strength (Fig.~\ref{fig_phase_multi}\textbf{a}), there are three pure phases as $r_1$ and $r_2$ increase, from $DD$, through $CD$, to $CC$. Between these pure phases, there are mixed phases, such as the $DD+CD$ phase (between $DD$ and $CC$) and the $CD+CC$ phase (between $CD$ and $CC$). However, in the weak selection limit, these mixed phases do not exist (Fig.~\ref{fig_phase_multi}\textbf{b}). On the presented phase diagrams, there are only three pure phases. This is similar to the phenomenon in the traditional public goods game, where only $C$ and $D$ phases exist under weak selection (Supplementary Note~\ref{sec_PGG_st}) and the mixed phase $C+D$ only exists under non-marginal selection~\cite{perc_jrsi13}. In the extended multi-stage public goods game, the predicted phase boundaries under weak selection (Fig.~\ref{fig_phase_multi}\textbf{b}) qualitatively match the ones under non-marginal selection (Fig.~\ref{fig_phase_multi}\textbf{a}), illustrating the utility of our analytical framework.

Our analytical framework also predicts a $DC$ phase, which exists when $r_1 r_2<k+1$ and $r_2>k+1$ (see Eq.~(\ref{sieq_J_multi_PC_JDC})). This region is outside the upper left corner of Fig.~\ref{fig_phase_multi}\textbf{b} and is invisible in the parameter space we present. In this region, $r_1<1$, which is not meaningful in the sense of public goods games. Therefore, neither the previous numerical work nor our analytical framework shows the region where the $DC$ phase exists.

\section{: Death-birth}\label{sec_DB}
Here, we supplement the corresponding results under the death-birth rule.

In a unit time, a random focal individual $A$ is selected to die, and $A$'s neighbors compete for the vacant position proportional to their fitness. The neighbor $B$ reproduces its strategy to the position of $A$ with the probability as follows:
\begin{equation}
    W=\frac{F_B}{\sum_k{F_X}},
\end{equation}
where $\sum_k{F_X}$ represents the total fitness of $A$'s neighbors. If individual $B$ has a higher fitness among $A$'s neighbors, then $B$ has a higher probability to reproduce its strategy to the position of $A$. It should be noted that the focal individual's fitness is completely ignored under death-birth.

Below, we analyze the dynamics of the death-birth rule rigorously.

\subsection{The increase of $i$-players}
The increase of $i$-players happens when a focal $j$-player ($j\neq i$) is selected to update its strategy and an $i$-player takes the position. Given the focal $j$-player's neighbor configuration $\mathbf{k}$, the probability that an $i$-player takes the $j$-player's position is
\begin{equation}\label{sieq_jgetDB}
    \mathcal{P}(j\gets i)=\frac{k_i F_{i|j}^\mathbf{k}}{\sum_{i'=1}^n k_{i'} F_{i'|j}^\mathbf{k}}
    =\frac{k_i}{k}+\frac{k_i}{k}\left(
    \pi_{i|j}^\mathbf{k}-\sum_{i'=1}^n \frac{k_{i'}}{k}\pi_{i'|j}^\mathbf{k}
    \right) \delta +\mathcal{O}(\delta^2).
\end{equation}
Then, we apply it to all possibilities for $j\neq i$ and neighbor configurations $\mathbf{k}$, obtaining the probability that the number of $i$-players increases by 1 during a unit time step,
\begin{align}\label{sieq_i+}
    \mathcal{P}\left(\Delta x_i=\frac{1}{N}\right)
    &=\sum_{j=1,j\neq i}^n x_j
    \sum_{\sum_{i'=1}^n k_{i'}=k}\frac{k!}{\prod_{i'=1}^n k_{i'}!} \left(\prod_{i'=1}^n {q_{i'|j}}^{k_{i'}}\right)
    \mathcal{P}(j\gets i) \nonumber\\
    &=\sum_{j=1,j\neq i}^n x_j q_{i|j}
    +\sum_{j=1,j\neq i}^n x_j 
    \sum_{\sum_{i'=1}^n k_{i'}=k}\frac{k!}{\prod_{i'=1}^n k_{i'}!} \left(\prod_{i'=1}^n {q_{i'|j}}^{k_{i'}}\right)
    \frac{k_i}{k}\left(
    \pi_{i|j}^\mathbf{k}-\sum_{i'=1}^n \frac{k_{i'}}{k}\pi_{i'|j}^\mathbf{k}
    \right) \delta +\mathcal{O}(\delta^2).
\end{align}

\subsection{The decrease of $i$-players}
The decrease of $i$-players happens when a focal $i$-player is selected to update its strategy and the player who takes the position is not an $i$-player. Unlike pairwise comparison, where the focal individual has a probability of keeping its own strategy, $\sum_{j=1}^n \mathcal{P}(i\gets j)\neq 1$, here, under death-birth, the focal individual must adopt the strategy of a neighbor, $\sum_{j=1}^n \mathcal{P}(i\gets j)=1$. Therefore, given the focal $i$-player's neighbor configuration $\mathbf{k}$, the probability that the player who takes the position is not an $i$-player can be written as
\begin{equation}\label{sieq_igetDB}
    1-\mathcal{P}(i\gets i)=1-\frac{k_i F_{i|i}^\mathbf{k}}{\sum_{i'=1}^n k_{i'} F_{i'|i}^\mathbf{k}}
    =\frac{k-k_i}{k}-\frac{k_i}{k}\left(
    \pi_{i|i}^\mathbf{k}-\sum_{i'=1}^n \frac{k_{i'}}{k}\pi_{i'|i}^\mathbf{k}
    \right) \delta +\mathcal{O}(\delta^2).
\end{equation}
Applying it to all possibilities for the neighbor configuration $\mathbf{k}$ after selecting a focal $i$-player with probability $x_i$, we obtain the probability that the number of $i$-players decreases by 1 during a unit time step,
\begin{align}\label{sieq_i-}
    \mathcal{P}\left(\Delta x_i=-\frac{1}{N}\right)
    &=x_i
    \sum_{\sum_{i'=1}^n k_{i'}=k}\frac{k!}{\prod_{i'=1}^n k_{i'}!} \left(\prod_{i'=1}^n {q_{i'|i}}^{k_{i'}}\right)
    \left[1-\mathcal{P}(i\gets i)\right] \nonumber\\
    &=x_i (1-q_{i|i})
    -x_i 
    \sum_{\sum_{i'=1}^n k_{i'}=k}\frac{k!}{\prod_{i'=1}^n k_{i'}!} \left(\prod_{i'=1}^n {q_{i'|i}}^{k_{i'}}\right)
    \frac{k_i}{k}\left(
    \pi_{i|i}^\mathbf{k}-\sum_{i'=1}^n \frac{k_{i'}}{k}\pi_{i'|i}^\mathbf{k}
    \right) \delta +\mathcal{O}(\delta^2).
\end{align}

\subsection{The replicator equation}\label{sec_repliactor_DB}
The instant change in the proportion $x_i$ of $i$-players consists of the increase and decrease of $i$-players. Applying Eqs.~(\ref{sieq_i+}) and (\ref{sieq_i-}), and considering that a full Monte Carlo step contains $N$ elementary steps, we have
\begin{align}\label{sieq_xi}
    \dot{x}_i&=N\times \left\{\frac{1}{N}\mathcal{P}\left(\Delta x_i=\frac{1}{N}\right)
    +\left(-\frac{1}{N}\right)\mathcal{P}\left(\Delta x_i=-\frac{1}{N}\right)\right\} \nonumber \\
    &=\sum_{j=1}^n x_j \sum_{\sum_{i'=1}^n k_{i'}=k}\frac{k!}{\prod_{i'=1}^n k_{i'}!} \left(\prod_{i'=1}^n {q_{{i'}|j}}^{k_{i'}}\right) \frac{k_i}{k}\left(
    \pi_{i|j}^\mathbf{k}-\sum_{i'=1}^n \frac{k_{i'}}{k}\pi_{i'|j}^\mathbf{k}
    \right)\delta +\mathcal{O}(\delta^2).
\end{align}
In Eq.~(\ref{sieq_xi}), the $\delta^0$ term has been eliminated: applying Eq.~(\ref{sieq_constraint}), we know $x_j q_{i|j}=p_{ij}=x_i q_{j|i}$, such that the $\delta^0$ term in Eq.~(\ref{sieq_i+}) can be expressed as $\sum_{j=1,j\neq i}^n x_j q_{i|j}=\sum_{j=1,j\neq i}^n x_i q_{j|i}$; similarly, we know $1-q_{i|i}=\sum_{j=1,j\neq i}^n q_{j|i}$ by Eqs.~(\ref{sieq_constraint}), which rewrites the $\delta^0$ term in Eq.~(\ref{sieq_i-}) as $x_i (1-q_{i|i})=x_i \sum_{j=1,j\neq i}^n q_{j|i}$, equal to the one in Eq.~(\ref{sieq_i+}).

The $\delta^0$ term being eliminated, the instant change in $x_i$ happens on the order of $\delta^1$. Meanwhile, the instant change in $q_{i|j}$ for $i,j=1,2,\dots,n$ happens on the order of $\delta^0$ since the $\delta^0$ is non-zero (see Supplementary Note~\ref{sec_edgeDB}). That is, the change in $q_{i|j}$ is much faster than $x_i$, so that $x_i$ changes on the basis of $q_{i|j}$ achieving equilibrium. According to Supplementary Note~\ref{sec_edgeDB}, we have the following solution when $q_{i|j}$ achieves stability.
\begin{equation}\label{sieq_qij}
    q_{i|j}=\begin{cases}
    \displaystyle{\frac{k-2}{k-1}x_i}, & j\neq i,\\
    \\
    \displaystyle{\frac{k-2}{k-1}x_i+\frac{1}{k-1}}, & j=i.
    \end{cases}
\end{equation}
Eq.~(\ref{sieq_qij}) is independently derived under death-birth but appear the same as the ones under pairwise comparison.

The primitive replicator equation of $x_i$ under death-birth is Eq.~(\ref{sieq_xi}),
where $\mathcal{O}(\delta^2)=0$, $\pi_j^\mathbf{k}$ and $\pi_i^\mathbf{k}$ are given by Eq.~(\ref{sieq_pi_j}), $\pi_{i|j}^\mathbf{k}$ and $\pi_{j|i}^\mathbf{k}$ are given by Eq.~(\ref{sieq_pi_i|j}), and $q_{{i'}|j}$ is given by Eq.~(\ref{sieq_qij}). The degrees of freedom of the replicator dynamics system are $n-1$, represented by independent variables $x_i$ ($i=1,2,\dots,n$, cancel one of them by $\sum_i^n x_i=1$).

\subsection{Simplification and discussion}
Next, we utilize the useful Theorem~\ref{1} to simplify the replicator equation under death-birth and discuss it.

\subsubsection{Decomposition to accumulated payoff}
First, we decompose the general replicator equations by keeping expected accumulated payoffs. Using Theorem~\ref{1}, we can calculate Eq.~(\ref{sieq_xi}) as follows.
\begin{align}\label{sieq_pilevel1_DB}
    \dot{x}_i
    =&~\delta\sum_{j=1}^n x_j \sum_{\sum_{i'=1}^n k_{i'}=k}\frac{k!}{\prod_{i'=1}^n k_{i'}!} \left(\prod_{i'=1}^n {q_{{i'}|j}}^{k_{i'}}\right) \frac{k_i}{k}\left(
    \pi_{i|j}^\mathbf{k}-\sum_{i'=1}^n \frac{k_{i'}}{k}\pi_{i'|j}^\mathbf{k}
    \right) 
    \nonumber\\
    =&~\delta\sum_{j=1}^n x_j q_{i|j} \sum_{\sum_{i'=1}^n k_{i'}=k-1}\frac{(k-1)!}{\prod_{i'=1}^n k_{i'}!} \left(\prod_{i'=1}^n {q_{{i'}|j}}^{k_{i'}}\right)
    \pi_{i|j}^{\mathbf{k}_{+i}} \nonumber\\
    &~ -\delta \sum_{j=1}^n x_j q_{i|j} \sum_{\sum_{i'=1}^n k_{i'}=k-1}\frac{(k-1)!}{\prod_{i'=1}^n k_{i'}!} \left(\prod_{i'=1}^n {q_{{i'}|j}}^{k_{i'}}\right)\left(\frac{k_i+1}{k}
    \pi_{i|j}^{\mathbf{k}_{+i}}
    +\sum_{i'=1,i'\neq i}^n \frac{k_{i'}}{k}\pi_{i'|j}^{\mathbf{k}_{+i}}
    \right) 
    \nonumber\\
    =&~\delta\sum_{j=1}^n x_j q_{i|j} \sum_{\sum_{i'=1}^n k_{i'}=k-1}\frac{(k-1)!}{\prod_{i'=1}^n k_{i'}!} \left(\prod_{i'=1}^n {q_{{i'}|j}}^{k_{i'}}\right)
    \pi_{i|j}^{\mathbf{k}_{+i}} \nonumber\\
    &~ -\delta \sum_{j=1}^n x_j q_{i|j} \sum_{\sum_{i'=1}^n k_{i'}=k-1}\frac{(k-1)!}{\prod_{i'=1}^n k_{i'}!} \left(\prod_{i'=1}^n {q_{{i'}|j}}^{k_{i'}}\right)\left(\frac{1}{k}
    \pi_{i|j}^{\mathbf{k}_{+i}}
    +\sum_{i'=1}^n \frac{k_{i'}}{k}\pi_{i'|j}^{\mathbf{k}_{+i}}
    \right) 
    \nonumber\\
    =&~\frac{\delta(k-1)}{k}x_i \sum_{j=1}^n q_{j|i} \sum_{\sum_{i'=1}^n k_{i'}=k-1}\frac{(k-1)!}{\prod_{i'=1}^n k_{i'}!} \left(\prod_{i'=1}^n {q_{{i'}|j}}^{k_{i'}}\right)
    \pi_{i|j}^{\mathbf{k}_{+i}} \nonumber\\
    &~-\frac{\delta(k-1)}{k}x_i \sum_{j=1}^n q_{j|i} \sum_{\sum_{i'=1}^n k_{i'}=k-2}\frac{(k-2)!}{\prod_{i'=1}^n k_{i'}!} \left(\prod_{i'=1}^n {q_{{i'}|j}}^{k_{i'}}\right)
    \sum_{i'=1}^n q_{i'|j}\pi_{i'|j}^{\mathbf{k}_{+i,+i'}}.
\end{align}

Using the aforementioned notations for $\langle \pi_{i|j}^{\mathbf{k}_{+i}} \rangle$ in Eq.~(\ref{sieq_<pi_i|j>}), as well as its extension for $\langle \pi_{i'|j}^{\mathbf{k}_{+i,+i'}} \rangle$,
\begin{equation}
    \langle \pi_{i'|j}^{\mathbf{k}_{+i,+i'}} \rangle
    =\sum_{\sum_{i'=1}^n k_{i'}=k-2}\frac{(k-2)!}{\prod_{i'=1}^n k_{i'}!} \left(\prod_{i'=1}^n {q_{{i'}|j}}^{k_{i'}}\right)
    \pi_{i'|j}^{\mathbf{k}_{+i,+i'}},
\end{equation}
we can write Eq.~(\ref{sieq_pilevel1_DB}) as 
\begin{align}\label{sieq_pilevel2_DB}
    \dot{x}_i=&~\frac{\delta(k-1)}{k} x_i\sum_{j=1}^n q_{j|i}\left(
    \langle \pi_{i|j}^{\mathbf{k}_{+i}}\rangle
    -\sum_{i'=1}^n q_{i'|j}
    \langle \pi_{i'|j}^{\mathbf{k}_{+i,+i'}}\rangle
    \right) \nonumber\\
    =&~\frac{\delta(k-1)}{k} x_i\left(
    \langle \pi_i^\mathbf{k}\rangle
    -\sum_{j=1}^n \sum_{i'=1}^n q_{i'|j} q_{j|i}
    \langle \pi_{i'|j}^{\mathbf{k}_{+i,+i'}}\rangle
    \right).
\end{align}
In Eq.~(\ref{sieq_pilevel2_DB}), we have used the relation $\sum_{j=1}^n q_{j|i}\langle \pi_{i|j}^{\mathbf{k}_{+i}} \rangle
=\langle \pi_i^{\mathbf{k}} \rangle$ as suggested by Theorem~\ref{theorem_sumpiij_pii}.

Again, the replicator equation expressed by expected accumulated payoffs given by Eq.~(\ref{sieq_pilevel2_DB}) bears an intuitive understanding if we use the following concepts:
\begin{itemize}
    \item $\pi_i^{(0)}=\langle \pi_i^\mathbf{k} \rangle$, the expected accumulated payoff of the $i$-player itself (zero-step away on the graph).
    \item $\pi_i^{(2)}=\sum_{j=1}^n \sum_{i'=1}^n q_{i'|j} q_{j|i} \langle \pi_{i'|j}^{\mathbf{k}_{+i,+i'}}\rangle$, the expected accumulated payoff of the $i$-player's second-order neighbors (two-step away on the graph). Intuitively, $q_{i'|j} q_{j|i}$ is used to find two-step away $i'$-players away from the $i$-player. The first step walks to a $j$-player, whose neighbor configuration is $\mathbf{k}_{+i,+i'}$. The second step walks to an $i'$-player.
\end{itemize}

Using these concepts, we know that $\dot{x}_i\propto x_i(\pi_i^{(0)}-\pi_i^{(2)})$. Under death-birth, the reproduction rate of $i$-players depends on how much their expected accumulated payoff higher than the one of neighbors' neighbors. The essence of replicator dynamics $\dot{x}_i$ is the competition between oneself and its second-order neighbors. This is also consistent with the result obtained by the identity-by-descent idea~\cite{allen2014games,allen2017evolutionary}, and here we further generalize it to $n$-strategy multiplayer systems.

\subsubsection{Decomposition to single-game payoff}\label{sec_DB_single}
Next, we further divide the replicator equation into expected single-game payoffs, stressing the convenience for actual calculation.

Similarly, we use the notation of $\langle a_{i|\mathbf{k}}\rangle_j$ defined in Eq.~(\ref{sieq_<a>}) to represent the payoff in a single game. We know the expression of $\langle \pi_i^\mathbf{k} \rangle$ from Eq.~(\ref{sieq_<pi_i>_<a>}), which is derived from the payoff calculation stage, independent of strategy update rules. Furthermore, we expand $\langle \pi_{i'|j}^{\mathbf{k}_{+i,+i'}} \rangle$ by Eq.~(\ref{sieq_pi_i|j}) and obtain
\begin{align}\label{sieq_<pi_i|j>_<a>}
    \langle \pi_{i'|j}^{\mathbf{k}_{+i,+i'}} \rangle
    =&\sum_{\sum_{i'=1}^n k_{i'}=k-2}\frac{(k-2)!}{\prod_{i'=1}^n k_{i'}!} \left(\prod_{i'=1}^n {q_{{i'}|j}}^{k_{i'}}\right)
    \Bigg\{a_{i'|\mathbf{k}_{+i,+j}}+\sum_{\sum_{l=1}^n k'_l=k-1}\frac{(k-1)!}{\prod_{l=1}^n k'_l!} \left(\prod_{l=1}^n {q_{l|i'}}^{k'_l}\right) \Bigg[
    a_{i'|\mathbf{k}'_{+j}}
    \nonumber\\
    &+\sum_{l=1}^n k'_l \sum_{\sum_{\ell=1}^n k''_\ell=k-1}\frac{(k-1)!}{\prod_{\ell=1}^n k''_\ell!} \left(\prod_{\ell=1}^n {q_{\ell|l}}^{k''_\ell}\right) a_{i'|\mathbf{k}''_{+l}}
    \Bigg]\Bigg\} \nonumber\\
    =&\sum_{\sum_{i'=1}^n k_{i'}=k-2}\frac{(k-2)!}{\prod_{i'=1}^n k_{i'}!} \left(\prod_{i'=1}^n {q_{{i'}|j}}^{k_{i'}}\right)
    a_{i'|\mathbf{k}_{+i,+j}} 
    +\sum_{\sum_{l=1}^n k'_l=k-1}\frac{(k-1)!}{\prod_{l=1}^n k'_l!} \left(\prod_{l=1}^n {q_{l|i'}}^{k'_l}\right)
    a_{i'|\mathbf{k}'_{+j}} \nonumber\\
    &+(k-1)\sum_{l=1}^n q_{l|i'} \sum_{\sum_{\ell=1}^n k''_\ell=k-1}\frac{(k-1)!}{\prod_{\ell=1}^n k''_\ell!} \left(\prod_{\ell=1}^n {q_{\ell|l}}^{k''_\ell}\right) a_{i'|\mathbf{k}''_{+l}}
    \nonumber\\
    =&~\langle a_{i'|\mathbf{k}_{+i,+j}}\rangle_j+
    \langle a_{i'|\mathbf{k}_{+j}}\rangle_{i'}+
    (k-1)\sum_{l=1}^n q_{l|i'}\langle a_{i'|\mathbf{k}_{+l}}\rangle_l.
\end{align}

Applying Eqs.~(\ref{sieq_<pi_i>_<a>}) and (\ref{sieq_<pi_i|j>_<a>}) to Eq.~(\ref{sieq_pilevel2_DB}), we have
\begin{equation}\label{sieq_alevel1_DB}
    \dot{x}_i=\frac{\delta(k-1)}{k} x_i \left[\left(
    \langle a_{i|\mathbf{k}}\rangle_i
    +k\sum_{j=1}^n q_{j|i}\langle a_{i|\mathbf{k}_{+j}}\rangle_j \right)
    -\sum_{j=1}^n \sum_{i'=1}^n q_{i'|j} q_{j|i}\left(
    \langle a_{i'|\mathbf{k}_{+j}}\rangle_{i'}
    +\langle a_{i'|\mathbf{k}_{+i,+j}}\rangle_j
    +(k-1)\sum_{l=1}^n q_{l|i'}
    \langle a_{i'|\mathbf{k}_{+l}}\rangle_l
    \right)\right].
\end{equation}
To decrease the number of elements, we transform $\langle a_{i|\mathbf{k}}\rangle_i$ into $\sum_{j=1}^{n} q_{j|i} \langle a_{i|\mathbf{k}_{+j}}\rangle_i$ according to Theorem~\ref{theorem_aiki_qjiaiji} and rewrite Eq.~(\ref{sieq_alevel1_DB}) as 
\begin{equation}\label{sieq_alevel2_DB}
    \dot{x}_i=\frac{\delta(k-1)}{k} x_i\sum_{j=1}^n q_{j|i}\left(
    \langle a_{i|\mathbf{k}_{+j}}\rangle_i
    +k\langle a_{i|\mathbf{k}_{+j}}\rangle_j
    -\sum_{i'=1}^n q_{i'|j}\left(
    \langle a_{i'|\mathbf{k}_{+j}}\rangle_{i'}
    +\langle a_{i'|\mathbf{k}_{+i,+j}}\rangle_j
    +(k-1)\sum_{l=1}^n q_{l|i'}
    \langle a_{i'|\mathbf{k}_{+l}}\rangle_l
    \right)\right).
\end{equation}
We do not bother giving the general expression of Eq.~(\ref{sieq_alevel2_DB}) with all $q_{j|i}$ quantities transformed into $x_j$, since the results would be complicated and unruly. Please note that the calculation inside each $\langle \cdot\rangle$ also contains $q_{j|i}$ quantities and we always need to transform them manually in applications.

According to Eq.~(\ref{sieq_alevel2_DB}), we can attribute everything about $\langle \cdot\rangle$ into three types, the `$\langle a_{i|\mathbf{k}_{+j}}\rangle_i$ type', the `$\langle a_{i|\mathbf{k}_{+j}}\rangle_j$ type', and the `$\langle a_{i'|\mathbf{k}_{+i,+j}}\rangle_j$ type'. The $\langle a_{i|\mathbf{k}_{+j}}\rangle_i$ and $\langle a_{i|\mathbf{k}_{+j}}\rangle_j$ types have been listed as matrices $\left[\langle a_{i|\mathbf{k}_{+j}}\rangle_i\right]_{ij}$ and $\left[\langle a_{i|\mathbf{k}_{+j}}\rangle_j\right]_{ij}$ in Eqs.~(\ref{sieq_aiji_type}) and (\ref{sieq_aijj_type}), respectively. The number of elements to calculate manually in the $\langle a_{i|\mathbf{k}_{+j}}\rangle_i$ and $\langle a_{i|\mathbf{k}_{+j}}\rangle_j$ types is $(2n-1)n$ as we have discussed.

The $\langle a_{i'|\mathbf{k}_{+i,+j}}\rangle_j$ type, however, contains three dimensions. It can be expressed by totaling $i'$ matrices, each with dimensions $i$ and $j$:
\begin{itemize}
    \item The $\langle a_{i'|\mathbf{k}_{+i,+j}}\rangle_j$ type: 
    \begin{equation}
        \left[\langle a_{i'|\mathbf{k}_{+i,+j}}\rangle_j\right]_{ij}=
        \displaystyle{\begin{pmatrix}
        \langle a_{i'|\mathbf{k}_{+1,+1}}\rangle_1 & \langle a_{i'|\mathbf{k}_{+1,+2}}\rangle_2 & \cdots & \langle a_{i'|\mathbf{k}_{+1,+n}}\rangle_n \\
        \langle a_{i'|\mathbf{k}_{+2,+1}}\rangle_1 & \langle a_{i'|\mathbf{k}_{+2,+2}}\rangle_2 & \cdots & \langle a_{i'|\mathbf{k}_{+2,+n}}\rangle_n \\
        \vdots & \vdots & \ddots & \vdots \\
        \langle a_{i'|\mathbf{k}_{+n,+1}}\rangle_1 & \langle a_{i'|\mathbf{k}_{+n,+2}}\rangle_2 & \cdots & \langle a_{i'|\mathbf{k}_{+n,+n}}\rangle_n
    \end{pmatrix}},
    \end{equation}
    for $i'=1,2,\dots,n$.
\end{itemize}
There are $n^3$ elements in all $\left[\langle a_{i'|\mathbf{k}_{+i,+j}}\rangle_j\right]_{ij}$ matrices. Therefore, the total computation involves $n^3+(2n-1)n=(n^2+2n-1)n$ elements. The computational complexity is $\mathrm{O}(n^3)$, which is feasible within polynomial time.

\subsubsection{Special linear system}\label{sec_db_linear}
We can also further simplify the calculation when faced with special payoff structures. Given a co-player configuration $\mathbf{k}$, a linear payoff structure can be determined by the coefficient matrix $\mathbf{b}$ and the constant vector $\mathbf{c}$, that is, $a_{i|\mathbf{k}}=\sum_{l=1}^n b_{il}k_l+c_i$, as previously shown in Eqs.~(\ref{sieq_b_c})--(\ref{sieq_linear_a_PC}).

The expressions of $\langle a_{i|\mathbf{k}_{+j}}\rangle_i$ and $\langle a_{i|\mathbf{k}_{+j}}\rangle_j$ as functions of $\mathbf{b}$ and $\mathbf{c}$ has been obtained in Eqs.~(\ref{sieq_aiji_bc}) and (\ref{sieq_aijj_bc}), respectively. We further calculate the elements of the $\langle a_{i'|\mathbf{k}_{+i,+j}}\rangle_j$ type,
\begin{equation}\label{sieq_ai'ij_bc}
    \langle a_{i'|\mathbf{k}_{+i,+j}}\rangle_j
    =(k-2)\sum_{l=1}^{n} b_{i'l}q_{l|j}+b_{i'i}+b_{i'j}+c_{i'}
    =\frac{(k-2)^2}{k-1}\sum_{l=1}^{n} b_{i'l}x_l+\frac{2k-3}{k-1}b_{i'j}+b_{i'i}+c_{i'}.
\end{equation}

We substitute Eqs.~(\ref{sieq_aiji_bc}), (\ref{sieq_aijj_bc}) and (\ref{sieq_ai'ij_bc}) into Eq.~(\ref{sieq_alevel2_DB}). Through a long but feasible calculation (transforming all $q_{j|i}$ quantities to $x_j$ quantities and organizing the result), we obtain
\begin{align}\label{sieq_linear_DB}
    \dot{x}_i=&~\frac{\delta(k-2)}{k(k-1)^2}x_i \Bigg\{
    \frac{(k^2-2)^2}{k}(\Bar{\pi}_i-\Bar{\pi})
    +\frac{3k^2-4}{k}\left((kb_{ii}+c_i)-\sum_{j=1}^n x_j(kb_{jj}+c_j) \right) \nonumber\\
    &~-(k^2+2k-4)\sum_{j=1}^n x_j\left(b_{ji}-\sum_{l=1}^n x_l b_{jl} \right) \Bigg\},
\end{align}
where $\Bar{\pi}_i=k\sum_{l=1}^n x_l b_{il}+c_i$ is the mean payoff of $i$-players in a well-mixed population, and $\Bar{\pi}=\sum_{i=1}^n x_i\Bar{\pi}_i$ is the mean payoff of all individuals in a well-mixed population.

We know that the replicator equation in a well-mixed population is $\dot{x}_i=x_i(\Bar{\pi}_i-\Bar{\pi})$. In this way, Eq.~(\ref{sieq_linear_DB}) clearly showed the additional terms brought by death-birth in a structured population compared to the well-mixed population.

\subsection{Comparison with two-strategy dynamics}\label{sec_DBn=2}
Let us discuss on how our replicator equations under death-birth for $n$-strategy systems reduce to the 2-strategy system. Similarly, the configuration $\mathbf{k}=(k_1,k_2)=(k_1,k-k_1)$ can be represented by only $k_1$, where $k_1=0,1,\dots,k$.

The $n$-strategy system simplified to expected accumulated payoffs shown in Eq.~(\ref{sieq_pilevel2_DB}) is for comparison with the identity-by-decent idea by Allen and Nowak~\cite{allen2014games}, that the essence of death-birth is the competition from the individuals from two-step away. However, in this section, we want to compare with the 2-strategy system for general multiplayer games in structured populations, which was studied by Li {\it et al.}~\cite{li2016evolutionary}. In this way, we need to reorganize the results in Eq.~(\ref{sieq_pilevel2_DB}) first.

According to Theorem~\ref{5}, we have $\langle \pi_i^{\mathbf{k}} \rangle=\sum_{j=1}^n \sum_{i'=1}^n q_{i'|j} q_{j|i}\langle \pi_{i|j}^{\mathbf{k}_{+i,+i'}}\rangle$. Substituting this into Eq.~(\ref{sieq_pilevel2_DB}), we have
\begin{equation}
    \dot{x}_i
    =\frac{\delta(k-1)}{k} x_i
    \sum_{j=1}^n \sum_{i'=1}^n q_{i'|j} q_{j|i}
    \left(\langle \pi_{i|j}^{\mathbf{k}_{+i,+i'}}\rangle
    -\langle \pi_{i'|j}^{\mathbf{k}_{+i,+i'}}\rangle
    \right).
\end{equation}
Further replacing all $q_{j|i}$ quantities with $x_j$ quantities by Eq.~(\ref{sieq_qij}) and organizing the results, we have
\begin{equation}\label{sieq_DB_n}
    \dot{x}_i=\frac{\delta(k-2)}{k(k-1)}x_i \sum_{j=1}^{n} x_j
    \left[
    \left(\langle \pi_{i|j}^{\mathbf{k}_{+i,+j}} \rangle
    -\langle \pi_{j|j}^{\mathbf{k}_{+i,+j}} \rangle \right)
    +\left(\langle \pi_{i|i}^{\mathbf{k}_{+i,+j}} \rangle
    -\langle \pi_{j|i}^{\mathbf{k}_{+i,+j}} \rangle \right)+(k-2)\sum_{i'=1}^{n} x_{i'} \left(\langle \pi_{i|j}^{\mathbf{k}_{+i,+i'}} \rangle
    -\langle \pi_{i'|j}^{\mathbf{k}_{+i,+i'}} \rangle \right)\right],
\end{equation}
which can be reduced to $2$-strategy system depicted by $\dot{x}_1$ with the consideration of $x_2=1-x_1$,
\begin{align}\label{sieq_DB_n2}
    \dot{x}_1=&~\frac{\delta(k-2)}{k(k-1)}x_1(1-x_1) \Big\{
    k\left(\langle \pi_{1|2}^{\mathbf{k}_{+1,+2}} \rangle
    -\langle \pi_{2|2}^{\mathbf{k}_{+1,+2}} \rangle \right)
    \nonumber\\
    &+[(k-2)x_1+1]\left[
    \left(\langle \pi_{1|1}^{\mathbf{k}_{+1,+2}} \rangle
    -\langle \pi_{2|1}^{\mathbf{k}_{+1,+2}} \rangle \right)
    -\left(\langle \pi_{1|2}^{\mathbf{k}_{+1,+2}} \rangle
    -\langle \pi_{2|2}^{\mathbf{k}_{+1,+2}} \rangle \right) \right]\Big\}.
\end{align}

Eq.~(\ref{sieq_DB_n2}), which describes general 2-strategy multiplayer games in structured populations under the death-birth rule, has the same form as the one given by Li {\it et al.}~\cite{li2016evolutionary}. As a supplement to their work, here we stress a detail as indicated in Eq.~(\ref{sieq_DB_n2}): the calculation should be made in the remaining $k-2$ individuals given the existence of at least a pair of 1-player and 2-player.

In particular, let us list the quantities in Eq.~(\ref{sieq_DB_n2}) and clarify the calculation. Referring to Eq.~(\ref{sieq_<pi_i|j>}), we have
\begin{subequations}\label{<pi_ij>n2forLi}
    \begin{align}
        \langle \pi_{1|1}^{\mathbf{k}_{+1,+2}} \rangle&=
        \sum_{k_1=0}^{k-2}\frac{(k-2)!}{k_1!(k-k_1-2)!}{q_{1|1}}^{k_1}{q_{2|1}}^{k-k_1-2} \pi_{1|1}^{\mathbf{k}_{+1,+2}}, \\
        \langle \pi_{2|1}^{\mathbf{k}_{+1,+2}} \rangle&=
        \sum_{k_1=0}^{k-2}\frac{(k-2)!}{k_1!(k-k_1-2)!}{q_{1|1}}^{k_1}{q_{2|1}}^{k-k_1-2} \pi_{2|1}^{\mathbf{k}_{+1,+2}}, \\
        \langle \pi_{1|2}^{\mathbf{k}_{+1,+2}} \rangle&=
        \sum_{k_1=0}^{k-2}\frac{(k-2)!}{k_1!(k-k_1-2)!}{q_{1|2}}^{k_1}{q_{2|2}}^{k-k_1-2} \pi_{1|2}^{\mathbf{k}_{+1,+2}}, \\
        \langle \pi_{2|2}^{\mathbf{k}_{+1,+2}} \rangle&=
        \sum_{k_1=0}^{k-2}\frac{(k-2)!}{k_1!(k-k_1-2)!}{q_{1|2}}^{k_1}{q_{2|2}}^{k-k_1-2} \pi_{2|2}^{\mathbf{k}_{+1,+2}},
    \end{align}
\end{subequations}
where, according to Eq.~(\ref{sieq_pi_i|j}), we have
\begin{subequations}\label{pi_ijn2forLi}
    \begin{align}
        \pi_{1|1}^{\mathbf{k}_{+1,+2}}=
        &~a_{1|\mathbf{k}_{+1,+2}}+
        \sum_{k'_1=0}^{k-1}\frac{(k-1)!}{k'_1!(k-k'_1-1)!}
        {q_{1|1}}^{k'_1}{q_{2|1}}^{k-k'_1-1}
        \Bigg(a_{1|\mathbf{k}'_{+1}}+
        k'_1 \sum_{k''_1=0}^{k-1}\frac{(k-1)!}{k''_1!(k-k''_1-1)!}
        {q_{1|1}}^{k''_1}{q_{2|1}}^{k-k''_1-1}
        a_{1|\mathbf{k}''_{+1}} \nonumber\\
        &+(k-k'_1-1) \sum_{k''_1=0}^{k-1}\frac{(k-1)!}{k''_1!(k-k''_1-1)!}
        {q_{1|2}}^{k''_1}{q_{2|2}}^{k-k''_1-1}
        a_{1|\mathbf{k}''_{+2}}
        \Bigg), \\
        \pi_{2|1}^{\mathbf{k}_{+1,+2}}=
        &~a_{2|\mathbf{k}_{+1,+1}}+
        \sum_{k'_1=0}^{k-1}\frac{(k-1)!}{k'_1!(k-k'_1-1)!}
        {q_{1|2}}^{k'_1}{q_{2|2}}^{k-k'_1-1}
        \Bigg(a_{2|\mathbf{k}'_{+1}}+
        k'_1 \sum_{k''_1=0}^{k-1}\frac{(k-1)!}{k''_1!(k-k''_1-1)!}
        {q_{1|1}}^{k''_1}{q_{2|1}}^{k-k''_1-1}
        a_{2|\mathbf{k}''_{+1}} \nonumber\\
        &+(k-k'_1-1) \sum_{k''_1=0}^{k-1}\frac{(k-1)!}{k''_1!(k-k''_1-1)!}
        {q_{1|2}}^{k''_1}{q_{2|2}}^{k-k''_1-1}
        a_{2|\mathbf{k}''_{+2}}
        \Bigg), \\
        \pi_{1|2}^{\mathbf{k}_{+1,+2}}=
        &~a_{1|\mathbf{k}_{+2,+2}}+
        \sum_{k'_1=0}^{k-1}\frac{(k-1)!}{k'_1!(k-k'_1-1)!}
        {q_{1|1}}^{k'_1}{q_{2|1}}^{k-k'_1-1}
        \Bigg(a_{1|\mathbf{k}'_{+2}}+
        k'_1 \sum_{k''_1=0}^{k-1}\frac{(k-1)!}{k''_1!(k-k''_1-1)!}
        {q_{1|1}}^{k''_1}{q_{2|1}}^{k-k''_1-1}
        a_{1|\mathbf{k}''_{+1}} \nonumber\\
        &+(k-k'_1-1) \sum_{k''_1=0}^{k-1}\frac{(k-1)!}{k''_1!(k-k''_1-1)!}
        {q_{1|2}}^{k''_1}{q_{2|2}}^{k-k''_1-1}
        a_{1|\mathbf{k}''_{+2}}
        \Bigg), \\
        \pi_{2|2}^{\mathbf{k}_{+1,+2}}=
        &~a_{2|\mathbf{k}_{+1,+2}}+
        \sum_{k'_1=0}^{k-1}\frac{(k-1)!}{k'_1!(k-k'_1-1)!}
        {q_{1|2}}^{k'_1}{q_{2|2}}^{k-k'_1-1}
        \Bigg(a_{2|\mathbf{k}'_{+2}}+
        k'_1 \sum_{k''_1=0}^{k-1}\frac{(k-1)!}{k''_1!(k-k''_1-1)!}
        {q_{1|1}}^{k''_1}{q_{2|1}}^{k-k''_1-1}
        a_{2|\mathbf{k}''_{+1}} \nonumber\\
        &+(k-k'_1-1) \sum_{k''_1=0}^{k-1}\frac{(k-1)!}{k''_1!(k-k''_1-1)!}
        {q_{1|2}}^{k''_1}{q_{2|2}}^{k-k''_1-1}
        a_{2|\mathbf{k}''_{+2}}
        \Bigg).
    \end{align}
\end{subequations}
In this way, the calculation of general 2-strategy multiplayer games given by Eq.~(\ref{sieq_DB_n2}) is completed.

Similarly, after presenting the 2-strategy case by expected accumulated payoffs, we can further compute the results, which are divided into expected single-game payoffs. Under death-birth, however, the results would seem unruly and not easier than Eq.~(\ref{sieq_DB_n2}). The 2-strategy case keeping expected single-game payoffs can be obtained by applying $n=2$ to Eq.~(\ref{sieq_alevel2_DB}), which leads to an over-complicated result. In conclusion, we always need to treat the payoff structure of different models case by case, and using Eq.~(\ref{sieq_DB_n2}) is an acceptable choice that we have.

In the case of special linear systems, however, we can refer to a well-organized general equation. Applying $n=2$ to Eq.~(\ref{sieq_linear_DB}), we have
\begin{align}\label{sieq_DB_linear_n2}
    \dot{x}_1=&~\frac{\delta(k-2)}{k(k-1)^2} \Bigg\{
    \frac{(k^2-2)^2}{k}x_1(\Bar{\pi}_i-\Bar{\pi})
    +\frac{3k^2-4}{k}x_1(1-x_1)\left[k(b_{11}-b_{22})+c_1-c_2\right] \nonumber\\
    &-(k^2+2k-4)x_1(1-x_1)\left[(b_{11}-b_{12})x_1+(b_{21}-b_{22})(1-x_1)\right] \Bigg\},
\end{align}
where $b_{ij}$ and $c_i$ are given by Eq.~(\ref{sieq_b_c}) to express the linear payoff structure $a_{i|\mathbf{k}}=\sum_{l=1}^n b_{il}k_l+c_i$, and $x_1(\Bar{\pi}_1-\Bar{\pi})$ corresponds to the replicator dynamics in a well-mixed population, which can be obtained by existing knowledge of well-mixed populations.

\section{: Edge dynamics}\label{sec_appen_edge}
As we know from Eq.~(\ref{sieq_constraint}) in Supplementary Note~\ref{sec_system}, the system can be described by $x_i$ and $q_{j|i}$. While we study the dynamics of $x_i$ in previous sections, the dynamics of $q_{j|i}$ is further supplemented here. In particular, we show how to obtain the relation between $x_i$ and $q_{j|i}$ as shown in Eqs.~(\ref{sieq_qij_PC}) and (\ref{sieq_qij}), for both pairwise comparison and death-birth. The same results have been declared in previous literature~\cite{ohtsuki2006replicator}, but the details of the calculation process were unclear. 

\subsection{Pairwise comparison}\label{sec_edgePC}
First, we study edge dynamics under pairwise comparison to explain the origin of Eq.~(\ref{sieq_qij_PC}) presented in Supplementary Note~\ref{sec_repliactor_PC}.

\subsubsection{The increase of $il$-edges}
The number of $il$-edges in the system increases by $k_l$ when a $j$-player is replaced by an $i$-player ($j\neq i$)---the new $i$-player connects to the previous $j$-player's neighbors $l$, thus increasing the number of $il$-edges by $k_l$. Since the total number of edges in the system is $kN/2$, the frequency of $il$-edges, $p_{il}$, is increased by $k_l/(kN/2)=2k_l/(kN)$. For a possible neighbor configuration $\mathbf{k}$ of the $j$-player, we go through all $j\neq i$, obtaining the probability that the configuration $\mathbf{k}$ is found around the $j$-player and thus the frequency of $il$-edges increases by $2k_l/(kN)$:
\begin{equation}
    \mathcal{P}\left(\Delta p_{il}=\frac{2k_l}{kN}\right)
    =\sum_{j=1,j\neq i}^{n} x_j
    \frac{k!}{\prod_{i'=1}^n k_{i'}!} \left(\prod_{i'=1}^n {q_{i'|j}}^{k_{i'}}\right)
    \mathcal{P}(j\gets i).
\end{equation}
Then, we go through all possible configurations $\mathbf{k}$. We put this step late here because the change in the frequency of $il$-edges, $2k_l/(kN)$, varies in different $\mathbf{k}$. As we see, the expected change in the frequency of $il$-edges is calculated as:
\begin{align}\label{sieq_PC_edgeincrease}
    \sum_{\sum_{i'=1}^n k_{i'}=k} \frac{2k_l}{kN}
    \mathcal{P}\left(\Delta p_{il}=\frac{2k_l}{kN}\right)
    &=\sum_{\sum_{i'=1}^n k_{i'}=k} \frac{2k_l}{kN}
    \sum_{j=1,j\neq i}^{n} x_j
    \frac{k!}{\prod_{i'=1}^n k_{i'}!} \left(\prod_{i'=1}^n {q_{i'|j}}^{k_{i'}}\right)
    \mathcal{P}(j\gets i) \nonumber\\
    &=\sum_{\sum_{i'=1}^n k_{i'}=k} \frac{2k_l}{kN}
    \sum_{j=1,j\neq i}^{n} x_j
    \frac{k!}{\prod_{i'=1}^n k_{i'}!} \left(\prod_{i'=1}^n {q_{i'|j}}^{k_{i'}}\right)
    \left(\frac{k_i}{2k}+\mathcal{O}(\delta)\right) \nonumber\\
    &=\begin{cases} 
    \displaystyle{
    \sum_{j=1,j\neq i}^{n} x_j
    \frac{q_{i|j}[(k-1)q_{i|j}+1]}{kN} +\mathcal{O}(\delta)
    },  & \mbox{$i=l$,} \\
    \displaystyle{
    \sum_{j=1,j\neq i}^{n} x_j
    \frac{(k-1)q_{l|j}q_{i|j}}{kN} +\mathcal{O}(\delta)
    }, & \mbox{$i\neq l$.}
    \end{cases} \nonumber\\
    &=\frac{x_i}{kN}
    \begin{cases} 
    \displaystyle{
    \sum_{j=1,j\neq i}^{n}
    q_{j|i}[(k-1)q_{i|j}+1] +\mathcal{O}(\delta)
    },  & \mbox{$i=l$,} \\
    \displaystyle{
    \sum_{j=1,j\neq i}^{n}
    (k-1)q_{l|j}q_{j|i} +\mathcal{O}(\delta)
    }, & \mbox{$i\neq l$.}
    \end{cases}
\end{align}
In Eq.~(\ref{sieq_PC_edgeincrease}), $\mathcal{P}(j\gets i)$ refers to Eq.~(\ref{sieq_jgetPC}). We only take the first-order Taylor expansion, because the first-order term would be non-zero as we will see in Eq.~(\ref{sieq_PC_edge}), thus not necessary to study the second-order term.

\subsubsection{The decrease of $il$-edges}
The number of $il$-edges in the system decreases by $k_l$ when an $i$-player is replaced by a $j$-player ($j\neq i$). The died $i$-player disconnects from the previous neighbors $l$, thus decreasing the number of $il$-edges by $k_l$. In other words, the frequency of $il$-edges $p_{il}$ is decreased by $k_l/(kN/2)=2k_l/(kN)$. For a possible neighbor configuration $\mathbf{k}$ of the $i$-player, we go through all possible $j\neq i$, obtaining the probability that the configuration $\mathbf{k}$ is found around the $i$-player and thus the frequency of $il$-edges decreases by $2k_l/(kN)$:
\begin{equation}
    \mathcal{P}\left(\Delta p_{il}=-\frac{2k_l}{kN}\right)
    =x_i\sum_{j=1,j\neq i}^{n}
    \frac{k!}{\prod_{i'=1}^n k_{i'}!} \left(\prod_{i'=1}^n {q_{i'|i}}^{k_{i'}}\right)
    \mathcal{P}(i\gets j).
\end{equation}
Then, we go through all possible configurations $\mathbf{k}$. The expected change in the frequency of $il$-edges is calculated as:
\begin{align}\label{sieq_PC_edgedecrease}
    \sum_{\sum_{i'=1}^n k_{i'}=k} \left(-\frac{2k_l}{kN}\right)
    \mathcal{P}\left(\Delta p_{il}=-\frac{2k_l}{kN}\right)
    &=\sum_{\sum_{i'=1}^n k_{i'}=k} \left(-\frac{2k_l}{kN}\right)
    x_i\sum_{j=1,j\neq i}^{n}
    \frac{k!}{\prod_{i'=1}^n k_{i'}!} \left(\prod_{i'=1}^n {q_{i'|i}}^{k_{i'}}\right)
    \mathcal{P}(i\gets j) \nonumber\\
    &=\sum_{\sum_{i'=1}^n k_{i'}=k} \left(-\frac{2k_l}{kN}\right)
    x_i\sum_{j=1,j\neq i}^{n}
    \frac{k!}{\prod_{i'=1}^n k_{i'}!} \left(\prod_{i'=1}^n {q_{i'|i}}^{k_{i'}}\right)
    \left(\frac{k_j}{2k}+\mathcal{O}(\delta)\right) \nonumber\\
    &=\sum_{\sum_{i'=1}^n k_{i'}=k} \left(-\frac{2k_l}{kN}\right)
    x_i
    \frac{k!}{\prod_{i'=1}^n k_{i'}!} \left(\prod_{i'=1}^n {q_{i'|i}}^{k_{i'}}\right)
    \left(\frac{k-k_i}{2k}+\mathcal{O}(\delta)\right) \nonumber\\
    &=\begin{cases} 
    \displaystyle{
    x_i \frac{q_{i|i}}{N}-
    x_i \frac{q_{i|i}[(k-1)q_{i|i}+1]}{kN} +\mathcal{O}(\delta)
    },  & \mbox{$i=l$,} \\
    \displaystyle{
    x_i \frac{q_{l|i}}{N}-
    x_i \frac{(k-1)q_{l|j}q_{i|j}}{kN} +\mathcal{O}(\delta)
    }, & \mbox{$i\neq l$.}
    \end{cases} \nonumber\\
    &=\frac{x_i}{kN}
    \begin{cases} 
    \displaystyle{
    kq_{i|i}-
    q_{i|i}[(k-1)q_{i|i}+1] +\mathcal{O}(\delta)
    },  & \mbox{$i=l$,} \\
    \displaystyle{
    kq_{l|i}-
    (k-1)q_{l|i}q_{i|i} +\mathcal{O}(\delta)
    }, & \mbox{$i\neq l$.}
    \end{cases}
\end{align}
In Eq.~(\ref{sieq_PC_edgedecrease}), $\mathcal{P}(i\gets j)$ refers to Eq.~(\ref{sieq_igetPC}). Again, we only take the first-order Taylor expansion, because the first-order term would be non-zero as we will see in Eq.~(\ref{sieq_PC_edge}), thus not necessary to study the second-order term.

\subsubsection{Separation of different time scales}
We combine the increase and decrease of $il$-edges given by Eqs.~(\ref{sieq_PC_edgeincrease}) and (\ref{sieq_PC_edgedecrease}), and consider that a full MC step contains $N$ elementary steps. In this way, we have the master equation of the change in the $il$-edge frequency:
\begin{align}\label{sieq_PC_edge}
    \dot{p}_{il}
    &=N\times \left(\sum_{\sum_{i'=1}^n k_{i'}=k} \frac{2k_l}{kN}
    \mathcal{P}\left(\Delta p_{il}=\frac{2k_l}{kN}\right)
    +\sum_{\sum_{i'=1}^n k_{i'}=k} \left(-\frac{2k_l}{kN}\right)
    \mathcal{P}\left(\Delta p_{il}=-\frac{2k_l}{kN}\right)
    \right)\nonumber\\
    &=\frac{x_i}{k}
    \begin{cases} 
    \displaystyle{
    \sum_{j=1,j\neq i}^{n}
    q_{j|i}[(k-1)q_{i|j}+1]
    -kq_{i|i}
    +q_{i|i}[(k-1)q_{i|i}+1] +\mathcal{O}(\delta)
    },  & \mbox{$i=l$,} \\
    \displaystyle{
    \sum_{j=1,j\neq i}^{n}
    (k-1)q_{l|j}q_{j|i}
    -kq_{l|i}
    +(k-1)q_{l|i}q_{i|i} +\mathcal{O}(\delta)
    }, & \mbox{$i\neq l$.}
    \end{cases} \nonumber\\
    &=\frac{x_i}{k}
    \begin{cases} 
    \displaystyle{
    \sum_{j=1}^{n} q_{j|i}
    +(k-1)\sum_{j=1}^{n} q_{l|j}q_{j|i}
    -kq_{l|i} +\mathcal{O}(\delta)
    },  & \mbox{$i=l$,} \\
    \displaystyle{
    (k-1)\sum_{j=1}^{n} q_{l|j}q_{j|i}
    -kq_{l|i} +\mathcal{O}(\delta)
    }, & \mbox{$i\neq l$.}
    \end{cases} \nonumber\\
    &=\frac{x_i}{k}\left(
    \theta_{il}+(k-1)\sum_{j=1}^n q_{l|j}q_{j|i}-kq_{l|i}
    \right)+\mathcal{O}(\delta),
\end{align}
where
\begin{equation}
    \theta_{il}
    =\begin{cases} 
    \displaystyle{1},  & \mbox{$i=l$,} \\
    \displaystyle{0}, & \mbox{$i\neq l$.}
    \end{cases}
\end{equation}

Previously, we conclude from Eq.~(\ref{sieq_constraint}) in Supplementary Note~\ref{sec_system} that the system can be expressed by $x_i$ and $q_{j|i}$, eliminating the need to use $p_{ij}$. Therefore, we consider $q_{l|i}=p_{il}/x_i$ and write $\dot{q}_{l|i}$ by $\dot{p}_{il}$ as follows.
\begin{equation}\label{sieq_qli_PC}
    \dot{q}_{l|i}
    =\frac{\mathrm{d}}{\mathrm{d}t}\left(\frac{p_{il}}{x_i}\right)
    =\frac{\dot{p}_{il}x_i-\dot{x}_i p_{il}}{{x_i}^2}
    =\frac{\dot{p}_{il}}{x_i}
    =\frac{1}{k}\left(
    \theta_{il}+(k-1)\sum_{j=1}^n q_{l|j}q_{j|i}-kq_{l|i}
    \right)+\mathcal{O}(\delta).
\end{equation}
In Eq.~(\ref{sieq_qli_PC}), we only take the first-order Taylor expansion, as the first-order term is non-zero and one can ignore the remaining higher order infinitesimals. It is noted that we have taken $\dot{x}_i=0$ in the calculation, because as seen in Eq.~(\ref{sieq_replicator_PC}), $\dot{x}_i$ is zero in its $\delta^0$ term and only becomes non-zero since the $\delta^1$ term, which is the ignored higher order infinitesimals when calculating $\dot{q}_{l|i}$.

Here, an important characteristic arises: the time scales of the changes in $\dot{x}_i$ and $\dot{q}_{l|i}$ are different. Comparing Eqs.~(\ref{sieq_replicator_PC}) and (\ref{sieq_qli_PC}), we observe that the change in $x_i$ happens at the $\delta^1$ scale, while the change in $q_{l|i}$ happens at the $\delta^0$ scale. In other words, the change of $q_{l|i}$ is much faster than in $x_i$.

Therefore, we can study the dynamics of $x_i$ on the basis of $q_{l|i}$ achieving equilibrium (rapidly). That is, we can solve $\dot{q}_{l|i}=0$, which is 
\begin{equation}\label{sieq_edge_eq_PC}
    \theta_{il}+(k-1)\sum_{j=1}^n q_{l|j}q_{j|i}-kq_{l|i}=0, 
\end{equation}
and represent all $q_{l|i}$ quantities by $x_i$ quantities. Considering $x_i q_{l|i}=x_l q_{i|l}$, the solution of Eq.~(\ref{sieq_edge_eq_PC}) is
\begin{equation}
    q_{l|i}=\frac{(k-2)x_l+\theta_{il}}{k-1},
\end{equation}
which is Eq.~(\ref{sieq_qij_PC}) presented in Supplementary Note~\ref{sec_repliactor_PC}, with the consideration of $\theta_{il}=1$ if $i=l$ and $\theta_{il}=0$ otherwise.

\subsection{Death-birth}\label{sec_edgeDB}
Similarly, we supplement edge dynamics under death-birth to explain the origin of Eq.~(\ref{sieq_qij}) presented in Supplementary Note~\ref{sec_repliactor_DB}.

\subsubsection{The increase of $il$-edges}
The number of $il$-edges in the system increases by $k_l$ when a $j$-player is replaced by an $i$-player ($j\neq i$). The frequency of $il$-edges, $p_{il}$, is increased by $2k_l/(kN)$. For a possible neighbor configuration $\mathbf{k}$ of the $j$-player, we go through all $j\neq i$, obtaining the probability that the configuration $\mathbf{k}$ is found around the $j$-player and thus the frequency of $il$-edges increases by $2k_l/(kN)$:
\begin{equation}
    \mathcal{P}\left(\Delta p_{il}=\frac{2k_l}{kN}\right)
    =\sum_{j=1,j\neq i}^{n} x_j
    \frac{k!}{\prod_{i'=1}^n k_{i'}!} \left(\prod_{i'=1}^n {q_{i'|j}}^{k_{i'}}\right)
    \mathcal{P}(j\gets i).
\end{equation}
Then, we go through all possible configurations $\mathbf{k}$. The expected change in the frequency of $il$-edges is calculated as:
\begin{align}\label{sieq_DB_edgeincrease}
    \sum_{\sum_{i'=1}^n k_{i'}=k} \frac{2k_l}{kN}
    \mathcal{P}\left(\Delta p_{il}=\frac{2k_l}{kN}\right)
    &=\sum_{\sum_{i'=1}^n k_{i'}=k} \frac{2k_l}{kN}
    \sum_{j=1,j\neq i}^{n} x_j
    \frac{k!}{\prod_{i'=1}^n k_{i'}!} \left(\prod_{i'=1}^n {q_{i'|j}}^{k_{i'}}\right)
    \mathcal{P}(j\gets i) \nonumber\\
    &=\sum_{\sum_{i'=1}^n k_{i'}=k} \frac{2k_l}{kN}
    \sum_{j=1,j\neq i}^{n} x_j
    \frac{k!}{\prod_{i'=1}^n k_{i'}!} \left(\prod_{i'=1}^n {q_{i'|j}}^{k_{i'}}\right)
    \left(\frac{k_i}{k}+\mathcal{O}(\delta)\right) \nonumber\\
    &=\frac{2x_i}{kN}
    \begin{cases} 
    \displaystyle{
    \sum_{j=1,j\neq i}^{n}
    q_{j|i}[(k-1)q_{i|j}+1] +\mathcal{O}(\delta)
    },  & \mbox{$i=l$,} \\
    \displaystyle{
    \sum_{j=1,j\neq i}^{n}
    (k-1)q_{l|j}q_{j|i} +\mathcal{O}(\delta)
    }, & \mbox{$i\neq l$.}
    \end{cases}
\end{align}
In Eq.~(\ref{sieq_DB_edgeincrease}), $\mathcal{P}(j\gets i)$ refers to Eq.~(\ref{sieq_jgetDB}). We only take the first-order Taylor expansion, because the first-order term would be non-zero as we will see in Eq.~(\ref{sieq_DB_edge}).

\subsubsection{The decrease of $il$-edges}
The number of $il$-edges in the system decreases by $k_l$ when an $i$-player is replaced by a $j$-player ($j\neq i$). The frequency of $il$-edges $p_{il}$ is decreased by $2k_l/(kN)$. For a possible neighbor configuration $\mathbf{k}$ of the $i$-player, we go through all possible $j\neq i$, obtaining the probability that the configuration $\mathbf{k}$ is found around the $i$-player and thus the frequency of $il$-edges decreases by $2k_l/(kN)$:
\begin{equation}
    \mathcal{P}\left(\Delta p_{il}=-\frac{2k_l}{kN}\right)
    =x_i
    \frac{k!}{\prod_{i'=1}^n k_{i'}!} \left(\prod_{i'=1}^n {q_{i'|i}}^{k_{i'}}\right)
    [1-\mathcal{P}(i\gets i)].
\end{equation}
Then, we go through all possible configurations $\mathbf{k}$. The expected change in the frequency of $il$-edges is calculated as:
\begin{align}\label{sieq_DB_edgedecrease}
    \sum_{\sum_{i'=1}^n k_{i'}=k} \left(-\frac{2k_l}{kN}\right)
    \mathcal{P}\left(\Delta p_{il}=-\frac{2k_l}{kN}\right)
    &=\sum_{\sum_{i'=1}^n k_{i'}=k} \left(-\frac{2k_l}{kN}\right)
    x_i
    \frac{k!}{\prod_{i'=1}^n k_{i'}!} \left(\prod_{i'=1}^n {q_{i'|i}}^{k_{i'}}\right)
    [1-\mathcal{P}(i\gets i)] \nonumber\\
    &=\sum_{\sum_{i'=1}^n k_{i'}=k} \left(-\frac{2k_l}{kN}\right)
    x_i
    \frac{k!}{\prod_{i'=1}^n k_{i'}!} \left(\prod_{i'=1}^n {q_{i'|i}}^{k_{i'}}\right)
    \left(1-\frac{k_i}{k}+\mathcal{O}(\delta)\right) \nonumber\\
    &=\frac{2x_i}{kN}
    \begin{cases} 
    \displaystyle{
    kq_{i|i}-
    q_{i|i}[(k-1)q_{i|i}+1] +\mathcal{O}(\delta)
    },  & \mbox{$i=l$,} \\
    \displaystyle{
    kq_{l|i}-
    (k-1)q_{l|i}q_{i|i} +\mathcal{O}(\delta)
    }, & \mbox{$i\neq l$.}
    \end{cases}
\end{align}
In Eq.~(\ref{sieq_DB_edgedecrease}), $[1-\mathcal{P}(i\gets i)]$ refers to Eq.~(\ref{sieq_igetDB}). Again, we only take the first-order Taylor expansion, because the first-order term would be non-zero as we will see in Eq.~(\ref{sieq_DB_edge}).

\subsubsection{Separation of different time scales}
We combine the increase and decrease of $il$-edges given by Eqs.~(\ref{sieq_DB_edgeincrease}) and (\ref{sieq_DB_edgedecrease}), and consider that a full MC step contains $N$ elementary steps. In this way, we have the master equation of the change in the $il$-edge frequency under death-birth:
\begin{align}\label{sieq_DB_edge}
    \dot{p}_{il}
    &=N\times \left(\sum_{\sum_{i'=1}^n k_{i'}=k} \frac{2k_l}{kN}
    \mathcal{P}\left(\Delta p_{il}=\frac{2k_l}{kN}\right)
    +\sum_{\sum_{i'=1}^n k_{i'}=k} \left(-\frac{2k_l}{kN}\right)
    \mathcal{P}\left(\Delta p_{il}=-\frac{2k_l}{kN}\right)
    \right)\nonumber\\
    &=\frac{2x_i}{k}\left(
    \theta_{il}+(k-1)\sum_{j=1}^n q_{l|j}q_{j|i}-kq_{l|i}
    \right)+\mathcal{O}(\delta),
\end{align}
where $\theta_{il}=1$ if $i=l$ and $\theta_{il}=0$ otherwise.

According to Eq.~(\ref{sieq_constraint}), we consider $q_{l|i}=p_{il}/x_i$ and write $\dot{q}_{l|i}$ by $\dot{p}_{il}$ as follows.
\begin{equation}\label{sieq_qli_DB}
    \dot{q}_{l|i}
    =\frac{\mathrm{d}}{\mathrm{d}t}\left(\frac{p_{il}}{x_i}\right)
    =\frac{2}{k}\left(
    \theta_{il}+(k-1)\sum_{j=1}^n q_{l|j}q_{j|i}-kq_{l|i}
    \right)+\mathcal{O}(\delta).
\end{equation}
In Eq.~(\ref{sieq_qli_DB}), we only take the first-order Taylor expansion, as the first-order term is non-zero and one can ignore the remaining higher order infinitesimals. Comparing Eqs.~(\ref{sieq_xi}) and (\ref{sieq_qli_DB}), we observe that the change in $x_i$ happens at the $\delta^1$ scale, while the change in $q_{l|i}$ happens at the $\delta^0$ scale. The change of $q_{l|i}$ are much faster than $x_i$.

Therefore, we can study the dynamics of $x_i$ on the basis of $q_{l|i}$ achieving equilibrium (rapidly). That is, we can solve $\dot{q}_{l|i}=0$, which is $\theta_{il}+(k-1)\sum_{j=1}^n q_{l|j}q_{j|i}-kq_{l|i}=0$, and represent all $q_{l|i}$ quantities by $x_i$ quantities. Considering $x_i q_{l|i}=x_l q_{i|l}$, the solution is
\begin{equation}
    q_{l|i}=\frac{(k-2)x_l+\theta_{il}}{k-1},
\end{equation}
which is Eq.~(\ref{sieq_qij}) presented in Supplementary Note~\ref{sec_repliactor_DB}, with the consideration of $\theta_{il}=1$ if $i=l$ and $\theta_{il}=0$ otherwise.

\section{: Theorem on computation and operators}\label{sec_theorem}
\newtheorem{theorem}{Theorem}
A summation relation central to the theoretical derivation in this work is given by Theorem~\ref{1}.
\begin{theorem}\label{1}
For any real number $z_j$, where $j=1,2,\dots,n$, satisfying $0\leq z_j\leq 1$ and $\sum_{j=1}^n z_j=1$, and any scalar multivariate function $g(\mathbf{k})$ of $\mathbf{k}$, we have the following relation:
\begin{equation}\label{sieq_simplify_theorem}
    \sum_{\sum_{j=1}^n k_j=k}
    \frac{k!}{\prod_{j=1}^n k_j!} \left(\prod_{j=1}^n {z_j}^{k_j}\right)
    k_i g(\mathbf{k})
    =
    kz_i
    \sum_{\sum_{j=1}^n k_j=k-1}
    \frac{(k-1)!}{\prod_{j=1}^n k_j!} \left(\prod_{j=1}^n {z_j}^{k_j}\right)
    g(\mathbf{k}_{+i}).
\end{equation}

\begin{proof}
    Recall the definitions given in the main text or \ref{sec_theory}: $\mathbf{k}=(k_1,k_2,\dots,k_n)$, where $\sum_{j=1}^n k_j =k$; and $\mathbf{k}_{+i}=(k_1,k_2,\dots,k_i+1,\dots,k_n)$, where $\sum_{j=1}^n k_j =k-1$. In this way, we have
    \begin{align}\label{sieq_simplify_theorem_proof}
        \sum_{\sum_{j=1}^n k_j=k}
        \frac{k!}{\prod_{j=1}^n k_j!} \left(\prod_{j=1}^n {z_j}^{k_j}\right)
        k_i g(\mathbf{k})
        &=kz_i\sum_{\sum_{j=1}^n k_j=k, k_i\neq 0}
        \frac{(k-1)!}{(k_i-1)!\prod_{j=1,j\neq i}^n k_j!} \left({z_i}^{k_i-1} \prod_{j=1,j\neq i}^n {z_j}^{k_j}\right)
        g(\mathbf{k}) \nonumber\\
        &=kz_i\sum_{\sum_{j=1}^n k_j=k-1}
        \frac{(k-1)!}{\prod_{j=1}^n k_j!} \left(\prod_{j=1}^n {z_j}^{k_j}\right)
        g(\mathbf{k}_{+i}).
    \end{align}

    The first step of Eq.~(\ref{sieq_simplify_theorem_proof}) uses $k!=k(k-1)!$, ${z_i}^{k_i}=z_i{z_i}^{k_i-1}$, $k_i/(k_i!)=1/[(k_i-1)!]$ and considers that the entire formula equals to zero when $k_i=0$. The second step of Eq.~(\ref{sieq_simplify_theorem_proof}) is a straightforward application of the $\mathbf{k}_{+i}$ concept. 
\end{proof}

\end{theorem}

Theorem~\ref{1} gives a general relation which helps us eliminate $k_i$ inside summations at the cost of transforming $g(\mathbf{k})$ into $g(\mathbf{k}_{+i})$. For convenience of our readers, there are several useful applications of Theorem~\ref{1}. 

For example, taking $g(\mathbf{k})=1$, we have
\begin{equation}
    \sum_{\sum_{j=1}^n k_j=k}
    \frac{k!}{\prod_{j=1}^n k_j!} \left(\prod_{j=1}^n {z_j}^{k_j}\right)
    k_i
    =
    kz_i.
\end{equation}

Another example is $g(\mathbf{k})=k_l$. If $l=i$, we have
\begin{equation}
    \sum_{\sum_{j=1}^n k_j=k}
    \frac{k!}{\prod_{j=1}^n k_j!} \left(\prod_{j=1}^n {z_j}^{k_j}\right)
    k_i k_l
    =
    kz_i
    \sum_{\sum_{j=1}^n k_j=k-1}
    \frac{(k-1)!}{\prod_{j=1}^n k_j!} \left(\prod_{j=1}^n {z_j}^{k_j}\right)
    (k_l+1)
    =
    kz_i[(k-1)z_l+1],
\end{equation}
and if $l\neq i$, it becomes
\begin{equation}
    \sum_{\sum_{j=1}^n k_j=k}
    \frac{k!}{\prod_{j=1}^n k_j!} \left(\prod_{j=1}^n {z_j}^{k_j}\right)
    k_i k_l
    =
    kz_i
    \sum_{\sum_{j=1}^n k_j=k-1}
    \frac{(k-1)!}{\prod_{j=1}^n k_j!} \left(\prod_{j=1}^n {z_j}^{k_j}\right)
    k_l
    =
    k(k-1)z_i z_l.
\end{equation}

Finally, a potentially misleading example is $g(\mathbf{k})=k'_l$, where $k'_l\in \mathbf{k}'\neq\mathbf{k}$ (recall that $\mathbf{k}'$ is an independent variable of $\mathbf{k}$). In this way, elements in $\mathbf{k}'$ are essentially constants when studying $g(\mathbf{k})$,
\begin{equation}
    \sum_{\sum_{j=1}^n k_j=k}
    \frac{k!}{\prod_{j=1}^n k_j!} \left(\prod_{j=1}^n {z_j}^{k_j}\right)
    k_i k'_l
    =
    kz_i k'_l.
\end{equation}

In this work, we have created many operators to convey various concepts. The main relations between these operators are summarized in Theorems~\ref{2}--\ref{5}.
\begin{theorem}\label{2}
    The following equation holds:
    \begin{equation}
        \langle \pi_{i|j}^{\mathbf{k}_{+i}} \rangle
        =\langle \pi_i^{\mathbf{k}_{+j}} \rangle.
    \end{equation}
    Intuitively speaking, these two concepts are equivalent: (1) The expected accumulated payoff of an $i$-player neighboring a $j$-player, where the $j$-player has at least one $i$-player neighbor. (2) The expected accumulated payoff of an $i$-player, where the $i$-player has at least one $j$-player neighbor. 
    \begin{proof}
        Let us expand $\pi_{i|j}^{\mathbf{k}_{+i}}$ according to Eq.~(\ref{sieq_pi_i|j}), 
    \begin{equation}\label{sieq_pi_i|j^+i}
        \pi_{i|j}^{\mathbf{k}_{+i}}=a_{i|\mathbf{k}_{+j}}
        +\sum_{\sum_{l=1}^n k'_l=k-1}\frac{(k-1)!}{\prod_{l=1}^n k'_l!} \left(\prod_{l=1}^n {q_{l|i}}^{k'_l}\right) 
        \left(a_{i|\mathbf{k}'_{+j}}
        +\sum_{l=1}^n k'_l \sum_{\sum_{\ell=1}^n k''_\ell=k-1}\frac{(k-1)!}{\prod_{\ell=1}^n k''_\ell!} \left(\prod_{\ell=1}^n {q_{\ell|l}}^{k''_\ell}\right) a_{i|\mathbf{k}''_{+l}} \right).
    \end{equation}
    By substituting Eq.~(\ref{sieq_pi_i|j^+i}) into Eq.~(\ref{sieq_<pi_i|j>}), we calculate
    \begin{align}\label{sieq_<pi_i|j^+i>}
        \langle \pi_{i|j}^{\mathbf{k}_{+i}} \rangle
        =&~\sum_{\sum_{i'=1}^n k_{i'}=k-1}\frac{(k-1)!}{\prod_{i'=1}^n k_{i'}!} \left(\prod_{i'=1}^n {q_{{i'}|j}}^{k_{i'}}\right)
        a_{i|\mathbf{k}_{+j}} \nonumber\\
        &~+\sum_{\sum_{i'=1}^n k_{i'}=k-1}\frac{(k-1)!}{\prod_{i'=1}^n k_{i'}!} \left(\prod_{i'=1}^n {q_{{i'}|j}}^{k_{i'}}\right)
        \sum_{\sum_{l=1}^n k'_l=k-1}\frac{(k-1)!}{\prod_{l=1}^n k'_l!} \left(\prod_{l=1}^n {q_{l|i}}^{k'_l}\right) 
        \Bigg[a_{i|\mathbf{k}'_{+j}} \nonumber\\
        &~+\sum_{l=1}^n k'_l \sum_{\sum_{\ell=1}^n k''_\ell=k-1}\frac{(k-1)!}{\prod_{\ell=1}^n k''_\ell!} \left(\prod_{\ell=1}^n {q_{\ell|l}}^{k''_\ell}\right) a_{i|\mathbf{k}''_{+l}} \Bigg] \nonumber\\
        =&~\sum_{\sum_{i'=1}^n k_{i'}=k-1}\frac{(k-1)!}{\prod_{i'=1}^n k_{i'}!} \left(\prod_{i'=1}^n {q_{{i'}|j}}^{k_{i'}}\right)
        a_{i|\mathbf{k}_{+j}}
        +\sum_{\sum_{l=1}^n k'_l=k-1}\frac{(k-1)!}{\prod_{l=1}^n k'_l!} \left(\prod_{l=1}^n {q_{l|i}}^{k'_l}\right) a_{i|\mathbf{k}'_{+j}} \nonumber\\
        &~+\sum_{\sum_{l=1}^n k'_l=k-1}\frac{(k-1)!}{\prod_{l=1}^n k'_l!} \left(\prod_{l=1}^n {q_{l|i}}^{k'_l}\right) 
        \sum_{l=1}^n k'_l \sum_{\sum_{\ell=1}^n k''_\ell=k-1}\frac{(k-1)!}{\prod_{\ell=1}^n k''_\ell!} \left(\prod_{\ell=1}^n {q_{\ell|l}}^{k''_\ell}\right) a_{i|\mathbf{k}''_{+l}}.
    \end{align}

    Similarly, we expand $\pi_{i}^{\mathbf{k}_{+j}}$ according to Eq.~(\ref{sieq_pi_j}), 
    \begin{equation}\label{sieq_pi_i^+j}
        \pi_i^{\mathbf{k}_{+j}}=a_{i|\mathbf{k}_{+j}}+
        \sum_{l=1}^n k_l \sum_{\sum_{\ell=1}^n k'_\ell=k-1}\frac{(k-1)!}{\prod_{\ell=1}^n k'_\ell!} \left(\prod_{\ell=1}^n {q_{\ell|l}}^{k'_\ell}\right) a_{i|\mathbf{k}'_{+l}}.
    \end{equation}
    By substituting Eq.~(\ref{sieq_pi_i^+j}) into Eq.~(\ref{sieq_<pi_j>}), we calculate
    \begin{align}\label{sieq_<pi_i^+j>}
        \langle \pi_i^{\mathbf{k}_{+j}} \rangle
        =&~\sum_{\sum_{i'=1}^n k_{i'}=k-1}\frac{(k-1)!}{\prod_{i'=1}^n k_{i'}!} \left(\prod_{i'=1}^n {q_{{i'}|i}}^{k_{i'}}\right)
        \pi_i^{\mathbf{k}_{+j}} \nonumber\\
        =&~\sum_{\sum_{i'=1}^n k_{i'}=k-1}\frac{(k-1)!}{\prod_{i'=1}^n k_{i'}!} \left(\prod_{i'=1}^n {q_{{i'}|i}}^{k_{i'}}\right)
        a_{i|\mathbf{k}_{+j}} 
        +
        \sum_{\sum_{\ell=1}^n k'_\ell=k-1}\frac{(k-1)!}{\prod_{\ell=1}^n k'_\ell!} \left(\prod_{\ell=1}^n {q_{\ell|j}}^{k'_\ell}\right) a_{i|\mathbf{k}'_{+j}} \nonumber\\
        &~+\sum_{\sum_{i'=1}^n k_{i'}=k-1}\frac{(k-1)!}{\prod_{i'=1}^n k_{i'}!} \left(\prod_{i'=1}^n {q_{{i'}|i}}^{k_{i'}}\right)
        \sum_{l=1}^n k_l \sum_{\sum_{\ell=1}^n k'_\ell=k-1}\frac{(k-1)!}{\prod_{\ell=1}^n k'_\ell!} \left(\prod_{\ell=1}^n {q_{\ell|l}}^{k'_\ell}\right) a_{i|\mathbf{k}'_{+l}}.
    \end{align}

    Comparing Eqs.~(\ref{sieq_<pi_i|j^+i>}) and (\ref{sieq_<pi_i^+j>}) and realizing the equivalence of auxiliary variables, we have $\langle \pi_{i|j}^{\mathbf{k}_{+i}} \rangle=\langle \pi_i^{\mathbf{k}_{+j}} \rangle$.
    \end{proof}
\end{theorem}

\begin{theorem}\label{theorem_sumpiij_pii}
    The following equation holds:
\begin{equation}\label{sieq_piij_to_pii}
    \sum_{j=1}^n q_{j|i}\langle \pi_{i|j}^{\mathbf{k}_{+i}} \rangle
    =\langle \pi_i^{\mathbf{k}} \rangle.
\end{equation}
    Intuitive interpretation: (1) the expected accumulated payoff of an $i$-player over all possible neighboring $j$-players found near an $i$-player, where the $j$-player has at least one $i$-player neighbor, is equivalent to (2) the expected accumulated payoff of an $i$-player.
\begin{proof}
    Further developing Eq.~(\ref{sieq_<pi_i|j^+i>}) results in
    \begin{align}\label{sieq_q_ji pi_i|j}
        \sum_{j=1}^n q_{j|i}\langle \pi_{i|j}^{\mathbf{k}_{+i}} \rangle
        =&\sum_{j=1}^n q_{j|i} \sum_{\sum_{i'=1}^n k_{i'}=k-1}\frac{(k-1)!}{\prod_{i'=1}^n k_{i'}!} \left(\prod_{i'=1}^n {q_{{i'}|j}}^{k_{i'}}\right)
        a_{i|\mathbf{k}_{+j}}
        +\sum_{j=1}^n q_{j|i} \sum_{\sum_{l=1}^n k'_l=k-1}\frac{(k-1)!}{\prod_{l=1}^n k'_l!} \left(\prod_{l=1}^n {q_{l|i}}^{k'_l}\right) a_{i|\mathbf{k}'_{+j}}
        \nonumber\\
        &+\sum_{j=1}^n q_{j|i} \sum_{\sum_{l=1}^n k'_l=k-1}\frac{(k-1)!}{\prod_{l=1}^n k'_l!} \left(\prod_{l=1}^n {q_{l|i}}^{k'_l}\right) 
        \sum_{l=1}^n k'_l \sum_{\sum_{\ell=1}^n k''_\ell=k-1}\frac{(k-1)!}{\prod_{\ell=1}^n k''_\ell!} \left(\prod_{\ell=1}^n {q_{\ell|l}}^{k''_\ell}\right) a_{i|\mathbf{k}''_{+l}} \nonumber\\
        =&\sum_{j=1}^n q_{j|i} \sum_{\sum_{i'=1}^n k_{i'}=k-1}\frac{(k-1)!}{\prod_{i'=1}^n k_{i'}!} \left(\prod_{i'=1}^n {q_{{i'}|j}}^{k_{i'}}\right)
        a_{i|\mathbf{k}_{+j}}
        +\sum_{j=1}^n \frac{k_j}{k} \sum_{\sum_{l=1}^n k'_l=k}\frac{k!}{\prod_{l=1}^n k'_l!} \left(\prod_{l=1}^n {q_{l|i}}^{k'_l}\right) a_{i|\mathbf{k}'}
        \nonumber\\
        &+(k-1) \sum_{\sum_{l=1}^n k'_l=k-2}\frac{(k-2)!}{\prod_{l=1}^n k'_l!} \left(\prod_{l=1}^n {q_{l|i}}^{k'_l}\right) 
        \sum_{l=1}^n q_{l|i} \sum_{\sum_{\ell=1}^n k''_\ell=k-1}\frac{(k-1)!}{\prod_{\ell=1}^n k''_\ell!} \left(\prod_{\ell=1}^n {q_{\ell|l}}^{k''_\ell}\right) a_{i|\mathbf{k}''_{+l}} \nonumber\\
        =&\sum_{j=1}^n q_{j|i} \sum_{\sum_{i'=1}^n k_{i'}=k-1}\frac{(k-1)!}{\prod_{i'=1}^n k_{i'}!} \left(\prod_{i'=1}^n {q_{{i'}|j}}^{k_{i'}}\right)
        a_{i|\mathbf{k}_{+j}}
        +\sum_{\sum_{l=1}^n k'_l=k}\frac{k!}{\prod_{l=1}^n k'_l!} \left(\prod_{l=1}^n {q_{l|i}}^{k'_l}\right) a_{i|\mathbf{k}'}
        \nonumber\\
        &+(k-1) \sum_{l=1}^n q_{l|i} \sum_{\sum_{\ell=1}^n k''_\ell=k-1}\frac{(k-1)!}{\prod_{\ell=1}^n k''_\ell!} \left(\prod_{\ell=1}^n {q_{\ell|l}}^{k''_\ell}\right) a_{i|\mathbf{k}''_{+l}} \nonumber\\
        =&\sum_{\sum_{l=1}^n k'_l=k}\frac{k!}{\prod_{l=1}^n k'_l!} \left(\prod_{l=1}^n {q_{l|i}}^{k'_l}\right) a_{i|\mathbf{k}'}
        +k \sum_{l=1}^n q_{l|i} \sum_{\sum_{\ell=1}^n k''_\ell=k-1}\frac{(k-1)!}{\prod_{\ell=1}^n k''_\ell!} \left(\prod_{\ell=1}^n {q_{\ell|l}}^{k''_\ell}\right) a_{i|\mathbf{k}''_{+l}}.
    \end{align}
    On the other hand, we apply the expression of $\pi_i^{\mathbf{k}}$ in Eq.~(\ref{sieq_pi_j}) to Eq.~(\ref{sieq_<pi_j>}) and calculate 
    \begin{align}\label{sieq_<pi_i>_calcu}
        \langle \pi_i^{\mathbf{k}} \rangle
        =&~\sum_{\sum_{i'=1}^n k_{i'}=k}\frac{k!}{\prod_{i'=1}^n k_{i'}!} \left(\prod_{i'=1}^n {q_{{i'}|i}}^{k_{i'}}\right)
        \pi_i^{\mathbf{k}} \nonumber\\
        =&~\sum_{\sum_{i'=1}^n k_{i'}=k}\frac{k!}{\prod_{i'=1}^n k_{i'}!} \left(\prod_{i'=1}^n {q_{{i'}|i}}^{k_{i'}}\right)
        a_{i|\mathbf{k}} \nonumber\\
        &~+\sum_{\sum_{i'=1}^n k_{i'}=k}\frac{k!}{\prod_{i'=1}^n k_{i'}!} \left(\prod_{i'=1}^n {q_{{i'}|i}}^{k_{i'}}\right)
        \sum_{l=1}^n k_l \sum_{\sum_{\ell=1}^n k'_\ell=k-1}\frac{(k-1)!}{\prod_{\ell=1}^n k'_\ell!} \left(\prod_{\ell=1}^n {q_{\ell|l}}^{k'_\ell}\right) a_{i|\mathbf{k}'_{+l}} \nonumber\\
        =&~\sum_{\sum_{i'=1}^n k_{i'}=k}\frac{k!}{\prod_{i'=1}^n k_{i'}!} \left(\prod_{i'=1}^n {q_{{i'}|i}}^{k_{i'}}\right)
        a_{i|\mathbf{k}}
        +k \sum_{l=1}^n q_{l|i} \sum_{\sum_{\ell=1}^n k'_\ell=k-1}\frac{(k-1)!}{\prod_{\ell=1}^n k'_\ell!} \left(\prod_{\ell=1}^n {q_{\ell|l}}^{k'_\ell}\right) a_{i|\mathbf{k}'_{+l}}.
    \end{align}
    
    Comparing Eqs.~(\ref{sieq_q_ji pi_i|j}) and (\ref{sieq_<pi_i>_calcu}) and noting that 
    $\mathbf{k}'$, $\mathbf{k}''_{+l}$, $\mathbf{k}$, and $\mathbf{k}'_{+l}$ are calculated separately here (i.e., we can treat $\mathbf{k}'$, $\mathbf{k}$ as $\mathbf{k}$ and treat $\mathbf{k}''_{+l}$, $\mathbf{k}'_{+l}$ as $\mathbf{k}_{+l}$), we have $\sum_{j=1}^n q_{j|i}\langle \pi_{i|j}^{\mathbf{k}_{+i}} \rangle=\langle \pi_i^{\mathbf{k}} \rangle$.
\end{proof}
\end{theorem}

\begin{theorem}\label{theorem_aiki_qjiaiji}
     The following equation holds: 
    \begin{equation}\label{sieq_<a_i>_i}
    \langle a_{i|\mathbf{k}}\rangle_i
    =\sum_{j=1}^{n} q_{j|i} \langle a_{i|\mathbf{k}_{+j}}\rangle_i.
\end{equation}
    Intuitive interpretation: (1) the expected single-game payoff of an $i$-player is equivalent to (2) the expected single-game payoff of an $i$-player conditioning on all possible neighboring $j$-players, where the $i$-player has at least one $j$-player neighbor.

\begin{proof}
According to the definition of $\langle a_{i|\mathbf{k}}\rangle_j$ in Eq.~(\ref{sieq_<a>}), we have 
    \begin{align}\label{sieq_<a_i>_i_proof}
    \langle a_{i|\mathbf{k}}\rangle_i
    &=\sum_{\sum_{i'=1}^n k_{i'}=k}\frac{k!}{\prod_{i'=1}^n k_{i'}!} \left(\prod_{i'=1}^n {q_{{i'}|i}}^{k_{i'}}\right)
    a_{i|\mathbf{k}} \nonumber\\
    &=\sum_{j=1}^{n}\sum_{\sum_{i'=1}^n k_{i'}=k}\frac{k!}{\prod_{i'=1}^n k_{i'}!} \left(\prod_{i'=1}^n {q_{{i'}|i}}^{k_{i'}}\right)
    \frac{k_j}{k} a_{i|\mathbf{k}} \nonumber\\
    &=\sum_{j=1}^{n} q_{j|i}\sum_{\sum_{i'=1}^n k_{i'}=k-1}\frac{(k-1)!}{\prod_{i'=1}^n k_{i'}!} \left(\prod_{i'=1}^n {q_{{i'}|i}}^{k_{i'}}\right)
    a_{i|\mathbf{k}_{+j}} \nonumber\\
    &=\sum_{j=1}^{n} q_{j|i} \langle a_{i|\mathbf{k}_{+j}}\rangle_i,
\end{align}
which completes the proof.
\end{proof}
\end{theorem}

\begin{theorem}\label{5}
    The following equation holds: 
    \begin{equation}
        \sum_{j=1}^n \sum_{i'=1}^n q_{i'|j} q_{j|i}\langle \pi_{i|j}^{\mathbf{k}_{+i,+i'}}\rangle
        =\langle \pi_i^{\mathbf{k}} \rangle.
    \end{equation}
    Intuitive interpretation: (1) the expected accumulated payoff of an $i$-player over all possible neighboring $j$-players found near an $i$-player and all possible neighboring $i'$-players found near an $j$-player, where the $j$-player has at least one $i$-player neighbor and one $i'$-player neighbor, is equivalent to (2) the expected accumulated payoff of an $i$-player. 
    
    \begin{proof}
        According to Theorem~\ref{theorem_sumpiij_pii}, we have $\langle \pi_i^{\mathbf{k}} \rangle=\sum_{j=1}^n q_{j|i}\langle \pi_{i|j}^{\mathbf{k}_{+i}} \rangle$. Furthermore, we calculate
        \begin{align}\label{sieq_piij+1_to_piij+2}
            \langle \pi_{i|j}^{\mathbf{k}_{+i}} \rangle
            =&\sum_{\sum_{i'=1}^n k_{i'}=k-1}\frac{(k-1)!}{\prod_{i'=1}^n k_{i'}!} \left(\prod_{i'=1}^n {q_{{i'}|j}}^{k_{i'}}\right)
            \pi_{i|j}^{\mathbf{k}_{+i}} \nonumber\\
            =&\sum_{\sum_{i'=1}^n k_{i'}=k-1}\frac{(k-1)!}{\prod_{i'=1}^n k_{i'}!} \left(\prod_{i'=1}^n {q_{{i'}|j}}^{k_{i'}}\right)
            \sum_{i'=1}^{n} \frac{k_{i'}}{k-1}
            \pi_{i|j}^{\mathbf{k}_{+i}} \nonumber\\
            =&\sum_{\sum_{i'=1}^n k_{i'}=k-2}\frac{(k-2)!}{\prod_{i'=1}^n k_{i'}!} \left(\prod_{i'=1}^n {q_{{i'}|j}}^{k_{i'}}\right)
            \sum_{i'=1}^{n} q_{i'|j}
            \pi_{i|j}^{\mathbf{k}_{+i,+i'}} \nonumber\\
            =&\sum_{i'=1}^{n} q_{i'|j}
            \langle \pi_{i|j}^{\mathbf{k}_{+i,+i'}}\rangle.
        \end{align}
        Applying Eq.~(\ref{sieq_piij+1_to_piij+2}) to Theorem~\ref{theorem_sumpiij_pii}, we obtain $\langle \pi_i^{\mathbf{k}} \rangle=\sum_{j=1}^n \sum_{i'=1}^n q_{i'|j} q_{j|i}\langle \pi_{i|j}^{\mathbf{k}_{+i,+i'}}\rangle$, which completes the proof.
    \end{proof}
    
\end{theorem}

\addtocontents{toc}{\protect\setcounter{tocdepth}{-1}}

\newpage

% \bibliography{refs}

\end{document}